\documentclass[dvipdfm,11pt]{revtex4}
\usepackage{graphicx}
\makeatother
\newcommand{\BF}[1]{\mbox{\boldmath$#1$}}
\def\al{\alpha}
\def\del{\delta}
\def\lam{\lambda}
\def\Gam{\Gamma}

\def\be{\begin{equation}}
\def\ee{\end{equation}}
\def\bea{\begin{eqnarray}}
\def\eea{\end{eqnarray}}
\def\la{\label}
\def\bsea{\begin{subeqnarray}}
\def\esea{\end{subeqnarray}}
\def\u{\underline}
\def\Tr{\mbox{Tr}}

\usepackage{amsmath, amssymb}
\begin{document}
\title{A FDR-preserving field theory for interacting Brownian particles: 
one-loop theory and MCT}
\author{Bongsoo Kim$^1$\footnote{Permanent address:
Department of Physics, Changwon National University, Changwon 641-773, Korea}
\footnote{bkim@ims.ac.jp}
 and Kyozi Kawasaki$^2$\footnote{tomo402000@yahoo.co.jp}}
\affiliation{${}^1$ Department of Theoretical Studies, 
Institute for Molecular Science, Okazaki 444-8585, Japan\\
${}^2$ Electronics Research Laboratory,
Fukuoka Institute Technology, Fukuoka 811-0295, Japan}
\date{\today}

\begin{abstract}
We develop a field theoretical treatment of a model of interacting Brownian particles. 
We pay particular attention to the requirement of the time reversal invariance 
and the fluctuation-dissipation relationship (FDR). 
The method used is a modified version of the auxiliary field method due 
originally to Andreanov, Biroli and Lefevre [J. Stat. Mech. P07008 (2006)]. 
We recover the correct diffusion law when the interaction is dropped as well as 
the standard mode coupling equation in the one-loop order calculation for 
interacting Brownian particle systems. 
\end{abstract}
\maketitle
\section{Introduction}
\setcounter{equation}{0}
The only existing successful first-principle theory of structural glass transition 
\cite{ediger,debene,andersen,sciortino,lubwoly}, the mode coupling theory (MCT) 
\cite{goetze,yip,das,reich,miya0}, is beset with absence of controllable approximation characterized by 
smallness parameter. Some years ago one of us attempted to remedy the situation by introducing and 
working out a dynamical fluid model  with a Kac-type long range interaction \cite{kk03} of appropriate form among elements of the reference fluid, 
which is anticipated to exhibit glassy behavior \cite{kkloh}. As is well-known, this model has a smallness parameter which 
is the inverse force range of the Kac potential measured in units of inverse microscopic length scale of the reference fluid.
However, the difficulty with this work is the inadequacy of the expansion scheme which violated the detailed balance originating from the time-reversal (TR) 
invariance of the model equation. 

Recently a great deal of attention 
\cite{szamel03,bb04,wucao,miya,bbw,cates,bbmr06,mayer,saltzman,iwss,mazenko,iwss2,bb07,bbbkmr,greenall,szamel07} 
is being paid to go beyond the standard MCT. 
This stimulated developments of satisfactory perturbative calculational methods. 
Particularly noteworthy is the work of Andreanov, Biroli and Lefevre (ABL) \cite{abl} 
where complications associated with the nonlinear TR transformation of 
the variable set (namely, the density field $[\rho]$ and its conjugate $[\hat\rho]$ 
\cite{note1}) are avoided by introducing auxiliary variable set 
$[\theta],\,[\hat\theta]$ which linearizes the TR transformation. 
The theory was applied to interacting Brownian particle system \cite{kk97}. 
The theory was also applied to continuum nonlinear fluid model\cite{dasmazenko} where the variable set had to be
 supplemented with the momentum density and velocity fields and their conjugate fields. 
In this case the theory becomes enormously complicated and will not be considered here. 
However, although ingenious is the whole approach, consequences of the theory worked out so far have yielded 
some unsatisfactory features as follows.
\begin{itemize}
\item The equation for the nonergodicity parameter gives non-trivial result 
even for non-interacting  Brownian particle systems.
\item The memory integrals entering the equation for the density-density correlation 
function are ill-behaved.
\end{itemize}

In our recent communication \cite{jpakk07}, we have proposed  a new set of auxiliary 
fields still denoted as $[\theta],\,[\hat\theta]$ which are defined slightly 
differently from ABL. However, consequences are drastically different so that 
the two unsatisfactory features mentioned above now disappear.
Here we present a detailed account of our short communication paper.
The paper is organized as follows. 
In Section II the dynamical density field model of interacting Brownian particle 
system is introduced which is expressed as an action integral 
containing the density field $[\rho]$ and its conjugate field $[\hat\rho]$. 
This action is shown to be invariant under a certain nonlinear TR transformation. 
This transformation can be converted into a linear one by introducing a conjugate 
pair of auxiliary fields $[\theta]$ and $[\hat\theta]$. 
The resulting action integral is divided into the Gaussian and non-Gaussian parts, 
each of which is separately TR-invariant. However, certain terms coming from 
the both parts cancel when summed over. 
Consideration of this fact is essential to recover a simple diffusion law 
for the nonequilibrium averaged density in non-interacting case. 
In Sections III and IV we develop a renormalized perturbation theory for interacting cases, 
and recover within one-loop order the standard MCT equation for the density-density 
correlation function by invoking the irreducible memory function approach.
Section IV summarizes the paper and gives discussion.

\section{The dynamic density functional theory}
\setcounter{equation}{0}
\subsection{The dynamic equation for the density fluctuations}
We start with the following Langevin equation for the density field
$\rho({\bf r},t)$ of interacting Brownian particles 
\be
\partial_t \rho({\bf r},t)=
\nabla \cdot \Big( \rho({\bf r},t) \nabla \frac{\delta F[\rho]}
{\delta \rho({\bf r},t)} \Big) +\eta({\bf r},t)
\la{eqn:2.1}
\ee
where the Gaussian thermal noise $\eta({\bf r},t)$
has zero mean and variance of the form
\be
<\eta ({\bf r},t)\eta ({\bf r}',t')> = 2T \nabla \cdot \nabla' \Big(
 \rho({\bf r},t)  \delta ({\bf r}-{\bf r}')\delta (t-t')\Big)
 \la{eqn:2.2}
\ee
where the Boltzmann constant $k_B$ is set to unity, and $T$ is the 
temperature of the system.
Note that the noise correlation depends on the density variable, i.e.,
the noise is multiplicative.  This is necessary for the Langevin equation (\ref{eqn:2.1}) to satisfy 
the detailed balance condition so that the system is guaranteed to evolve into the equilibrium state governed by
the free energy $F[\rho]$. In (\ref{eqn:2.1}), $F[\rho]$ is the free energy density functional
which takes the following form:
\be
F[\rho]= T \int d{\bf r} \,
 \rho ({\bf r}) \Big(\ln \frac{\rho ({\bf r})}{\rho_0}-1 \Big) 
+ \frac{1}{2} \int  d {\bf r} \int  d {\bf r}'
\, \delta \rho ({\bf r})\, U({\bf r}-{\bf r}')\,  \delta \rho ({\bf r}')
\la{eqn:2.3}
\ee
where $\delta \rho ({\bf r},t) \equiv \rho({\bf r},t)-\rho_0$
is the density fluctuation around the equilibrium density $\rho_0$.
In (\ref{eqn:2.3}) the first term is the ideal gas part \cite{ideal} of the free energy,
$F_{id}[\rho]$, and the second term the interaction part of the free energy,
$F_{int}[\rho]$.
Using Ito calculus, Dean \cite{dean} has derived
the above nonlinear Langevin equation for the (microscopic) density 
of system of interacting Brownian particles 
with pair potential $ U({\bf r})$. Earlier, Kawasaki \cite{kk94} has also obtained
the same form of Langevin equation for the coarse-grained density 
with $U({\bf r})$ replaced by $-T c ({\bf r}),\,\, c({\bf r})$ being 
the direct correlation function, by adiabatically eliminating the momentum field 
in the fluctuating hydrodynamic equations \cite{dasmazenko} of the simple dense liquids. 
For this case, the (effective) free energy density functional (\ref{eqn:2.3})
takes the Ramakrishnan-Yussouff (RY) form \cite{ry}. 
For further discussions regarding the nature (and controversy) of  the dynamic equation 
(\ref{eqn:2.1})-(\ref{eqn:2.3}), we refer to Ref. \cite{kk07}.
  
 \subsection{The dynamic action and the time-reversal invariance}
We consider the corresponding action integral ${\cal S}[\rho, \hat \rho]$ 
\cite{msr} which governs the stochastic dynamics of the coarse-grained density variable.
The dynamic action can be derived as follows.
The average of a dynamic quantity of density variable, $A[\rho]$, 
should be taken over the thermal-noise driven density fluctuations satisfying the dynamic equation 
(\ref{eqn:2.1}) and (\ref{eqn:2.2}):
\bea
&&< A[\rho] > = \int {\cal D}\rho \,\, A[\rho] \,\, \Big<
\delta \Big[ \partial_t \rho({\bf r},t)-
\nabla \cdot \Big(\rho\nabla\frac{\delta F}{\delta \rho} \Big)
-\eta({\bf r},t) \Big] \Big>_{\eta} \nonumber \\
&=& \int {\cal D}\rho \int {\cal D}\hat\rho \,\, A[\rho] \,\,
\exp \Big( \int d{\bf r} \int dt \,\,  i\hat\rho 
 \big[\partial_t \rho-\nabla \cdot \big(\rho\nabla\frac{\delta F}{\delta \rho}\big)
 \big] \Big) \Big< \exp \Big(-\int d{\bf r} \int dt \,\,
i\hat\rho({\bf r},t) \eta ({\bf r},t) \Big) \Big>_{\eta} \nonumber \\
&=& \int {\cal D} \rho \int {\cal D}\hat\rho \,\, A[\rho] \,\,
\exp\Big( {\cal S}[\rho,\hat\rho]\Big), \nonumber \\
{\cal S}[\rho, \hat \rho] &\equiv& \int \, d {\bf r} \int dt \,
 \Big\{ i\hat \rho \Big[ \partial_t \rho
-\nabla \cdot \Big( \rho \nabla \frac{\delta F}{\delta \rho} \Big)  \Big]
 -T \rho ( \nabla \hat \rho)^2 \Big\}
 \la{eqn:2.4}
\eea
where the auxiliary field $\hat\rho$ is real and the
last term involving quadratic $\hat\rho$ comes from the average over
 multiplicative thermal noise $\eta$. 
 In the first line of (\ref{eqn:2.4}), employing  the Ito calculus  
makes the Jacobian of transformation constant.  
The dynamic action of this form appearing in the above equation with the RY free energy functional was 
first written down in \cite{kk97}. 

The TR symmetry of the dynamics should be manifested in the dynamic
action. In particular, under the TR the two fields $\rho$ and $\hat\rho$ should transform in such a way that the dynamic action (\ref{eqn:2.4}) remains invariant under these transformations. 
In order to see this TR invariance of the action,  
one can rearrange the dynamic action (\ref{eqn:2.4}) as 
\bea
{\cal S}[\rho, \hat \rho] &=& \int \, d {\bf r} \int dt \,
 \Big\{ i\hat \rho \Big[ \partial_t \rho
-\nabla \cdot \Big( \rho \nabla \frac{\delta F}{\delta \rho} \Big)  \Big]
 -T i  {\hat\rho} \nabla \cdot ( \rho \nabla i\hat \rho) \Big\} \nonumber \\
&=&  \int \, d {\bf r} \int dt \,
 \Big\{ i\hat \rho \Big[ \partial_t \rho
+iT\nabla \cdot \Big( \rho \nabla \Big(-\hat \rho
+\frac{i}{T}\frac{\delta F}{\delta \rho} \Big)\Big)  \Big] \Big\}
\la{eqn:2.5}
\eea
This last form of the action suggests that the dynamic action becomes 
invariant under the field transformation 
\bea
\quad \rho ({\bf r}, -t)& = & \rho ({\bf r}, t) \nonumber \\
\hat \rho ({\bf r}, -t) & = & -\hat \rho ({\bf r}, t)
+ \frac{i}{T} \frac{\delta F}{\delta \rho ({\bf r},t)}
\la{eqn:2.6}
\eea
The invariance of the action under the transformation (\ref{eqn:2.6}) can be shown as follows. Using the transformation of $\hat \rho $ in (\ref{eqn:2.6})
one can rewrite (\ref{eqn:2.5}) as 
\be
{\cal S}[\rho, \hat \rho]  
= \int \, d {\bf r} \int dt \, i\hat\rho({\bf r},t) \partial_t \rho({\bf r},t)
+ T \int \, d {\bf r} \int dt \, i\hat\rho({\bf r},t) 
\nabla \cdot \Big( \rho({\bf r},t) \nabla i \hat\rho( {\bf r}, -t)\Big)
\la{eqn:2.7}
\ee
With the transformation (\ref{eqn:2.6}) the first integral in (\ref{eqn:2.7}) is shown to be TR invariant as 
\bea
&&\int \, d {\bf r} \int dt \,\, i\hat\rho({\bf r},t) \partial_t \rho({\bf r},t)
=-\int \, d {\bf r} \int dt \,\, i\hat\rho({\bf r},-t) \partial_t \rho({\bf r},t)
\nonumber \\
&=&-\int \, d {\bf r} \int dt \,\, i \Big( -\hat \rho ({\bf r}, t)
+ \frac{i}{T} \frac{\delta F}{\delta \rho ({\bf r},t)}\Big) \partial_t \rho({\bf r},t)
= \int \, d {\bf r} \int dt \,\, i \hat\rho({\bf r},t) \partial_t \rho({\bf r},t)
+ \frac{1}{T}\int dt \, \partial_t F[\rho] \nonumber \\
&=& \int \, d {\bf r} \int dt \, \,i \hat\rho({\bf r},t) \partial_t \rho({\bf r},t)
\la{eqn:2.8}
\eea
where the first line results from the change of integration variable $t \rightarrow -t$, and  we used the fact that the surface term $\int dt \, \partial_t F[\rho]/T $ vanishes. The second term in (\ref{eqn:2.7}) is manifestly TR invariant since
when the time is reversed the spatial integration by parts twice recovers its original form. This proves the invariance of the action (\ref{eqn:2.4})
under the TR transformation (\ref{eqn:2.6}).
Note that the TR invariance  holds for the generalized form of dynamic action
\be
{\cal S}[\rho, \hat \rho] = \int \, d {\bf r} \int dt \,
 \Big\{ i\hat \rho \Big[ \partial_t \rho
-\nabla \cdot \Big( {\cal D}(\rho) \nabla \frac{\delta F}{\delta \rho} \Big)  \Big]
 -T i  {\hat\rho} \nabla \cdot ({\cal D}(\rho) \nabla i\hat \rho) \Big\} 
\la{eqn:2.9}
\ee
where ${\cal D}\big(\rho({\bf r},t)\big)$ is a local function of the density field.
This action corresponds to the dynamic equation of the form \cite{mazenko}
\bea
\partial_t \rho({\bf r},t)&=&
\nabla \cdot \Big( {\cal D} \big(\rho({\bf r},t)\big)) \nabla \frac{\delta F[\rho]}
{\delta \rho({\bf r},t)} \Big) +\eta({\bf r},t), \nonumber \\
<\eta ({\bf r},t)\eta ({\bf r}',t')> &=& 2T \nabla \cdot \nabla' \Big(
 {\cal D}\big(\rho({\bf r},t)\big)  \delta ({\bf r}-{\bf r}')\delta (t-t')\Big)
 \la{eqn:2.10}
\eea

There exists another field-transformation  leading to the time-reversal invariance of 
the action. To see this, one can again rearrage the dynamic action 
(\ref{eqn:2.4}) as 
\be
{\cal S}[\rho, \hat \rho] = \int \, d {\bf r} \int dt \,
 \Big\{ i\hat \rho({\bf r},t) (-iT)\Big( \frac{i}{T}\partial_t \rho({\bf r},t)
 + \nabla \cdot ( \rho({\bf r},t) \nabla \hat \rho({\bf r},t)) \Big)
-i\hat \rho({\bf r},t) \nabla \cdot \Big( \rho({\bf r},t) 
\nabla \frac{\delta F}{\delta \rho} \Big)  \Big\} 
\la{eqn:2.11}
\ee
This form of the action suggests the following transformation  
\bea
\rho({\bf r},-t)&=&\rho({\bf r},t) \nonumber \\
\nabla \cdot ( \rho({\bf r},t) \nabla \hat\rho({\bf r},-t))&=&
\frac{i}{T} \partial_t \rho({\bf r},t)
 + \nabla \cdot ( \rho({\bf r},t) \nabla \hat\rho({\bf r},t))
\la{eqn:2.12}
\eea
or equivalently, 
\bea
\rho({\bf r},-t)&=&\rho({\bf r},t) \nonumber \\
\hat\rho({\bf r},-t)&=& \hat\rho({\bf r},t)+i A(\rho({\bf r},t))
\la{eqn:2.13}
\eea
with the local function $A(\rho({\bf r},t))$ defined as 
\be
\nabla \cdot ( \rho({\bf r},t) \nabla A(\rho({\bf r},t)))
=\frac{1}{T}\partial_t \rho({\bf r},t)
\la{eqn:2.14}
\ee
One can rewrite  (\ref{eqn:2.11}) using the transformation (\ref{eqn:2.12}) as 
\be
{\cal S}[\rho, \hat \rho] = \int \, d {\bf r} \int dt \,
 \Big\{ i\hat \rho({\bf r},t)(-T) \Big( 
 \nabla \cdot ( \rho({\bf r},t) \nabla i\hat \rho({\bf r},-t)) \Big)
-i\hat \rho({\bf r},t) \nabla \cdot \Big( \rho({\bf r},t) 
\nabla \frac{\delta F}{\delta \rho} \Big)  \Big\} 
\la{eqn:2.15}
\ee
We have previously noted that the first term is manifestly TR invariant.
We thus only need to look at the last term in (\ref{eqn:2.15}):
\bea
&& -\int \, d {\bf r} \int dt \, 
i\hat \rho({\bf r},t) \nabla \cdot \Big( \rho({\bf r},t) 
\nabla \frac{\delta F}{\delta \rho} \Big) 
=-\int \, d {\bf r} \int dt \, 
i\hat \rho({\bf r},-t) \nabla \cdot \Big( \rho({\bf r},t) 
\nabla \frac{\delta F}{\delta \rho} \Big) \nonumber \\
&=& -\int \, d {\bf r} \int dt \, 
 i\frac{\delta F}{\delta \rho} \nabla \cdot \Big( \rho({\bf r},t)
\nabla \hat \rho({\bf r},-t) \Big) 
= -\int \, d {\bf r} \int dt \, 
 i\frac{\delta F}{\delta \rho} \Big[\frac{i}{T}\partial_t \rho({\bf r},t)
 + \nabla \cdot ( \rho({\bf r},t) \nabla \hat\rho({\bf r},t)) \Big] \nonumber \\
&=& -\int \, d {\bf r} \int dt \, 
i\hat \rho({\bf r},t) \nabla \cdot \Big( \rho({\bf r},t) 
\nabla \frac{\delta F}{\delta \rho} \Big)
\la{eqn:2.16}
\eea 
Therefore the dynamic action (\ref{eqn:2.4}) is TR invariant under 
the second type of the transformation (\ref{eqn:2.12}) as well.
The invariance of the action under the second transformation will also hold for 
the generalized form of the dynamic action (\ref{eqn:2.9}).
As  will be shown in the next subsection, the FDR is readily derived from
these TR transformations. 

Note that both  (\ref{eqn:2.6}) and 
(\ref{eqn:2.12}) are intrinsically nonlinear transformations. 
The equation (\ref{eqn:2.6}) is nonlinear because of the noninteracting
contribution $F_{id}$, whereas the transformation (\ref{eqn:2.12}) is nonlinear 
  due to the 'extra' factor of the density field in our dynamic equation 
(i.e. the multiplicative nature of the Langevin equation). 
As discussed in the work of ABL \cite{abl},  
this nonlinearity is the underlying reason why the FDR, obeyed by the action, is not preserved 
order by order in the renormalized perturbation theory developed 
for  the dynamic action (\ref{eqn:2.4}). 

\subsection{Fluctuation-dissipation relation (FDR)}
The FDR is a hallmark of the equilibrium dynamics, which provides 
a fundamental relationship between the correlation of equilibrium fluctuations and the linear response 
to external perturbation. Here it is shown that the FDR is a direct consequence of the TR symmetry.

The response function $R({\bf r},t; {\bf r}' t')$ is defined as a link between
induced density change $\Delta <\rho({\bf r},t)>$ and
an external  infinitesimal field  $h_e({\bf r}',t')$ added to $F$ (not to
the Langevin equation):
\be
\Delta <\rho( {\bf r},t)> \equiv \int d {\bf r}' \,
\int dt'  \,  R({\bf r},t; {\bf r}' t') \,  h_e({\bf r}',t')
\la{eqn:2.17}
\ee
The contribution of the external field to the free energy,
$\Delta F \equiv -\int d{\bf r} \int dt \,
\delta \rho({\bf r},t) h_e({\bf r},t)$, 
 will bring the corresponding change in the action, $\Delta {\cal S}$, 
which is given by
\be
\Delta {\cal S}= \int d {\bf r} \int dt \, i \hat\rho({\bf r},t)
\nabla \cdot \big(\rho({\bf r},t) \nabla h_e({\bf r},t)  \big)
\la{eqn:2.18}
\ee
The induced density change is then given by
\bea
\Delta <\rho( {\bf r},t)> 
&=& \Big< \rho( {\bf r},t) \Delta {\cal S} \Big>
= i\int d {\bf r}' \int dt'\,
\Big<\rho( {\bf r},t) 
\nabla' \cdot \Big( \rho( {\bf r}',t')
\nabla' {\hat \rho}( {\bf r}',t') \Big) \Big>  h_e( {\bf r}',t')
\la{eqn:2.19}
\eea 
where the integration by parts was performed twice.
We see from (\ref{eqn:2.17}) and (\ref{eqn:2.19}) that 
the response function $R( {\bf r},t; {\bf r}' t')$ is given by \cite{miya}
\be
R({\bf r},t; {\bf r}' t') =i\Big<  \rho( {\bf r},t) \nabla' 
\cdot \Big( \rho( {\bf r}',t') \nabla' {\hat \rho}( {\bf r}',t') \Big) \Big>
\la{eqn:2.20}
\ee
Note that the form of the response function differs from the conventional form of the response function 
$i <\rho({\bf r},t)\nabla'^2{\hat \rho}( {\bf r}',t')>$ which holds
 for the Langevin dynamics with additive noise: (\ref{eqn:2.20})  reflects the multiplicative nature of 
the original Langevin equation (\ref{eqn:2.1}) and (\ref{eqn:2.2}).
 %\cite{miya,abl}.
 
 Now we show that the FDR follows directly from  the TR transformations.
 We use the following identity 
\bea 
0&=&\Big< \rho ({\bf r},t) \frac{\delta {\cal S}}
{\delta {\hat \rho}({\bf r}',t')} \Big>
= \Big< \rho ({\bf r},t)\Big[i\partial_{t'} \rho ({\bf r}',t')
-i\nabla' \cdot \Big(  \rho ({\bf r}',t')  \nabla'
\frac{\delta F}{\delta \rho ({\bf r}',t')}  \Big) \nonumber \\
 &+& T \nabla' \cdot \Big( \rho ({\bf r}',t') \nabla'
{\hat \rho} ({\bf r}',t') \Big) + T \nabla' \cdot \Big( \rho ({\bf r}',t') \nabla'
{\hat \rho} ({\bf r}',t') \Big) \Big] \Big> \nonumber \\
&=&\Big< \rho ({\bf r},t)\Big[i\partial_{t'} \rho ({\bf r}',t')
-T\nabla' \cdot \Big(  \rho ({\bf r}',t')  \nabla'
{\hat \rho} ({\bf r}',-t') \Big) 
+ T \nabla' \cdot \Big( \rho ({\bf r}',t') \nabla'
{\hat \rho} ({\bf r}',t') \Big) \Big] \Big>
\la{eqn:2.21}
\eea
where the TR transformation (\ref{eqn:2.6}) was used.
The equations (\ref{eqn:2.20}) and  (\ref{eqn:2.21}) give the FDR \cite{deker}
\be
-\frac{1}{T} \partial_{t} G_{\rho \rho}({\bf r}-{\bf r}', t-t')
=-R({\bf r}-{\bf r}', t'-t) + R({\bf r}-{\bf r}', t-t')
\la{eqn:2.22}
\ee
where $G_{\rho \rho}({\bf r}-{\bf r}', t-t')\equiv
\big< \delta \rho ({\bf r},t) \delta \rho ({\bf r}',t')  \big>$ 
is the density correlation function. 
Since the causality requires $ R({\bf r}-{\bf r}', t'-t)=0$ for $t > t'$,  
(\ref{eqn:2.22}) leads to the standard form of the FDR
\be
R({\bf r}-{\bf r}', t-t')=-\Theta(t-t')\frac{1}{T}\partial_t
G_{\rho \rho}({\bf r}-{\bf r}', t-t')
\la{eqn:2.23}
\ee
where $\Theta(t)$ is the Heaviside step function.
The FDR (\ref{eqn:2.22}) is more directly obtained from the second transformation 
(\ref{eqn:2.12}) 
when the second member of  (\ref{eqn:2.12}) (denoting the space and 
time coordinates ${\bf r}'$ and $t'$, respectively) is multiplied 
 by $\rho({\bf r},t)$ and is taken average.

\subsection{Linearization of the time-reversal transformation} 
From now on, we focus on the first TR transformation (\ref{eqn:2.6}). 
Since the nonlinearity of the TR transformation makes the perturbation expansion inconsistent with
the FDR \cite{abl}, this inconsistency would be  resolved if the transformation is properly linearized. 
With the form of the free energy given in (\ref{eqn:2.3}), 
one can explicitly write $(1/T)\delta F/\delta \rho$  as
\bea
\frac{1}{T} \frac{\delta F_{id}[\rho]}{\delta \rho ({\bf r},t)}
 &=& \ln \frac{\rho ({\bf r},t)}{\rho_0}
 \equiv \frac{\delta \rho ({\bf r},t)}{\rho_0}+ f(\delta \rho ({\bf r},t)), \nonumber \\
\frac{1}{T} \frac{\delta F_{int}[\rho]}{\delta \rho ({\bf r},t)}
&=& \frac{1}{T} \int \, d {\bf r}'
 U({\bf r}-{\bf r}') \delta \rho ({\bf r}',t),  \nonumber \\
 f(\delta \rho({\bf r},t)) &\equiv& -\sum_{n=2}^{\infty}
\frac{1}{n} \big(-\delta \rho ({\bf r},t)/\rho_0 \big)^n
\la{eqn:2.24}
\eea 
where $f(\delta \rho({\bf r},t))$ is the contribution of the non-Gaussian (higher than the quadratic)  part of $F_{id}[\rho]$. 
The equation (\ref{eqn:2.24}) leads to an explicit form of the TR transformation 
(\ref{eqn:2.6}) as
\bea
\rho ({\bf r}, -t) &=&  \rho ({\bf r}, t) \nonumber \\
\hat \rho ({\bf r}, -t) &=&  -\hat \rho ({\bf r}, t) 
+i{\hat K} \ast \delta \rho({\bf r},t)+if(\delta \rho({\bf r},t)), \nonumber \\
{\hat K} \ast \delta \rho({\bf r},t)&\equiv& 
\int d {\bf r}'\, K({\bf r}-{\bf r}')\delta \rho({\bf r}',t)
\la{eqn:2.25}
\eea
where the kernel $ K({\bf r})$ is defined as 
 $K({\bf r})\equiv \big(\delta({\bf r})/\rho_0+U({\bf r})/T \big)$.
Note that the transformation (\ref{eqn:2.25}) is nonlinear due to 
the non-Gaussian nature of $F_{id}[\rho]$, the ideal-gas  part of the free energy.
%\cite{ideal}.

\subsubsection{The Gaussian approximation}
If one entirely neglects $f(\delta \rho({\bf r},t))$ in (\ref{eqn:2.24}), 
this is tantamount to approximating $F_{id}[\rho]$ to the Gaussian form 
\be
F^G_{id}[\rho] \simeq \frac{T}{2\rho_0} \int d{\bf r} \, (\delta \rho ({\bf r},t))^2
\la{eqn:2.26}
\ee
In this case, the transformation (\ref{eqn:2.25}) becomes linear and consequently the FDR would be 
preserved by the perturbation theory  order by order. 
However, when (\ref{eqn:2.26}) is substituted to the original equation (\ref{eqn:2.1}), the dynamic equation 
generates the following two terms
\be
\nabla \cdot \Big( \rho({\bf r},t) \nabla \frac{\delta F^G_{id}[\rho]}
{\delta \rho({\bf r},t)} \Big) 
=T\nabla^2 \rho({\bf r},t)+
\frac{T}{\rho_0} \nabla \cdot \big( \delta \rho  \nabla \rho \big)
\la{eqn:2.27}
\ee 
While the first term is the linear diffusion term, the term which is expected 
for the noninteracting case, the second nonlinear term gives rise to a spurious 
contribution, incorrectly yielding a nontrivial result even in the absence of 
particle interaction \cite{miya,abl,sdd}. In order to obtain the correct behavior for the noninteracting system,
 one really should take into account the full logarithmic form of $F_{id}[\rho]$. 

\subsubsection{Introduction of the auxiliary fields}
Fully incorporating $F_{id}[\rho]$ without making truncation, 
a natural way to make the transformation (\ref{eqn:2.25}) linear is to 
 introduce a new field $\theta({\bf r},t)$ defined as
\be
\theta ({\bf r},t) \equiv f(\delta \rho({\bf r},t)) =
\frac{1}{T} \frac{\del F_{id}}{\del \rho}-\frac{\del \rho}{\rho_0}
\la{eqn:2.28}
\ee
Note that the definition (\ref{eqn:2.28}) differs from that of ABL in that whereas 
in the work of ABL, the auxiliary field is defined as the functional derivative 
of the full free energy with respect to density
\be
\theta_{ABL}({\bf r},t)
 \equiv \frac{1}{T} \delta F/\delta \rho ({\bf r},t)=
{\hat K} \ast \delta \rho({\bf r},t)+f(\delta \rho({\bf r},t)),
 \la{eqn:2.29}
 \ee
(\ref{eqn:2.28}) limits the new variable $\theta({\bf r},t)$ to the
nonlinear part of the transformation. 
%As shown in subsection G, there is a fundamental difference 
%in the physical consequence of the two approaches (\ref{eqn:2.28}) and 
%(\ref{eqn:2.29}): ABL approach does not lead to the anticipated correct dynamic 
%behavior for the noninteracting system.

With the nonlinear constraint (\ref{eqn:2.28}), using the first member of
 (\ref{eqn:2.24}) we obtain the ideal-gas contribution to  the body force as
\be
\nabla \cdot \Big( \rho \nabla \frac{\delta F_{id}}
{\delta \rho} \Big)=T\nabla \cdot \Big(\rho \nabla \Big(\frac{\del \rho}
{\rho_0}+\theta \Big)  \Big) 
 = T\nabla^2 \rho +  \frac{T}{\rho_0} 
\nabla \cdot \big( \delta \rho  \nabla \rho \big)
 +\rho_0 T \nabla^2 \theta 
+T \nabla \cdot \big( \delta \rho  \nabla \theta \big)
\la{eqn:2.30}
\ee
We have seen that the first two terms are the contributions from the Gaussian part of $F_{id}$.
On the other hand, since due to cancellation of the two nonlinear effects the entire ideal-gas contribution 
to the dynamics should be of pure diffusion 
$$
\nabla \cdot \Big( \rho \nabla \delta F_{id}/\delta \rho \Big)
=T\nabla \cdot \Big( \rho \nabla \ln (\rho /\rho_0) \Big)= T\nabla^2 \rho, 
$$
 the sum of the last three terms in (\ref{eqn:2.30}) should vanish:
\be
 \frac{T}{\rho_0} \nabla \cdot \big( \delta \rho  \nabla \rho \big)
 +\rho_0 T \nabla^2 \theta +
T \nabla \cdot \big( \delta \rho  \nabla \theta \big)=0
\la{eqn:2.31}
\ee
As shown in the subsection G, this cancellation is used to obtain
the correct dynamic behavior for the noninteracting case.

\subsection{New form of the dynamic action}
With the constraint (\ref{eqn:2.28}), the action (\ref{eqn:2.4}) takes the following 
new form 
\bea 
{\cal S}[\rho, \hat \rho]
 &=& \int \, d {\bf r} \int dt \,
 \Big\{ i\hat \rho \Big[ \partial_t \rho 
-\nabla \cdot \Big( \rho \nabla \frac{\delta F_{id}}{\delta \rho}\Big)
  -\nabla \cdot \Big( \rho \nabla \frac{\delta F_{int}}{\delta \rho}\Big)\Big]
 -T \rho ( \nabla \hat \rho)^2 \Big\} \nonumber \\
 &{}& \nonumber \\
 \rightarrow  \quad {\cal S}[\rho, \hat \rho, \theta, \hat\theta]
&=& \int \, d {\bf r} \int dt \,
 \Big\{ i\hat\rho \Big[ \partial_t \rho 
-T\nabla \cdot \Big( \rho \nabla \Big(\frac{\delta \rho}{\rho_0}+\theta \Big) \Big)
  -T\nabla \cdot \Big( \rho \nabla {\hat U}\ast \delta \rho\Big)\Big] \nonumber \\
  &-&T \rho ( \nabla \hat \rho)^2 + i\hat\theta \Big(\theta-f(\delta \rho)\Big)
  \Big\}
\la{eqn:2.32}
\eea
where the last  term comes from the exponentiation of the delta functional
$\delta \Big[ \theta({\bf r},t)-f(\rho({\bf r},t)\Big]$. 

We now identify the {\em linear}  transformation 
under which the new action ${\cal S}[\rho, \hat \rho, \theta, \hat\theta]$ 
becomes invariant. The transformation of $\rho({\bf r},t)$
 and $\hat\rho({\bf r},t)$ are already given in (\ref{eqn:2.25}) with $f(\delta \rho({\bf r},t))$ being replaced by $\theta({\bf r},t)$.  
The transform of $\theta({\bf r},t)$ must be the same as that of 
$\rho({\bf r},t) $ since $\theta$ is a local function of $\rho$.
 Thus we only need to know how $\hat\theta({\bf r},t)$ should transform. 
Since as shown in (\ref{eqn:2.7}), the terms involving 
$\hat \rho ({\bf r},-t)$ becomes manifestly TR invariant, we only have to consider
the term $v_a({\bf r},t) \equiv i\hat\rho ({\bf r},t)\partial_t \rho({\bf r},t)$,
 together with the term $v_b({\bf r},t) \equiv i\hat\theta({\bf r},t)\theta({\bf r},t)$. When time is reversed, $v_a({\bf r},-t)$ generates a term 
$\theta({\bf r},t) \partial_t \rho({\bf r},t)$.
This term should be cancelled by the term generated by the $v_b({\bf r},-t)$.
This requires that  $\hat\theta$ transform as
$\hat\theta({\bf r},-t)=\hat\theta({\bf r},t)+i\partial_t \rho ({\bf r},t)$. 
Thus the linear transformation under which the new 
action (\ref{eqn:2.32}) becomes invariant is given by 
\bea
 \quad \rho ({\bf r}, -t)& = & \rho ({\bf r}, t) \nonumber \\
\hat \rho ({\bf r}, -t) & = & -\hat \rho ({\bf r}, t) +
 i {\hat K} \ast \delta \rho({\bf r},t)+i\theta({\bf r},t)  \nonumber \\
\theta ({\bf r}, -t) & = & \theta ({\bf r}, t) \nonumber \\
\hat \theta ({\bf r}, -t) & = & \hat \theta ({\bf r}, t)
+i \partial_t \rho ({\bf r},t)
\la{eqn:2.33}
\eea
It is easy to show that  the modulus of the associated transformation
 matrix $O$ is unity ($\det O=-1$). 
 
The new action ${\cal S}[\rho, \hat \rho, \theta, \hat\theta] $ can be 
decomposed into the Gaussian part ${\cal S}_g[\rho, \hat \rho, \theta, \hat\theta]$ and the non-Gaussian part
${\cal S}_{ng}[\rho, \hat \rho, \theta, \hat\theta]$ \cite{free}:
 \bea 
{\cal S}[\rho, \hat \rho, \theta, \hat\theta]
 &\equiv& {\cal S}_g[\rho, \hat \rho, \theta, \hat\theta] 
+{\cal S}_{ng}[\rho, \hat \rho, \theta, \hat\theta] \nonumber \\
{\cal S}_g[\rho, \hat \rho, \theta, \hat\theta]  &\equiv&
\int d{\bf r} \int dt \, \Big\{ i\hat\rho \Big[ \partial_t \rho
-T \nabla^2 \rho -\u{\rho_0 T \nabla^2 \theta}
-\rho_0 \nabla^2 {\hat U} \ast \delta \rho \Big]
 -T \rho_0 ( \nabla \hat \rho)^2 +  i\hat \theta \theta \Big\} \nonumber \\
{\cal S}_{ng} [\rho, \hat \rho, \theta, \hat\theta]  &\equiv&
\int d{\bf r} \int dt \, \Big\{ i\hat \rho \Big[
 -\nabla \cdot \Big(\delta \rho \nabla {\hat U} \ast \delta \rho \Big)
- \u{\frac{T}{\rho_0} \nabla \cdot \big( \delta \rho \nabla \rho \big)}
- \u{T \nabla \cdot \big( \delta \rho \nabla \theta \big)} \Big] \nonumber \\
&{}&- \, T \delta \rho ( \nabla \hat \rho)^2 - i\hat \theta f(\delta\rho) \Big\}
\la{eqn:2.34}
\eea
The actions  ${\cal S}_g[\rho, \hat \rho, \theta, \hat\theta]$
and ${\cal S}_{ng}[\rho, \hat \rho, \theta, \hat\theta]$ are 
shown to be {\em separately} invariant under the transformation (\ref{eqn:2.33}). 
This is apparent since the transformation (\ref{eqn:2.33}) is linear in 
$\delta \rho$, $\hat \rho$, $\theta$, and $\hat \theta$.
Though with the constraint (\ref{eqn:2.28}) the three underlined terms in (\ref{eqn:2.34}) vanish when 
summed together, one should keep each of them in the explicit calculation of renormalized perturbation theory 
since their presence is crucial for the separate invariance of the  actions 
${\cal S}_g[\rho, \hat \rho, \theta, \hat\theta]$ and ${\cal S}_{ng}[\rho, \hat \rho, \theta, \hat\theta]$ which
  enables one to construct the FDR-preserving renormalized perturbation theory 
from these actions. Nevertheless we explicitly show in the one-loop order that 
the ultimate effect of these three underlined terms is their cancellation. 
We also note that the linearization of the TR transformation inevitably brings back 
the non-polynomial nonlinearity $-i\hat\theta f(\delta \rho) $ in the action. 

\subsection{New form of the response function}
The presence of the external infinitesimal field $h_e({\bf r},t)$ leads to
shifting $f(\delta \rho({\bf r},t))$ to $f(\delta \rho({\bf r},t))-h_e({\bf r},t)/T$ in (\ref{eqn:2.25}). 
This gives rise to a change in action $\Delta {\cal S}[\rho, \hat \rho, \theta, \hat\theta]= 
\int d{\bf r} \int dt \, i{\hat \theta}( {\bf r},t)  h_e({\bf r},t)/T$.
Then
 \be
 \Delta <\rho ( {\bf r},t)> = \Big< \rho ( {\bf r},t)\Delta 
{\cal S}[\rho, \hat \rho, \theta, \hat\theta] \Big>
  = \int d{\bf r}' \int dt' \, \frac{i}{T} \Big< \delta \rho( {\bf r},t)\,
 {\hat \theta}( {\bf r}',t') \Big> h_e({\bf r}',t')
 \la{eqn:2.35}
 \ee
Therefore with the new form of the action (\ref{eqn:2.32}),   
the response function takes the form 
\be
R({\bf r},t; {\bf r}',t')=\frac{i}{T} \Big< \delta \rho( {\bf r},t)\,
 {\hat \theta}( {\bf r}',t') \Big>
 \la{eqn:2.36}
\ee
Taking correlation of the last member of (\ref{eqn:2.33}) 
with $i\delta \rho({\bf r},t)/T$ we obtain
 \be
\frac{i}{T} \Big< \delta \rho( {\bf r},t)\,
 {\hat \theta}( {\bf r}',-t') \Big> =
\frac{i}{T} \Big< \delta \rho( {\bf r},-t)\,
 {\hat \theta}( {\bf r}',-t') \Big> = 
 \frac{i}{T} \Big< \delta \rho( {\bf r},t)\,
 {\hat \theta}( {\bf r}',t') \Big>-\frac{1}{T}\partial_{t'} C_{\rho \rho}
 ({\bf r},t; {\bf r}',t')
\la{eqn:2.37}
\ee
which is the FDR (\ref{eqn:2.22}).

\subsection{Dynamics for the noninteracting case: nonperturbative result}
The noninteracting case ($U=0$) is an important guide in dealing with the $\theta$ field
 since one has to recover the linear diffusion law when $U=0$. 
%This case shows a fundamental difference between our approach and the ABL approach.  
In the absence of the particle interaction, the full action (\ref{eqn:2.32}) 
reduces to ${\cal S}_{id}[\psi] \equiv {\cal S}[\psi;  U=0]$
 \bea 
{\cal S}_{id}[\psi]  &\equiv&
\int d{\bf r} \int dt \, \Big\{ i\hat\rho \Big[ \partial_t \rho
-T \nabla^2 \rho -\u{\rho_0 T \nabla^2 \theta} 
- \u{\frac{T}{\rho_0} \nabla \cdot \big( \delta \rho \nabla \rho \big)}
- \u{T \nabla \cdot \big( \delta \rho \nabla \theta \big)}\Big] \nonumber \\
 &-& T \rho_0 ( \nabla \hat \rho)^2 - T \delta \rho ( \nabla \hat \rho)^2
 +i\hat\theta \theta
- i\hat \theta f(\delta\rho) \Big\}
\la{eqn:2.38}
\eea
where $\psi \equiv (\delta \rho, \hat\rho, \theta, \hat\theta)$.
We show that the action ${\cal S}_{id}[\psi]$ 
yields the dynamic behavior consistent for the noninteracting system.
We make use of the following identities
\be
\Big<\delta \rho({\bf 2}) \frac{\delta {\cal S}_{id}[\psi]}
{\delta {\hat \rho} ({\bf 1})} \Big> =0,
\qquad
\Big< \delta \rho({\bf 2}) \frac{\delta {\cal S}_{id}[\psi]}
{\delta  \theta ({\bf 1})} \Big> =0
\la{eqn:2.39}
\ee
where ${\bf 1} \equiv ({\bf r}, t)$ and ${\bf 2} \equiv ({\bf 0}, 0)$.
The first identity can be written explicitly as
\be
0=\Big<\delta \rho({\bf 2}) \frac{\delta {\cal S}_{id}[\psi]}
{\delta \hat \rho ({\bf 1})} \Big>
= i\Big(\frac{\partial}{\partial t} - T \nabla^2  \Big)G_{\rho \rho}
({\bf1}-{\bf 2}) +2T\rho_0 \nabla^2 \big< \hat \rho({\bf 1})
 \delta \rho({\bf 2}) \big> 
+ 2T \big< \delta \rho ({\bf 2}) \nabla \cdot \big(\delta \rho ({\bf 1}) \nabla
\hat \rho ({\bf 1}) \big)\big>
\la{eqn:2.40}
\ee
where we used the fact that the sum of the three underlined terms in
(\ref{eqn:2.38}) vanishes (see (\ref{eqn:2.31})).
Similarly, using the second identity in (\ref{eqn:2.39}),
we obtain
\be
0=\Big<\delta \rho({\bf 2}) \frac{\delta
{\cal S}_{id}[\psi]}{\delta \theta ({\bf 1})} \Big>
=-i\rho_0 T \nabla^2 \big< \hat \rho ({\bf 1}) \delta \rho({\bf 2}) \big>
+i\big<\hat \theta({\bf 1}) \delta \rho({\bf 2}) \big> 
-i T \big< \delta \rho ({\bf 2}) \nabla \cdot \big(\delta \rho ({\bf 1}) \nabla
\hat \rho ({\bf 1}) \big)\big>
\la{eqn:2.41}
\ee
where cancellation of the underlined terms was not used.
Since in (\ref{eqn:2.41}), 
$\big< \hat \rho ({\bf 1}) \delta \rho({\bf 2}) \big>
=\big<\hat \theta({\bf 1}) \delta \rho({\bf 2}) \big>=0$ 
for $t > 0$ by causality, we obtain
\be
\big< \delta \rho ({\bf 2}) \nabla \cdot \big(\delta \rho ({\bf 1}) \nabla
\hat \rho ({\bf 1}) \big)\big>=0   \quad \mbox{for} \quad t >0
\la{eqn:2.42}
\ee
The eqs. (\ref{eqn:2.40}), (\ref{eqn:2.42}), and causality lead to 
\be
\partial_t G_{\rho \rho}({\bf r},t)=T\nabla^2 G_{\rho \rho}({\bf r},t), \quad
\mbox{for} \quad t>0.
\la{eqn:2.43}
\ee
This result, the diffusion equation for the density correlation function, 
is the anticipated dynamic behavior for the noninteracting system$\diamondsuit $\cite{ideal2}$\diamondsuit $.  

On the other hand, if one employs the ABL approach ((\ref{eqn:2.29})), 
$$\theta_{ABL}({\bf r},t)\equiv\frac{1}{T}\frac{\del F}{\del \rho({\bf r},t)}
=\hat K*\del \rho({\bf r},t)+f(\delta \rho({\bf r},t)),$$ 
one obtains the following action
\be
{\cal S}_{ABL}[\psi]
= \int \, d {\bf r} \int dt \,
 \Big\{ i\hat\rho \Big[ \partial_t \rho 
-T\nabla \cdot \Big( \rho \nabla \theta \Big)\Big]
  -T \rho ( \nabla \hat \rho)^2 + i\hat\theta 
\Big(\theta-\hat K*\del \rho-f(\delta \rho)\Big) \Big\}
\la{eqn:2.44}
\ee
where the subscript ABL of $\theta$ is dropped for simplicity.
In the absence of interaction, $K({\bf r})=\del ({\bf r})/\rho_0$, and 
hence $\hat K* \del \rho({\bf r},t)=\del \rho({\bf r},t)/\rho_0$. 
Therefore 
\be
{\cal S}_{ABL, id}[\psi]
= \int \, d {\bf r} \int dt \,
 \Big\{ i\hat\rho \Big[ \partial_t \rho 
-T\nabla \cdot \Big( \rho \nabla \theta \Big)\Big]
  -T \rho ( \nabla \hat \rho)^2 + i\hat\theta 
\Big(\theta-\frac{\del \rho}{\rho_0}-f(\delta \rho)\Big) \Big\}
\la{eqn:2.45}
\ee
Applying   the above two identities and additional one
$\Big<\delta \rho({\bf 2}) \delta {\cal S}_{ABL, id}[\psi]
/\delta {\hat\theta}({\bf 1}) \Big> =0$ to $ {\cal S}_{ABL, id}[\psi]$, we 
obtain for $t>0$
\be
\partial_t G_{\rho \rho}({\bf r},t)=T\nabla^2 G_{\rho \rho}({\bf r},t)
+\rho_0 T \nabla^2 \Big<f(\del \rho({\bf r},t)\del \rho({\bf 0},0) \Big>
+T\nabla \cdot 
\Big<\del \rho({\bf r},t)\nabla \theta ({\bf r},t) \del \rho({\bf 0},0)\Big> 
\la{eqn:2.46}
\ee
The ABL action (\ref{eqn:2.45}) thus does not appear to yield the anticipated dynamic behavior for 
the noninteracting particles. This discrepancy is puzzling since no approximation has been made to reach 
(\ref{eqn:2.46}). 
In fact, however, a careful treatment as detailed in Appendix A leads to the diffusion law (\ref{eqn:2.43}).

\section{The renormalized perturbation theory: the loop expansion}
\setcounter{equation}{0}
We are now ready to develop a renormalized perturbation theory \cite{msr} for 
the new form of the action (\ref{eqn:2.34}), which preserves the FDR order by order.
\subsection{Action}
Four field variables $\psi_{\al}({\bf r},t)$, $\al=1,2,3,4$ are defined where
\be
\psi_1({\bf r},t)\equiv \delta \rho({\bf r},t)
\equiv\rho({\bf r},t)-\rho_0,\quad \psi_2({\bf r},t)
\equiv\hat \rho({\bf r},t),\quad \psi_3({\bf r},t)
\equiv\theta({\bf r},t),\quad \psi_4({\bf r},t)\equiv\hat\theta({\bf r},t)
\la{eqn:3.1}
\ee
We arrange fields $\psi_{\al}({\bf r},t)$ in column vector $\Psi(j)$ 
and row vector $\Psi^T(j)$ 
where $j$ stands for a set of indices and space-time variable: 
$j\equiv \al_j,{\bf x}_j$ with  ${\bf x}_j\equiv ({\bf r}_j, t_j)$ . 
Thus we have 
\be
\Psi(j)=\Big(\psi_{\al_j}({\bf x}_j)\Big)=\left(\begin{array}{r}
\delta \rho({\bf x}_j)\\ \hat\rho({\bf x}_j)\\ \theta({\bf x}_j)\\ 
\hat\theta({\bf x}_j)\end{array}\right),  \qquad
\Psi^T(j)=\Big(\psi_{\al_j}({\bf x}_j)\Big)^T=\left(\begin{array}{rrrr}
\delta \rho({\bf x}_j)& \hat\rho({\bf x}_j)& \theta({\bf x}_j)& 
\hat\theta({\bf x}_j)\end{array}\right)
\la{eqn:3.2}
\ee
Also introduce the notation
\be
\partial_j\equiv\frac{\partial}{\partial t_j},\quad {\BF\nabla}_j\equiv \frac{\partial}{\partial {\bf r}_j}
\la{eqn:3.3}
\ee

We formally write the action integral (\ref{eqn:2.32}) as
\be
S[\psi]=-\frac{1}{2}G^{-1}_0(12)\psi(1)\psi(2)+\sum_{n=3}^{\infty}
\frac{1}{n!}V_s(12\cdots n)\psi(1)\psi(2)\cdots \psi(n)
\la{eqn:3.4}
\ee
where $V_s(12\cdots n)$ is the fully symmetrized vertices, and we have used the convention that repeated arguments $j=\alpha_j,{\bf x}_j$ imply summation over $j=\alpha_j$ and integration over ${\bf x}_j$. 

\subsection{The one-particle irreducible (1PI) loop expansion} 
\subsubsection{Generating functionals and Legendre transform}
We introduce a generating functional defined as
\be
W[J] \equiv\ln <e^{J(1)\psi(1)}>,\qquad
<\cdots> \equiv \frac{\int d[\psi] \cdots e^{{\cal S}[\psi]}} 
{\int d[\psi] e^{{\cal S}[\psi]}}
\la{eqn:3.5}
\ee
where $J(1)$ is the local source field. The functional $W[J]$ is the generating functional for the connected 
(cumulant) correlation functions: averages and pair correlations without source field are obtained from 
functional derivatives of $W[J]$ with respect to the source field $J$ in the limit of zero source
\bea 
G(12 \cdots n) &\equiv& \frac{\del^n W[J]}
{\delta J(1) \delta J(2)  \cdots \delta J(n)}\Big |_{J=0}, \nonumber \\
G(1) &\equiv& \phi(1) =\frac{\delta W[J]}{\delta J(1)}\Big |_{J=0}= <\psi(1)>, \nonumber \\
G(12) &\equiv&  
\frac{\delta^2 W[J]}{\delta J(1) \delta J(2)}\Big |_{J=0}=
  \Big( <\psi(1)\psi(2)>-\phi(1)\phi(2)\Big), \quad \mbox{etc.}
\la{eqn:3.6}
\eea
where response functions are included by using hatted variables, 
and will not be explicitly mentioned.

We define a new generating functional known as the vertex functional 
$\Gamma[\phi]$ via the Legendre transform 
\be
\Gamma[\phi]\equiv W[J]-J(1)\phi(1)
\la{eqn:3.7}
\ee
We then obtain 
\be
\frac{\delta \Gam[\phi]}{\del \phi(1)}=\frac{\del W[J]}{\del J(2)} 
\frac{\del J(2)}{\del \phi(1)}  
-\frac{\del J(2)}{\del \phi(1)}\phi(2) -J(1)=-J(1) 
\la{eqn:3.8}
\ee
where the first two terms in the rhs cancels since 
$\del W[J]/\del J(2)=\phi(2)$.

The vertex functions $\Gam_n (12 \cdots n)$ are defined as the derivatives of 
the  vertex functional $\Gam[\phi]$ with respect to $\phi$'s:
\be
\Gam_n (12 \cdots n) \equiv -\frac{\del^n \Gam[\phi]}
{\del \phi(1) \del \phi(2)  \cdots \del\phi(n)}
\la{eqn:3.9}
\ee
Let us compute the first few vertex functions.
We already obtained the first one, 
$$\Gam_1 (1) =-\frac{\delta \Gam[\phi]}{\del \phi(1)}=J(1).$$
The second vertex function is given by the inverse of the propagator:
\be
\Gam_2(12)=-\frac{\del^2 \Gam[\phi]}
{\del \phi(1) \del \phi(2)}=\frac{\del J(1)}{\del \phi(2)}
=\Big[\frac{\del \phi(2)}{\del J(1)} \Big]^{-1}
=\Big[ \frac{\del^2 W[J]}{\del J(1)\del J(2)}\Big]^{-1}
=G^{-1}(12)
\la{eqn:3.10}
\ee
In order to obtain the higher-order vertex functions, 
we use the following general equation
\be
\del G^{-1}(12)=-G^{-1}(11')\del G(1'2') G^{-1}(2'2)
=-\Gam_2(11')\del G(1'2') \Gam_2(2'2)
\la{eqn:3.11}
\ee
Now the third vertex function $\Gam_3(123)$ is given by
\be
\Gam_3(123)\equiv\frac{\del \Gam_2(12)}{\del \phi(3)}=
\frac{\del G^{-1}(12)}{\del \phi(3)}
=-\Gam_2(11')\frac{\del G(1'2')}{\del \phi(3)} \Gam_2(2'2)
\la{eqn:3.12}
\ee
Using 
\be
\frac{\del G(12)}{\del \phi(3)}=\frac{\del G(12)}{\del J(3')}
 \frac{\del J(3')}{\del \phi(3)}=G(123')\Gam_2(3'3), 
 \la{eqn:3.13}
 \ee
we obtain 
\be
\Gam_3(123)=-\Gam_2(11')\Gam(22')\Gam_2(3'3)G(1'2'3') 
\la{eqn:3.14}
\ee
or equivalently
\be
G(123)=-G(11')G(22')G(33')\Gam_3(1'2'3') 
\la{eqn:3.15}
\ee
Likewise, the 4-point correlation function is given by
\be
G(1234)=GG\Gam_3 G \Gam_3 GG +(2 \, sym.  terms)
-GGGG \Gam_4
\la{eqn:3.16}
\ee
where the indices are suppressed for brevity.
Therefore one can obtain the higher order correlation functions from
the vertex functions, which in turn can be obtained from the vertex functional
$\Gam[\phi]$ by functional differentiation with respect to $\phi(1)$.

\subsubsection{1PI Loop expansion}
One can systematically calculate the vertex functional $\Gamma[\phi]$ via
the loop expansion \cite{zjustin,vasil,dm,jackiw}.
We first rewrite (\ref{eqn:3.5}) as 
\be
W[J]=\lam \ln \Big(c \int d[\psi] \exp \Big[ \frac{1}{\lam}
\Big( S[\psi]+J(1)\psi(1) \Big)\Big] \Big)
\la{eqn:3.17}
\ee
where $c$ is the normalization factor: it is determined by 
the requirement $W[J=0]=0$ which leads to $c^{-1}=\int d\psi 
\exp\Big( S[\psi]/\lam \Big)$. 
One can perform a formal expansion of $W[J]$ in powers of $\lam$ 
(and can set $\lam=1$ at the end of calculation). 
This is an expansion in the strength of fluctuations and is called the loop expansion.
We expand the integrand in (\ref{eqn:3.17}) around the field $\psi_c$ which extremizes
the action $S[\psi]+J(1)\psi(1)$ for a given source field $J(1)$:
\be
\frac{\del S[\psi_c]}{\del \psi_c(1)}=-J(1)
\la{eqn:3.18}
\ee
Writing $\psi(1)=\psi_c(1)+\sqrt{\lam}\chi(1)$, we expand  $S[\psi]+J(1)\psi(1)$ as
\bea
S[\psi]+J(1)\psi(1)&=&S[\psi_c+\sqrt{\lam}\chi]
+J(1)\big(\psi_c(1)+\sqrt{\lam}\chi(1) \big) \nonumber \\
&=& \big( S[\psi_c]+J(1)\psi_c(1)\big)+\sum_{n=2}^{\infty}
\frac{\lam^{n/2}}{n!}S^{(n)}_c(12 \cdots n)\chi(1)\chi(2)\cdots \chi(n) 
\la{eqn:3.19}
\eea 
where $S^{(n)}_c(12 \cdots n) \equiv \del^n S[\psi_c]/\del \psi_c(1) \del \psi_c(2)
\cdots \del\psi_c(n)$, and the term linear in $\chi$ vanishes due to the relation
(\ref{eqn:3.18}). 

The first two terms in (\ref{eqn:3.19}) give the leading order results 
\bea
W[J]&=& W_0[J]\equiv S[\psi_c]+J(1)\psi_c(1), \nonumber \\
\phi(1)&=& \phi_0(1) \equiv \frac{\del W_0[J]}{\del J(1)}=
\frac{\del \psi_c(2)}{\del J(1)}\frac{\del}{\del \psi_c(2)}
\Big(S[\psi_c]+J(1')\psi_c(1') \Big) \nonumber \\
&=&\frac{\del \psi_c(2)}{\del J(1)}\psi_c(1')\frac{ \del J(1')}{\del \psi_c(2)} 
=\psi_c(1) \nonumber \\
\Gam[\phi]&=&\Gam_0[\phi] \equiv W_0[J]-J(1)\phi_0(1)=S[\psi_c]=S[\phi]
\la{eqn:3.20}
\eea
The functional $W_0[J]$ contains the connected tree (no loop) diagrams only:
it is the generating functional of the connected tree diagrams.
The zeroth order result of $\Gam[\phi]$ is therefore given by the 'average' action $S[\phi]$. 

Considering the expansion (\ref{eqn:3.19}) up to the term proportional to $\lam$, 
we have the Gaussian integral
\be 
\int d \chi \exp \Big[-\frac{1}{2}|S^{(2)}_c(12)| \chi(1)\chi(2)  \Big] 
=\big[ \det |S^{(2)}_c(12)|\big]^{-1/2}=\exp \Big(-\frac{1}{2} 
\Tr \ln |S^{(2)}_c(12)|\Big)
\la{eqn:3.21}
\ee
where the identity $\ln \det M =\Tr (\ln M)$ is used for the last equality.
This means that 
\bea
W[J]&=& W_0[J]+\lam W_1[J]+O(\lam^2) \nonumber \\
W_1[J]&=& -\frac{1}{2} \Big[ \Tr \ln |S^{(2)}_c(12)|- 
\Tr \ln |S^{(2)}_c(12)|_{J=0}  \Big] 
\la{eqn:3.22}
\eea
where the last term comes from the normalization. 

With the general form of the action (\ref{eqn:3.4}), 
$S_c^{(2)}(12)$ is given by
\be
S_c^{(2)}(12)= -{\cal G}^{-1}(12;\psi_c) \equiv -G^{-1}_0(12)+V_s(123)\psi_c(3)
+\frac{1}{2!}V_s(1234)\psi_c(3)\psi_c(4)+\cdots 
\la{eqn:3.23}
\ee
Since from (\ref{eqn:3.18}) $\psi_c=0$ for $J=0$ (excluding any
symmetry breaking solution of (\ref{eqn:3.18}) with $J=0$), 
we obtain
\bea
 W_1[J] &=& -\frac{1}{2} \Tr \ln \Big( {\cal G}^{-1}(11';\psi_c) G_0(1'2)  
\Big) \nonumber \\
 &=&-\frac{1}{2} \Tr \ln \Big( \del(12)-V_s(11'3)\psi_c(3)
 G_0(1'2)-\frac{1}{2!}V_s(11'34)\psi_c(3)\psi_c(4)G_0(1'2)+\cdots  \Big) \nonumber \\
 &=&\frac{1}{2}\Big\{  V_s(11'3)\psi_c(3) G_0(1'1)
+\frac{1}{2}V_s(11'34)\psi_c(3)\psi_c(4)G_0(1'1) \nonumber \\
&-& \frac{1}{2}V_s(11'2) \psi_c(2)G_0(1'2')
V_s(2'3'4) \psi_c(4)G_0(3'1)+\cdots \Big\}
 \la{eqn:3.24}
 \eea
Thus the trace operation generates a set of 1PI \cite{onepi} one-loop diagrams.
This expression leads to 
\be
\Gam[\phi]=W_0[J]+\lam W_1[J] -J(1)\phi(1)
=S[\psi_c]+J(1)\big( \psi_c(1)-\phi(1) \big)+\lam W_1[J]
\la{eqn:3.25}
\ee
where we have not yet included the second order contribution to $W[J]$.
We need to express $\psi_c(1)$ and $J(1)=-\del S[\psi_c]/\del \psi_c(1)$ 
in terms of $\phi$ field.  Noting that 
\bea
 \psi_c(1)&=&\phi(1)+O(\lam) \nonumber \\
  -J(1)&=& S^{(1)}(1;\psi_c)\equiv \frac{\del S[\psi_c]}{\del \psi_c(1)} 
=S^{(1)}(1;\phi)+S^{(2)}(12;\phi)\big( \psi_c(2)-\phi (2)\big)+O(\lam^2), \nonumber \\
S[\psi_c]&=& S[\phi]+S^{(1)}(1;\phi)\big( \psi_c(1)-\phi (1)\big)+O(\lam^2),
\la{eqn:3.26}
\eea
we obtain for the first two terms in (\ref{eqn:3.25})
\bea
 S[\psi_c]&+& J(1)\big( \psi_c(1)-\phi(1) \big)
= S[\phi]+S^{(1)}(1;\phi) \big( \psi_c(1)-\phi(1) \big) 
+ \frac{1}{2} S^{(2)}(12;\phi) \big( \psi_c(1)-\phi(1) \big) 
\big( \psi_c(2)-\phi(2) \big) \nonumber \\ 
&-& \Big(S^{(1)}(1;\phi)+S^{(2)}(12;\phi)\big( \psi_c(2)-\phi (2)\big)  \Big) 
\big( \psi_c(1)-\phi(1) \big)+O(\lam^3) \nonumber \\
&=& S[\phi]-\frac{1}{2} S^{(2)}(12;\phi) \big( \psi_c(1)-\phi(1) \big) 
\big( \psi_c(2)-\phi(2)\big) +O(\lam^3)
\la{eqn:3.27}
\eea
where the two linear contributions cancel. 
Therefore, up to the one-loop order, the eq. (\ref{eqn:3.25}) leads to
\be
\Gam[\phi]=S[\phi]+\lam W_1[\psi_c=\phi]=
S[\phi]-\frac{\lam}{2}\Tr \ln \Big( {\cal G}^{-1}(\phi) \cdot G_0 \Big)+O(\lam^2)
\la{eqn:3.28}
\ee
where the last term in the last line of (\ref{eqn:3.27}) was not yet
included because of its being $O(\lam^2)$.

We now come to the two-loop calculation.
Considering (\ref{eqn:3.17}) and (\ref{eqn:3.19}), we have 
the following two-loop contributions for $W[J]$ 
\bea
W[J]&=& W_0[J]+\lam W_1[J]+ \lam^2 W_2[J]+O(\lam^3) \nonumber \\
W_2[J]&=&\frac{1}{4!}S^{(4)}_c(1234)<\chi(1)\chi(2)\chi(3)\chi(4)>_0 \nonumber \\
&+& \frac{1}{2!}\Big(\frac{1}{3!}\Big)^2 
S^{(3)}_c(123)S^{(3)}_c(456)<\chi(1)\chi(2)\chi(3)\chi(4)\chi(5)\chi(6)>_0
\la{eqn:3.29}
\eea
where $W_0[J]$ and $W_1[J]$ are given respectively by (\ref{eqn:3.20}) and 
(\ref{eqn:3.24}). In (\ref{eqn:3.29}), 
 $<\cdots>_0$ denotes the average over the Gaussian distribution
$$<\cdots>_0 \equiv \frac{\int d[\chi] (\cdots)\, \exp \Big(-\frac{1}{2}|S^{(2)}_c(12)| 
\chi(1) \chi(2) \Big)}
{\int d[\chi] \exp \Big(-\frac{1}{2}|S^{(2)}_c(12)| 
\chi(1) \chi(2) \Big)}.$$
Using the Wick's theorem and symmetry of $S^{(3)}_c(123)$ and 
$S^{(4)}_c(1234)$, we obtain 
\bea
W_2[J]&=& \frac{1}{8} S^{(4)}_c(1234){\cal G}(12){\cal G}(34)+\frac{1}{12}
S^{(3)}_c(123)S^{(3)}_c(456) {\cal G}(14){\cal G}(25){\cal G}(36) \nonumber \\
&+& \frac{1}{8} S^{(3)}_c(123)S^{(3)}_c(456)
{\cal G}(12){\cal G}(56){\cal G}(34)
\la{eqn:3.30}
\eea
The diagrammatic expression for $W_2[J]$ is shown in Fig.~1.
While the first two diagrams in Fig.~1 are 1PI diagrams, the last one in Fig.~1 is 1PR (one-particle reducible)
 diagram. We will see that the Legendre transform to $\Gamma [\phi]$ eliminates this 1PR diagram, 
which makes $\Gam[\phi]$ diagrammatically simpler than $W[J]$. 

Now going back to (\ref{eqn:3.25}) and adding $W_2[J]$, the second-order contribution of $W[J]$, 
we have for the Legendre transform of $W[J]$
\be
\Gam[\phi]=S[\psi_c]+J(1)\big( \psi_c(1)-\phi(1) \big)+\lam W_1[J]+\lam^2 W_2[J]
\la{eqn:3.31}
\ee
We have seen in (\ref{eqn:3.27}) that the first two terms in the rhs. of 
(\ref{eqn:3.31}) has no $O(\lam)$-contribution. 
Up to the second-order in $\lam$, they are then given by
\be
S[\psi_c]+J(1)\big( \psi_c(1)-\phi(1))
=S[\phi]-\frac{1}{2}S^{(2)}(12;\phi)(\psi_c(1)-\phi(1)) (\psi_c(2)-\phi(2))
+O(\lam^3)
\la{eqn:3.32}
\ee 
Now we compute $\psi_c(1)-\phi(1)$. 
Note that
\bea
\phi(1)&=&\frac{\del W[J]}{\del J(1)}=\psi_c(1)
+\lam \frac{\del W_1[J]}{\del J(1)}+\cdots \nonumber \\
\psi_c(1)-\phi(1)&=& -\lam \frac{\del W_1[J]}{\del J(1)} 
= -\lam \frac{\del \psi_c(2)}{\del J(1)} \frac{\del W_1[J]}{\del \psi_c(2)} 
=\lam [S^{(2)}_c]^{-1}(12)\frac{\del W_1[J]}{\del \psi_c(2)} \nonumber \\
&=& -\lam {\cal G}(12;\phi)\frac{\del W_1[J;\phi]}{\del \phi(2)}+O(\lam^2) 
\la{eqn:3.33}
\eea 
where we used (\ref{eqn:3.18}) and (\ref{eqn:3.23}).
Using 
\bea
\frac{\del W_1[J;\phi]}{\del \phi(2)} &=&
-\frac{1}{2} \frac{\del}{\del \phi(2)} 
\Big[\Tr \ln \Big( {\cal G}^{-1}(\phi) \cdot G_0 \Big)  \Big]
 \nonumber\\ &=& -\frac{1}{2}\Tr G_0^{-1}{\cal G}\frac{{\delta\cal G}^{-1}}{\delta\phi(2)}G_0=
-\frac{1}{2}G_0^{-1}(11'){\cal G}(1'3)\frac{\delta{\cal G}^{-1}(33')}{\delta\phi(2)}
G_0(3'1) \nonumber \\
&=& -\frac{1}{2}\delta(1'3'){\cal G}(1'3)\frac{\delta{\cal G}^{-1}(33')}
{\delta\phi(2)} =-\frac{1}{2}{\cal G}(3'3)\frac{\delta{\cal G}^{-1}(33')}{\delta\phi(2)} \nonumber \\
&=& \frac{1}{2} {\cal G}(3'3)\frac{S^{(2)}(33';\phi)}{\delta\phi(2)} 
= \frac{1}{2} S^{(3)}(233';\phi){\cal G}(3'3),
\la{eqn:3.34}
\eea
we obtain
\be
\psi_c(1)-\phi(1)=-\frac{\lam}{2} {\cal G}(12;\phi) S^{(3)}(234;\phi){\cal G}(34;\phi)
+O(\lam^2)
\la{eqn:3.35}
\ee
Substitution of (\ref{eqn:3.35}) into (\ref{eqn:3.32}) gives
\bea
&&S[\psi_c]+J(1)\big( \psi_c(1)-\phi(1)) \nonumber \\
&=& S[\phi]-\frac{1}{2}S^{(2)}(12;\phi) \cdot 
\Big( -\frac{\lam}{2}{\cal G}(13;\phi ) S^{(3)}(345;\phi){\cal G}(45;\phi) \Big)
\cdot \Big(- \frac{\lam}{2}{\cal G}(26;\phi ) 
S^{(3)}(678;\phi){\cal G}(78;\phi) \Big) \nonumber \\
&=& S[\phi]+\frac{\lam^2}{8} S^{(3)}(245;\phi){\cal G}(45;\phi) 
{\cal G}(26;\phi ) S^{(3)}(678;\phi){\cal G}(78;\phi)+O(\lam^3) 
\la{eqn:3.36}
\eea
where $S^{(2)}(12;\phi) {\cal G}(13;\phi )=-\delta(23)$ is used.
The last term in (\ref{eqn:3.36}) has the same structure as 
the 1PR diagram in (\ref{eqn:3.30}).
There is one additional term of the same structure which comes from 
$\lam W_1[J]$ in (\ref{eqn:3.31}).
Since $W_1[J]$ is a functional of $\psi_c(J)$ via 
$J(1)=-\del S[\psi_c]/\del \psi_c(1)$, one expands $\lam W_1[J]$ as 
\bea
\lam W_1[J;\psi_c]&=& \lam W_1[\phi]+\lam\frac{\del W_1[\phi]}{\del \phi(1)}
(\psi_c(1)-\phi(1))+O(\lam^3) \nonumber \\
&=& \lam W_1[\phi]-\frac{\lam^2}{4}S^{(3)}(123;\phi){\cal G}(23;\phi) 
{\cal G}(14;\phi ) S^{(3)}(456;\phi){\cal G}(56;\phi) +O(\lam^3)
\la{eqn:3.37}
\eea
where the last equality is obtained by use of (\ref{eqn:3.34}) and (\ref{eqn:3.35}).
Finally, we consider the last term in (\ref{eqn:3.31}):
\bea
&&\lam^2 W_2[J;\psi_c]= \lam^2W_2[\phi]+O(\lam^3) \nonumber \\
&=&\frac{\lam^2}{8} S^{(4)}(1234;\phi){\cal G}(12;\phi){\cal G}(34;\phi)
+\frac{\lam^2}{12}
S^{(3)}(123;\phi)S^{(3)}(456;\phi) {\cal G}(14;\phi){\cal G}(25;\phi)
{\cal G}(36;\phi) \nonumber \\
&+& \frac{\lam^2}{8} S^{(3)}(123;\phi)S^{(3)}(456;\phi)
{\cal G}(12;\phi){\cal G}(56;\phi){\cal G}(34;\phi)+O(\lam^3)
\la{eqn:3.38}
\eea
where (\ref{eqn:3.30}) was used.
Adding up (\ref{eqn:3.36})-(\ref{eqn:3.38}), we obtain up to the two-loop order
\bea
  \Gam [\phi]&=& S[\phi]
-\frac{\lam}{2}\Tr \ln \Big({\cal G}^{-1}(\phi) \cdot G_0 \Big)+\Gam_{1PI}[\phi], 
\nonumber \\
\Gam_{1PI}[\phi] &\equiv&  \lam^2 \Big\{\frac{1}{8} S^{(4)}(1234;\phi)
{\cal G}(12;\phi) {\cal G}(34;\phi) 
+ \frac{1}{12} S^{(3)}(123;\phi)S^{(3)}(456;\phi) 
{\cal G}(14;\phi) {\cal G}(25;\phi) {\cal G}(36;\phi) \Big\}  \nonumber \\
\la{eqn:3.39}
\eea
Note that the 1PR diagram was eliminated in $\Gamma[\phi]$.
Therefore, up to the second order in $\lam$, 
$\Gam_{1PI}[\phi]$ turns out to be  the sum of the {\em two-loop} 1PI diagrams. 
It was shown that this feature is indeed the general structure: 
$\Gam_{1PI}[\phi]$ is the sum of all (two and higher loop) 
1PI-diagrams with propagator ${\cal G}(\phi)$ and 
vertices dictated by the interaction potential. 
Shown in Fig.~2 are the diagrams for $\Gamma_1[\phi]$ up to the three-loop order
\cite{vasil,dm,jackiw,carrington}.

\subsection{The two-particle irreducible (2PI) loop expansion} 
Although $\Gam_{1PI}[\phi]$ consists of 1PI diagrams only,  some of the diagrams in $\Gam_{1PI}[\phi]$ are
2PR diagrams (the diagrams which are disconnected by cutting two lines).
Introduction of the bilocal source field $K(12)$ can eliminate all 2PR
diagrams in the vertex functional obtained via the double Legendre transform 
\cite{vasil,dm,cjt,carrington,abc}. 
Thus the resulting vertex functional $\Gam[\phi, G]$ has simpler 
structure than $\Gam[\phi]$.  
A specific example is the three-loop result for $\Gam_1[\phi]$. 
As shown in Fig.~2, in the three-loop order $\Gam_1[\phi]$ has six 1PI diagrams.  
Among these diagrams, three (the third, fifth, and sixth) diagrams  are 2PI ones
(the diagrams which are not disconnected by cutting the two lines), 
and the remaining three are 2PR ones. 
The two-loop diagrams in Fig.~2 are 2PI diagrams as well.
These 2PR diagrams were shown \cite{vasil,dm,cjt,carrington} 
to be eliminated in the functional $\Gam [\phi, G]$ 
obtained via the double Legendre transform.

The generating functional $W[J,K]$ for the connected correlation functions is 
defined as
\be
W[J,K]= \ln \Big( c\int d  [\psi] \exp 
\big( S[\psi]+J(1)\psi(1)+\frac{1}{2}\psi(1)K(12)\psi(2) \big) \Big)
\la{eqn:3.40}
\ee
where the normalization constant is determined by the condition
$W[J=0,K=0]=0$.
Averages and correlations are generated from $W[J,K]$ as
\bea
\phi(1) &\equiv& <\psi(1)> = \frac{\delta W[J,K]}{\delta J(1)}, \nonumber \\
G(12) &\equiv& \frac{\delta^2 W[J,K]}{\delta J(1) \delta J(2)}=
  \Big( <\psi(1)\psi(2)>-\phi(1)\phi(2)\Big), \nonumber \\
  \frac{\delta W[J,K]}{\delta K(12)}&=&\frac{1}{2}<\psi(1)\psi(2)>
  =\frac{1}{2}\Big( G(12)+\phi(1)\phi(2) \Big) \nonumber \\
  \frac{\delta W[J,K]}{\del K(12) \del K(34)}&=&
\frac{1}{4}<\psi(1)\psi(2)\psi(3)\psi(4)>
  -\frac{1}{4}<\psi(1)\psi(2)> <\psi(3)\psi(4)>
 \la{eqn:3.41}
\eea
The double Legendre transform is then given by
\be
\Gam[\phi,G]=W[J,K]-J(1)\phi(1)
-\frac{1}{2}K(12)\Big( G(12)+\phi(1)\phi(2) \Big) 
\la{eqn:3.42}
\ee
where the source fields $J$ and $K$ are should be eliminated in favor of 
$\phi$ and $G$. 
We obtain the derivatives  
\bea
\frac{\delta \Gam[\phi, G]}{\del \phi(1)}&=&\frac{\del W[J,K]}{\del J(2)} 
\frac{\del J(2)}{\del \phi(1)} 
+\frac{\del W[J,K]}{\del K(23)}\frac{\del K(23)}{\del \phi(1)} 
-\frac{\del J(2)}{\del \phi(1)}\phi(2) -J(1) \nonumber \\
&-& \frac{1}{2}\frac{\del K(23)}{\del \phi(1)}\Big( G(23)+\phi(2)\phi(3) \Big)
-K(12)\phi(2)= -J(1)-K(12)\phi(2), \nonumber \\
&{}& \nonumber \\ 
 \frac{\delta \Gam[\phi, G]}{\del G(12)}&=&\frac{\del W[J,K]}{\del J(3)} 
\frac{\del J(3)}{\del G(12)} 
+\frac{\del W[J,K]}{\del K(34)}\frac{\del K(34)}{\del G(12)} 
-\frac{\del J(3)}{\del G(12)}\phi(3) 
-\frac{1}{2}\frac{\del K(34)}{\del G(12)}\Big( G(34)+\phi(3)\phi(4) \Big) \nonumber \\
&-& \frac{1}{2} K(12) = -\frac{1}{2} K(12)
\la{eqn:3.43}
\eea

As before, we first define $W[J,K]$ as 
\be
W[J,K]= \lam \ln \Big( c\int d[\psi] \exp \Big[ \frac{1}{\lam}
\Big( S[\psi]+J(1)\psi(1)+\frac{1}{2}\psi(1)K(12)\psi(2) \Big)\Big] \Big)
\la{eqn:3.44}
\ee
We then expand the integrand (\ref{eqn:3.44}) around the minimum field
$\psi_c(1)$ of the exponential for a given $J$ and $K$:
\be
\frac{\del S[\psi_c]}{\del \psi_c(1)}=- \big( J(1)+K(12)\psi_c(2) \big)
\la{eqn:3.45}
\ee
Writing $\psi(1)=\psi_c(1)+\sqrt{\lam}\chi(1)$, we obtain
\bea
&{}&S[\psi]+J(1)\psi(1)+\frac{1}{2}K(12)\psi(1)\psi(2)
= \big( S[\psi_c]+J(1)\psi_c(1)+
\frac{1}{2}K(12)\psi_c(1) \psi_c(2) \big) \nonumber \\
&+&\frac{\lam}{2}\Big( S^{(2)}_c(12)+ K(12) \Big) \chi(1)\chi(2) 
+\sum_{n=3}^{\infty}\frac{\lam^{n/2}}{n!}S^{(n)}_c(12\cdots n)
\chi(1)\chi(2)\cdots \chi(n)
\la{eqn:3.46}
\eea 
where the linear terms in $\chi$ are eliminated due to (\ref{eqn:3.45}).
We see from (\ref{eqn:3.45}) and (\ref{eqn:3.46}) that 
the formal structure of the expansion is the same as before, provided that 
the following replacements are made:
\bea
S[\psi_c] &\rightarrow& S[\psi_c]+\frac{1}{2}K(12) \psi_c(1) \psi_c(2), \nonumber \\
S^{(2)}_c(12) &\rightarrow& S^{(2)}_c(12)+K(12) \nonumber \\
{\cal G}(12;\psi_c) &\equiv& -\big(S^{(2)}_c(12)+K(12) \big)^{-1}
\la{eqn:3.47}
\eea
There is no change in the higher order terms in (\ref{eqn:3.46}).
Therefore the analysis in the previous section  still holds, and we have
\bea
W[J,K]&=& W_0[J,K]+W_1[J,K]+W_2[J,K]+\cdots \nonumber \\
W_0[J,K]&=&S[\psi_c]+J(1)\psi_c(1)+\frac{1}{2}K(12)\psi_c(1)\psi_c(2) \nonumber \\
W_1[J,K]&=& -\frac{\lam}{2}\Tr \ln \Big({\cal G}^{-1}(\psi_c)\cdot G_0 \Big)
\nonumber \\
W_2[J,K] &=& \lam^2 \Big[\frac{1}{8} S^{(4)}_c(1234){\cal G}(12;\psi_c){\cal G}(34;\psi_c)
+\frac{1}{12}
S^{(3)}_c(123)S^{(3)}_c(456) {\cal G}(14;\psi_c){\cal G}(25;\psi_c)
{\cal G}(36;\psi_c) \nonumber \\
&+& \frac{1}{8} S^{(3)}_c(123)S^{(3)}_c(456)
{\cal G}(12;\psi_c){\cal G}(56;\psi_c){\cal G}(34;\psi_c) \Big] 
\la{eqn:3.48}
\eea
where the normalization constant is left out. 

The double Legendre transform is now given by
\bea
\Gam[\phi,G]&=&-J(1)\phi(1)-\frac{1}{2}K(12)
\Big(\lam G(12)+\phi(1)\phi(2)\Big)+W_0[J,K]+W_1[J,K]+W_2[J,K]+\cdots \nonumber \\
&=& -J(1)\phi(1)-\frac{1}{2}K(12)\Big(\lam G(12)+\phi(1)\phi(2)\Big) \nonumber \\
&+& S[\psi_c]+J(1)\psi_c(1)+\frac{1}{2}K(12)\psi_c(1)\psi_c(2)
+W_1[J,K]+W_2[J,K]+\cdots \nonumber \\
&=&S[\psi_c]+\big( J(1)+K(12)\psi_c(2)\big)\big( \psi_c(1)-\phi(1) \big)
-\frac{1}{2}K(12) \big( \psi_c(1)-\phi(1) \big)\big( \psi_c(2)-\phi(2)\big) \nonumber \\
&-&\frac{\lam}{2}K(12)G(12)+W_1[J,K]+W_2[J,K]+\cdots
\la{eqn:3.49}
\eea
Let us look at the first three terms of the rhs in the last line of (\ref{eqn:3.49}).
Using (\ref{eqn:3.45}) and expanding them around $\phi$, we obtain
\bea
&& S[\psi_c]+\big( J(1)+K(12)\psi_c(2)\big)\big( \psi_c(1)-\phi(1) \big)
-\frac{1}{2}K(12)\big( \psi_c(1)-\phi(1) \big)\big( \psi_c(2)-\phi(2)\big) \nonumber \\
&=& S[\psi_c]-\frac{\del S[\psi_c]}{\del \psi_c(1)} 
\big( \psi_c(1)-\phi(1) \big)
-\frac{1}{2}K(12)\big( \psi_c(1)-\phi(1) \big)\big( \psi_c(2)-\phi(2)\big)\nonumber \\
&=&S[\phi]+S^{(1)}(1;\phi)\big( \psi_c(1)-\phi(1) \big)
+\frac{1}{2}S^{(2)}(12;\phi)\big( \psi_c(1)-\phi(1) \big)\big( \psi_c(2)-\phi(2) \big)
+\cdots \nonumber \\
&-&\Big( S^{(1)}(1;\phi)+S^{(2)}(12;\phi)\big( \psi_c(2)-\phi(2) \big)+\cdots \Big)
 \big( \psi_c(1)-\phi(1) \big) \nonumber \\
&-& \frac{1}{2}K(12)
\big( \psi_c(1)-\phi(1) \big)\big( \psi_c(2)-\phi(2)\big) \nonumber \\
&=& S[\phi]-\frac{1}{2}\Big(S^{(2)}(12;\phi)+K(12)\Big)
\big( \psi_c(1)-\phi(1) \big)\big( \psi_c(2)-\phi(2) \big)+\cdots 
\la{eqn:3.50}
\eea
We see that there is no one-loop contribution from these three terms since
the terms linear in $(\psi_c-\phi)$ cancel in (\ref{eqn:3.50});
they only have the zero-loop, two-loop and higher contributions for $\Gam[\phi,G]$.
Substituting (\ref{eqn:3.50}) into (\ref{eqn:3.49}), we obtain  
\bea
\Gam[\phi,G]&=&S[\phi]-\frac{1}{2}\Big(S^{(2)}(12;\phi)+K(12)\Big)
\big( \psi_c(1)-\phi(1) \big)\big( \psi_c(2)-\phi(2) \big) \nonumber \\
&-&\frac{\lam}{2}K(12)G(12)+W_1[J,K]+W_2[J,K]+\cdots
\la{eqn:3.51}
\eea
In order to simplify (\ref{eqn:3.51}), we first consider 
the two-point correlations function. It is given by using (\ref{eqn:3.35})
and (\ref{eqn:3.41}) 
\bea
\phi(1)&=&\psi_c(1)+\frac{\lam}{2}  
{\cal G}(12;\phi) S^{(3)}(234;\phi) {\cal G}(34;\phi)+O(\lam^2) \nonumber \\
G(12)&=& \frac{\del \phi(1)}{\del J(2)}=
\frac{\del \psi_c(1)}{\del J(2)} + \frac{\lam}{2}  \frac{\del \phi(1')}{\del J(2)}
  \frac{\del}{\del \phi(1')}\Big[ {\cal G}(12';\phi)S^{(3)}(2'34;\phi){\cal G}(34;\phi) 
 \Big] \nonumber \\
&=& {\cal G}(12;\psi_c)+\frac{\lam}{2} F(11';\phi)G(1'2)+O(\lam^2)
\label{eqn:3.52}
\eea
where $F(11';\phi)\equiv \frac{\del}{\del \phi(1')}\Big[
{\cal G}(12';\phi)S^{(3)}(2'34;\phi){\cal G}(34;\phi)\Big]$. 
From (\ref{eqn:3.52}), the inverse of ${\cal G}(12;\psi_c)$ is given by 
\be
{\cal G}^{-1}(12;\psi_c)= G^{-1}(12)+\frac{\lam}{2} G^{-1}(11') F(1'2;\phi)+O(\lam^2)
\la{eqn:3.53}
\ee
Using (\ref{eqn:3.53}), we obtain the following expression for 
 the term $-\frac{\lam}{2}K(12)G(12)$ in (\ref{eqn:3.51}).
\bea
&&-\frac{\lam}{2}K(12)G(12)=\frac{\lam}{2}G(12)\big(S^{(2)}_c(12)
+{\cal G}^{-1}(12;\psi_c )\big) \nonumber \\
&=& \frac{\lam}{2}G(12)\Big[ S^{(2)}(12;\phi)+S^{(3)}(123;\phi)(\psi_c(3)-\phi(3)) 
+ G^{-1}(12)+\frac{\lam}{2} G^{-1}(11') F(1'2;\phi) \Big]+O(\lam^3)\nonumber \\
&=& \frac{\lam}{2}G(12) \Big(  S^{(2)}(12;\phi)+G^{-1}(12) \Big) 
+ \frac{\lam}{2}G(12) S^{(3)}(123;\phi)(\psi_c(3)-\phi(3))
+\frac{\lam^2}{4}  F(22;\phi) +O(\lam^3) \nonumber \\
\la{eqn:3.54}
\eea
We now look at the term $W_1[J,K]$, (\ref{eqn:3.48}).
Using (\ref{eqn:3.53}), we have
\bea
W_1[J,K]&=& -\frac{\lam}{2} \Tr\ln \Big({\cal G}^{-1}(\psi_c) \cdot G_0 \Big) 
= -\frac{\lam}{2} \Tr \Big[ \ln \Big( G^{-1}\cdot G_0\Big) 
+\ln \big( I+\frac{\lam}{2} F \big) \Big] \nonumber \\
&=& -\frac{\lam}{2} \Tr \ln \Big( G^{-1}\cdot G_0\Big)
-\frac{\lam^2}{4} \Tr F  +O(\lam^3)
\la{eqn:3.55}
\eea
where $I$ is the unit matrix.
We find that when (\ref{eqn:3.54}) and (\ref{eqn:3.55}) are added, 
the last terms in (\ref{eqn:3.54}) and (\ref{eqn:3.55}) cancel against each other. 

Thus we have obtained the following expression for $\Gam[\phi,G]$ up to the two loop order 
\bea
\Gam[\phi,G]&=&S[\phi]-\frac{\lam}{2} \Tr\ln \Big(G^{-1}\cdot G_0\Big) 
+\frac{\lam}{2} \Tr \Big( G \cdot S^{(2)}(\phi)+ I \Big) \nonumber \\
&-& \frac{1}{2}\Big(S^{(2)}(12;\phi)+K(12)\Big)
\big( \psi_c(1)-\phi(1) \big)\big( \psi_c(2)-\phi(2) \big) \nonumber \\
&+&\frac{\lam}{2}G(12)S^{(3)}(123;\phi)(\psi_c(3)-\phi(3))+W_2[J,K]
+O(\lam^3)
\la{eqn:3.56}
\eea
For the last three terms, the same analysis in the previous section 
that led to (\ref{eqn:3.39}) is applied, 
and the 1PR diagram in $W_2[J,K]$ is eliminated.
Consequently, the final expression for $\Gam[\phi,G]$ up to the two-loops 
with $\lam=1$ is given by
\bea
\Gam[\phi,G]&=& S[\phi]-\frac{1}{2} \Tr\ln \Big(G^{-1}\cdot G_0\Big) 
+\frac{1}{2} \Tr \Big( G \cdot S^{(2)}(\phi)+ I \Big) +\Gam_{2PI}[\phi,G] 
\nonumber \\
\Gam_{2PI}[\phi,G] &\equiv&  \frac{1}{8} S^{(4)}(1234;\phi)G(12)G(34)
+ \frac{1}{12} S^{(3)}(123;\phi) S^{(3)}(456;\phi) G(14)G(25)G(36)  +O(\lam^3) 
\nonumber \\
\la{eqn:3.57}
\eea
where $G$ appears in place of ${\cal G}(\phi)$ since from (\ref{eqn:3.52})
$G={\cal G}(\phi)$ at the lowest order. 
We can obtain the three-loop calculation to explicitly show that the 2PR diagrams 
contained in $\Gam_{2PI}[\phi,G]$ are eliminated. 
It has also been shown \cite{dm,cjt} that 
$\Gam_{2PI}[\phi,G]$ is the sum of the two-loop and higher 2PI diagrams. 
Figure~3 shows the diagrams for $\Gam_{2PI}[\phi,G]$ up to the three-loop order.

In the absence of the source fields $J=K=0$, we have $\phi=0$, $S[\phi]=0$, 
$S^{(2)}(\phi)=-G^{-1}_0$, and $\del \Gam[\phi,G]/\del G(12)=0$. 
Also $S^{(3)}(123;\phi)$ etc. reduce to the ordinary vertices 
$V_s(123)$ etc. 
Therefore we have in the absence of sources
\bea
\Gam[0,G]&=& \frac{1}{2} \Tr\ln \Big(G \cdot G^{-1}_0\Big) 
-\frac{1}{2} \Tr \Big( G \cdot G^{-1}_0 - I \Big) +\Gam_{2PI}[0,G] \nonumber \\
0&=& \frac{\del \Gam[0,G]}{\del G(12)} =\frac{1}{2}G^{-1}(12)-
\frac{1}{2} G^{-1}_0(12) + \frac{\del \Gam_{2PI}[0,G]}{\del G(12)} 
\la{eqn:3.58}
\eea
The last line is the Schwinger-Dyson equation which can be cast into
the form  with the self-energy $\Sigma$
\bea
G^{-1}(12)&=& G^{-1}_0-\Sigma(12), \qquad \Sigma(12)\equiv 2 
\frac{\del \Gam_{2PI}[0,G]}{\del G(12)} \nonumber \\
\Sigma(12) &=& \frac{1}{2} V_s(1234) G(34)+ \frac{1}{2} V_s(134)
V_s(256) G(35)G(46)
\la{eqn:3.59}
\eea
where the last line, the one-loop expression for the self-energy $\Sigma$, 
is obtained by use of (\ref{eqn:3.57}). 
Figure~4 shows the self-energy diagrams up to the two-loop order.
These diagrams are obtained from the diagrams of $\Gamma_2[\phi,G]$ (Fig.~4)
by cutting a line (corresponding to taking a derivative of $\Gamma_{2PI}[\phi,G]$
with respect to $G$).

\subsection{Unperturbed propagator}
Recalling we wrote the Gaussian part of the action as 
$$S_g[\psi]=-\frac{1}{2}G^{-1}_0(12)\psi(1)\psi(2)
=-\frac{1}{2}\Psi^T({\bf 1})\cdot G^{-1}_0({\bf 12})\cdot \Psi({\bf 2}),$$ 
we read off the inverse of the unperturbed propagator from 
$S_g[\psi]$ in (\ref{eqn:2.34}) as follows.
\be 
G_0^{-1}({\bf 12})=\left(\begin{array}{rrrr}
0\qquad\qquad&iD_1\delta({\bf 12})+i\rho_0\nabla_1^2U({\bf 12})
&0\qquad\qquad&0\\
i{\tilde D}_1 \delta({\bf 12})+i\rho_0\nabla_1^2U({\bf 12})&
-2\rho_0T\nabla_1^2\delta({\bf 12})&i\rho_0T\nabla_1^2\delta({\bf 12})&0\\
0\qquad\qquad&iT\rho_0\nabla_1^2\delta({\bf 12})&0\qquad\qquad&-i\delta({\bf 12})\\
0\qquad\qquad&0\qquad\qquad&-i\delta({\bf 12})\qquad\qquad&0
\end{array}\right)
\la{eqn:3.60}
\ee
where $D_1 \equiv (\partial/\partial_1 + T \nabla^2_1)$ and 
${\tilde D}_1 \equiv (-\partial/\partial_1 + T \nabla^2_1)$.

\subsection{Vertices}
The non-Gaussian part of the action ${\cal S}_{ng}[\psi] $ given in (\ref{eqn:2.34})
can be written as 
\be
{\cal S}_{ng}[\psi]= \frac{1}{3!}V^S_{abc}({\bf 123})\psi_a({\bf 1})
\psi_b({\bf 2}) \psi_c({\bf 3})
+\frac{1}{4!}V^S_{abcd}({\bf 1234})\psi_a({\bf 1})
\psi_b({\bf 2}) \psi_c({\bf 3})\psi_d({\bf 4})+\cdots
\la{eqn:3.61}
\ee
The fully symmetrized vertices $V^S_{abc}({\bf 123})$ etc. are given by 
\bea
V_{abc}({\bf 123})&=& \frac{\del^3 {\cal S}_{ng}[\psi]}
{\del \psi_a({\bf 1})\del \psi_b({\bf 2})\del \psi_c({\bf 3})}, \nonumber \\
V^S_{abc}({\bf 123})&=& \Big[V_{abc}({\bf 123})+V_{acb}({\bf 132})+V_{bca}({\bf 231})
+V_{bac}({\bf 213})+V_{cab}({\bf 312})+V_{cba}({\bf 321}) \Big]
\la{eqn:3.62}
\eea

From the explicit expression of $S_{ng}$, we have the following 5 types of 
cubic vertices
\bea
V_{\hat\rho \rho\rho}({\bf 123})
&=& \frac{\del^3 {\cal S}_{ng}[\psi]}
{\del\hat\rho({\bf 1})\del\rho({\bf 2})\del\rho({\bf 3})}
\equiv V^{int}_{\hat\rho \rho \rho}({\bf 123})
+V^{id}_{\hat\rho \rho\rho}({\bf 123}) \nonumber \\
V^{int}_{\hat\rho \rho \rho}({\bf 123})&=&(-i)\nabla_1 \cdot 
\Big[\del({\bf 12})\nabla_1 U({\bf 13})+\del({\bf 13})\nabla_1 U({\bf 12})\Big], 
\nonumber \\
V^{id}_{\hat\rho \rho \rho}({\bf 123})&=&-\frac{iT}{\rho_0}\nabla^2_1  
\Big[\del({\bf 12})\del({\bf 13})\Big], \nonumber \\
V_{\hat\rho \rho \hat\rho}({\bf 123})&=& 
\frac{\del^3 {\cal S}_{ng}[\psi]}
{\del\hat\rho({\bf 1})\del\rho({\bf 2})\del\hat\rho({\bf 3})}
= 2T \nabla_1 \cdot \Big[ \del({\bf 12}) \nabla_1 \del({\bf 13})\Big], \nonumber \\
V_{\hat\rho \rho \theta}({\bf 123})&=& 
\frac{\del^3 {\cal S}_{ng}[\psi]}
{\del\hat\rho({\bf 1})\del\rho({\bf 2})\del\theta({\bf 3})}
= (-iT) \nabla_1 \cdot \Big[ \del({\bf 12}) \nabla_1 \del({\bf 13})\Big], \nonumber \\
V_{\hat\theta \rho \rho}({\bf 123})&=& 
\frac{\del^3 {\cal S}_{ng}[\psi]}
{\del\hat\theta({\bf 1})\del\rho({\bf 2})\del\rho({\bf 3})}
= \frac{i}{\rho^2_0} \del({\bf 12})\del({\bf 13})
\la{eqn:3.63}
\eea
where $\del({\bf 12})\equiv \del({\bf r}_1-{\bf r}_2)\del (t_1-t_2)$, etc. and the indices {\it int} and {\it id}  stand for particle interaction and ideal gas, respectively.
Note that the vertices $V^{int}_{\hat\rho \rho \rho}({\bf 123})$, 
$V^{id}_{\hat\rho \rho \rho}({\bf 123})$, and $V_{\hat\theta \rho \rho}({\bf 123})$ 
are already symmetric under exchange of ${\bf 2}$ and ${\bf 3}$, whereas 
$ V_{\hat\rho \rho \hat\rho}({\bf 123})$ and $V_{\hat\rho \rho \theta}({\bf 123})$
 are not. This fact should be kept in mind when we compute the 
one-loop diagrams in order  to avoid double-counting of the symmetry factor 
generated from the diagrams.
Also $U({\bf 12})$  in (\ref{eqn:3.63}) actually means 
$$U({\bf 12})\equiv U({\bf r}_1-{\bf r}_2)\del(t_1-t_2).$$
Thus the above 3-point vertices are nonzero only at the same time 
$t_1=t_2=t_3$. 
Finally, we observe an interesting relationship between the two vertices
$V^{id}_{\hat\rho \rho \rho}({\bf 123})$ 
and $V_{\hat\theta \rho \rho}({\bf 123})$:
\be
\rho_0 T \nabla^2_1 V_{\hat\theta \rho \rho}({\bf 123})
+V^{id}_{\hat\rho \rho \rho}({\bf 123})=0
\la{eqn:3.64}
\ee
which will be useful in computing the one-loop diagrams. 

The quartic and higher order vertices come from the term 
$ -i\hat\theta({\bf 1}) f(\del \rho({\bf 1}))$ in ${\cal S}_{ng}[\psi]$. 
For example, the nonvanishing quartic vertex is given by
\be
V_{\hat\theta \rho\rho\rho}({\bf 1234})=-2\frac{i}{\rho^3_0}\del({\bf 12})\del({\bf 13})
\del({\bf 14})
\la{eqn:3.65}
\ee

\subsection{Time-reversal symmetry and FDR for $G$ and $\Sigma$}
The time-reversed varible set, (\ref{eqn:2.33}), 
now denoted by $\tilde\Psi(j)$, is given by
\be
\tilde\Psi(j)=\Big(\tilde\psi_{\al_j}({\bf x}_j)\Big)=\left(\begin{array}{r}
\delta\rho({\bf x}_j)\qquad\\ -\hat\rho({\bf x}_j)+
i\theta({\bf x}_j)+i{\hat K}\ast \delta\rho({\bf x}_j) \\
 \theta({\bf x}_j)\qquad\\ \hat\theta({\bf x}_j)+i\partial_t\rho({\bf x}_j)
\end{array}\right) \equiv O(\partial_t) \cdot \Psi(j)
\la{eqn:3.66}
\ee
which also defines $O(\partial_t)$.
The transformation property of the propagator $G$ under TR 
with the spatial coordinates being suppressed is then given by
\bea
G(t-t')&=&\Big<\Psi(t) \Psi^T(t') \Big> \nonumber \\
G(t'-t)&=&\Big<\tilde\Psi(t) \tilde\Psi^T(t') \Big>
=\Big< \Big( O(\partial_t)\Psi(t)\Big) \, \Big( O(\partial_{t'})\Psi(t')\Big)^T \Big>
=O(\partial_t) \cdot G(t-t') \cdot O^T(\overleftarrow{\partial_{t'}}) \nonumber \\
&=& O(\partial_t) \cdot G(t-t') \cdot O^T(-\overleftarrow{\partial_t})
\la{eqn:3.67}
\eea
By setting $t'=0$ in (\ref{eqn:3.67}), we have 
\be
G(-t)=O(\partial_t) \cdot G(t) \cdot O^T(-\overleftarrow{\partial_t})
\la{eqn:3.68}
\ee

As for $\Sigma$, using the fact that $\Sigma$ transforms like $G^{-1}$
we obtain
\be 
\Sigma(-t)=\big[O^T(-\partial_t)\big]^{-1} 
\cdot \Sigma(t) \cdot O^{-1}(\overleftarrow{\partial_t}) 
= O^T(\partial_t) \cdot \Sigma(t) \cdot O(-\overleftarrow{\partial_t})
\la{eqn:3.69}
\ee
where we have used the fact that $O,\,O^T$ represent time reversal, 
$O^{-1},\,(O^T)^{-1}$ represent just another time reversal 
which is obtained by changing signs of $t$. 
The transformation matrices appearing in (\ref{eqn:3.68}) and 
(\ref{eqn:3.69}) are explicitly given by 
\bea
&&\left.\begin{array}{ll}
O(\partial_t) \equiv \left(\begin{array}{rrrr}
1\qquad&0&0&0\\ i{\hat K}\ast\qquad&-1&i&0\\0\qquad&0&1&0 \\ i\partial_t\qquad&0&0&1
\end{array}\right), &\quad
O^T(-\overleftarrow{\partial_t}) \equiv \left(\begin{array}{rrrr}
1&i\ast \overleftarrow{{\hat K}}&0&-i\overleftarrow{\partial_t}\\ 0&-1\quad&0&0\\
0&i\quad&1&0 \\ 0&0\quad&0&1
\end{array}\right)
\end{array}\right. \nonumber \\
&&\left.\begin{array}{ll}
O(-\overleftarrow{\partial_t})\equiv \left(\begin{array}{rrrr}
1\quad&0&0&0\\ 
i\ast \overleftarrow{\hat K}\quad&-1&i&0\\
0\quad&0&1&0 \\ 
-i\overleftarrow{\partial_t}\quad&0&0&1
\end{array}\right),&\quad
O^T(\partial_t)\equiv \left(\begin{array}{rrrr}
1&i \hat K*&0&i{\partial_t}\\ 
0&-1\quad&0&0\\
0&i\quad&1&0 \\ 
0&0\quad&0&1
\end{array}\right)
\end{array}\right.
\la{eqn:3.70}
\eea
where $\overleftarrow{\partial_t}$ is a differential operator acting to the left, 
and ${\hat K}\ast (\,  \ast \overleftarrow{\hat K})$ 
means the spatial convolution of  $K({\bf r})$ 
to whatever comes to the right (left).

\subsubsection{FDR for the propagator $G$}
Straightforward algebra gives the expressions for
the elements of the matrices on both sides of 
(\ref{eqn:3.68}):
\bea
G_{\rho\rho}({\bf r},-t)&=&G_{\rho\rho}({\bf r},t) \nonumber\\
G_{\rho\hat\rho}({\bf r},-t)&=&i\hat K* G_{\rho\rho}({\bf r},t)
-G_{\rho\hat\rho}({\bf r},t) +i G_{\rho\theta}({\bf r},t)\nonumber\\
G_{\rho\theta}({\bf r},-t)&=& G_{\rho\theta}({\bf r},t)\nonumber\\
G_{\rho\hat\theta}({\bf r},-t)&=&-i\partial_t G_{\rho\rho}({\bf r},t)
+G_{\rho\hat\theta}({\bf r},t) \nonumber\\
&{}& \nonumber \\
G_{\hat\rho\rho}({\bf r},-t)&=& i\hat K*G_{\rho\rho}({\bf r},t)
-G_{\hat\rho\rho}({\bf r},t) +iG_{\theta\rho}({\bf r},t)\nonumber\\
G_{\hat\rho \hat\rho}({\bf r},-t)&=&       
i\hat K*\Big(i\hat K*G_{\rho\rho}({\bf r},t)
-G_{\rho\hat\rho}({\bf r},t)+iG_{\rho\theta}({\bf r},t) \Big)
-i\Big(\hat K* G_{\hat\rho\rho}({\bf r},t)+G_{\hat\rho\theta}({\bf r},t)\Big) 
\nonumber\\
&+&i\Big(i\hat K*G_{\theta\rho}({\bf r},t)-G_{\theta \hat\rho}({\bf r},t)
+iG_{\theta\theta}({\bf r},t)\Big)=0\nonumber\\ 
G_{\hat\rho\theta}({\bf r},-t)&=& i\hat K* G_{\rho\theta}({\bf r},t)
-G_{\hat\rho\theta}({\bf r},t) + iG_{\theta\theta}({\bf r},t)\nonumber\\
G_{\hat\rho\hat\theta}({\bf r},-t)&=&
i\hat K*\Big(-i\partial_t G_{\rho\rho}({\bf r},t)+G_{\rho\hat\theta}({\bf r},t)\Big)
+i\partial_t G_{\hat\rho\rho}({\bf r},t)+\partial_t G_{\theta\rho}({\bf r},t)
+i G_{\theta\hat\theta}({\bf r},t)=0 \nonumber\\
&{}& \nonumber \\
G_{\theta\rho}({\bf r},-t)&=&G_{\theta\rho}({\bf r},t) \nonumber\\
G_{\theta\hat\rho}({\bf r},-t)&=& i\hat K*G_{\theta\rho}({\bf r},t)
-G_{\theta\hat\rho}({\bf r},t) +iG_{\theta\theta}({\bf r},t)\nonumber\\
G_{\theta\theta}({\bf r},-t)&=& G_{\theta\theta}({\bf r},t)\nonumber\\
G_{\theta\hat\theta}({\bf r},-t)&=& -i\partial_t G_{\theta\rho}({\bf r},t)
+G_{\theta\hat\theta}({\bf r},t)\nonumber\\
&{}& \nonumber \\
G_{\hat\theta\rho}({\bf r},-t)&=& i\partial_t G_{\rho\rho}({\bf r},t)
+G_{\hat\theta\rho}({\bf r},t)\nonumber\\
G_{\hat\theta \hat\rho}({\bf r},-t)&=&i
\partial_t \Big(i\hat K*G_{\rho\rho}({\bf r},t)-G_{\rho\hat\rho}({\bf r},t)
+iG_{\rho\theta}({\bf r},t) \Big)+i\hat K*G_{\hat\theta \rho}({\bf r},t)
+iG_{\hat\theta \theta}({\bf r},t)=0  \nonumber\\
G_{\hat\theta\theta}({\bf r},-t)&=& i\partial_t G_{\rho\theta}({\bf r},t)
+G_{\hat\theta\theta}({\bf r},t)\nonumber\\
G_{\hat\theta\hat\theta}({\bf r},-t)&=&i\partial_t 
\Big(-i\partial_t G_{\rho\rho}({\bf r},t)+G_{\rho\hat\theta}({\bf r},t)\Big)
-i\partial_t G_{\hat\theta \rho}({\bf r},t)=0
\la{eqn:3.71}
\eea
where we used the property of $G_{\hat\alpha \hat\beta}(t)=0$.

Now, various FDRs are obtained from relevant members of (\ref{eqn:3.71}). 
First we look at  the  $[\rho \hat\theta]$ element:
$$G_{\rho\hat\theta}({\bf r},-t)-G_{\rho\hat\theta}({\bf r},t)
=-i\partial_tG_{\rho\rho}({\bf r},t)$$
Since $G_{\rho\hat\theta}({\bf r},-t)=0$ for $t>0$ by causality 
we obtain the standard FDR in view of the response function (\ref{eqn:2.36}) as
\be
R({\bf r},t)=\frac{i}{T}G_{\rho\hat\theta}({\bf r},t)=
-\Theta(t)\frac{1}{T}\partial_tG_{\rho\rho}({\bf r},t)
%\quad [ABL\,\,(76)]
\la{eqn:3.72}
\ee

In the same manner we derive the following FDR's for $G$: 
\bea
G_{\rho\hat\rho}({\bf r},t) &=&
 i\Theta (t)\Big(\hat K*G_{\rho\rho}({\bf r},t)+G_{\rho\theta}({\bf r},t) \Big)
\nonumber \\
 G_{\theta\hat\rho}({\bf r},t)&=&
 i\Theta(t)\Big(\hat K*G_{\theta\rho}({\bf r},t)+G_{\theta\theta}({\bf r},t)\Big) 
 \nonumber \\
G_{\theta\hat\theta}({\bf r},t)&=& i\Theta(t) \partial_tG_{\theta\rho}({\bf r},t)
\la{eqn:3.73}
\eea
%[Caution: there is a singularity at $t=0$.]
Using the causality requirement $G_{\hat\al \beta}(t)=0$ for $t>0$, 
we obtain the FDRs for the adjoint elements:
\bea
G_{\hat\rho\rho}({\bf r},-t)&=& i\Theta(t) 
\Big(\hat K*G_{\rho\rho}({\bf r},t)+G_{\theta\rho}({\bf r},t)\Big)
= G_{\rho \hat\rho}({\bf r},t)\nonumber \\
G_{\hat\theta\rho}({\bf r},-t)&=& i\Theta(t)\partial_t G_{\rho\rho}({\bf r},t)
=G_{\rho \hat\theta}({\bf r},t) \nonumber \\
G_{\hat\rho\theta}({\bf r},-t)&=& i\Theta(t)\Big(\hat K* G_{\rho\theta}({\bf r},t)
 + G_{\theta\theta}({\bf r},t)\Big) 
=G_{\theta \hat\rho}({\bf r},t)\nonumber\\
 G_{\hat\theta\theta}({\bf r},-t)&=& i\Theta(t) \partial_t G_{\rho\theta}({\bf r},t)
 =G_{\theta\hat\theta}({\bf r},t)
 \la{eqn:3.74}
\eea
where $G_{\rho\theta}({\bf r},t)=G_{\theta\rho}({\bf r},t)$ was used.
Four correlations involving only hatted variables vanish. \\
 The second line of (\ref{eqn:3.74}) is nothing but the FDR (\ref{eqn:2.37}). 

\subsubsection{FDR for $\Sigma$}
As before, we first write out 16 matrix elements of (\ref{eqn:3.69}):
\bea
\Sigma_{\rho\rho}({\bf r},-t)&=&
i\hat K*\Sigma_{\rho\hat\rho}({\bf r},t)
-i\partial_t\Sigma_{\rho\hat\theta}({\bf r},t)
+i\hat K*\Big( \Sigma_{\hat\rho\rho}({\bf r},t)
+i\hat K* \Sigma_{\hat\rho \hat\rho}({\bf r},t)
-i\partial_t \Sigma_{\hat\rho \hat\theta}({\bf r},t) \Big) \nonumber \\
&+& i\partial_t \Big(\Sigma_{\hat\theta \rho}({\bf r},t)+i\hat K*
\Sigma_{\hat\theta \hat\rho}({\bf r},t)-i\partial_t 
\Sigma_{\hat\theta \hat\theta}({\bf r},t) \Big)=0 \nonumber \\
\Sigma_{\rho\hat\rho}({\bf r},-t)&=&-\Sigma_{\rho\hat\rho}({\bf r},t)
-i\hat K*\Sigma_{\hat\rho\hat\rho}({\bf r},t)
-i\partial_t\Sigma_{\hat\theta\hat\rho}({\bf r},t)\nonumber\\
\Sigma_{\rho\theta}({\bf r},-t)&=&i\Sigma_{\rho\hat\rho}({\bf r},t)
+ i\hat K*\Big(i\Sigma_{\hat\rho\hat\rho}({\bf r},t)
+\Sigma_{\hat\rho\theta}({\bf r},t)\Big)
+i\partial_t \Big(i\Sigma_{\hat\theta\hat\rho}({\bf r},t)
+\Sigma_{\hat\theta\theta}({\bf r},t)\Big)\nonumber\\
\Sigma_{\rho\hat\theta}({\bf r},-t)&=&\Sigma_{\rho\hat\theta}({\bf r},t)
+i\hat K*\Sigma_{\hat\rho\hat\theta}({\bf r},t)
+i\partial_t\Sigma_{\hat\theta \hat\theta}({\bf r},t)\nonumber\\
&{}& \nonumber \\
\Sigma_{\hat\rho\rho}({\bf r},-t)&=& -\Sigma_{\hat\rho\rho}({\bf r},t)
-i\hat K*\Sigma_{\hat\rho\hat\rho}({\bf r},t)
+i\partial_t \Sigma_{\hat\rho\hat\theta}({\bf r},t) \nonumber\\
\Sigma_{\hat\rho\hat\rho}({\bf r},-t)&=&\Sigma_{\hat\rho\hat\rho}({\bf r},t)\nonumber\\
\Sigma_{\hat\rho\theta}({\bf r},-t)&=&-i\Sigma_{\hat\rho\hat\rho}({\bf r},t)
-\Sigma_{\hat\rho\theta}({\bf r},t)\nonumber\\
\Sigma_{\hat\rho\hat\theta}({\bf r},-t)&=&-\Sigma_{\hat\rho\hat\theta}({\bf r},t)
\nonumber\\
%\mbox{End of upper half}\nonumber\\
&{}& \nonumber \\
\Sigma_{\theta\rho}({\bf r},-t)&=&i\Sigma_{\hat\rho\rho}({\bf r},t)
+i\hat K*\Big(i\Sigma_{\hat\rho\hat\rho}({\bf r},t)
+\Sigma_{\theta\hat\rho}({\bf r},t)\Big)
-i\partial_t\Big(i\Sigma_{\hat\rho\hat\theta}({\bf r},t)
+\Sigma_{\theta\hat\theta}({\bf r},t)\Big)
\nonumber\\
\Sigma_{\theta\hat\rho}({\bf r},-t)&=&-i\Sigma_{\hat\rho\hat\rho}({\bf r},t)
-\Sigma_{\theta\hat\rho}({\bf r},t)\nonumber\\
\Sigma_{\theta\theta}({\bf r},-t)&=&-\Sigma_{\hat\rho\hat\rho}({\bf r},t)
+i\Sigma_{\theta\hat\rho}({\bf r},t)+i\Sigma_{\hat\rho\theta}({\bf r},t)=0\nonumber\\
\Sigma_{\theta\hat\theta}({\bf r},-t)&=&i\Sigma_{\hat\rho\hat\theta}({\bf r},t)
+\Sigma_{\theta \hat\theta}({\bf r},t)\nonumber\\
%\mbox{End of row 3}\nonumber\\
&{}& \nonumber \\
\Sigma_{\hat\theta\rho}({\bf r},-t)&=&\Sigma_{\hat\theta\rho}({\bf r},t)
+i\hat K*\Sigma_{\hat\theta\hat\rho}({\bf r},t)
-i\partial_t \Sigma_{\hat\theta\hat\theta}({\bf r},t)
\nonumber\\
\Sigma_{\hat\theta\hat\rho}({\bf r},-t)&=&-\Sigma_{\hat\theta\hat\rho}({\bf r},t)
\nonumber\\
\Sigma_{\hat\theta\theta}({\bf r},-t)&=&i\Sigma_{\hat\theta\hat\rho}({\bf r},t)
+\Sigma_{\hat\theta\theta}({\bf r},t)
\nonumber\\
\Sigma_{\hat\theta\hat\theta}({\bf r},-t)&=&\Sigma_{\hat\theta\hat\theta}({\bf r},t)
\la{eqn:3.75}
\eea
where we used the property $\Sigma_{\alpha \beta}({\bf r},t)=0$ with unhatted 
$\alpha$ and $\beta$ indices.

Using the causality property $\Sigma_{\hat\al \beta}(-t)=0$ for $t>0$, 
we obtain from (\ref{eqn:3.75}) the following FDRs:
\bea
 \Sigma_{\hat\rho\rho}({\bf r},t)&=& i\Theta(t)
\Big(-\hat K*\Sigma_{\hat\rho\hat\rho}({\bf r},t)
+\partial_t \Sigma_{\hat\rho\hat\theta}({\bf r},t)\Big) \nonumber \\
\Sigma_{\hat\rho\theta}({\bf r},t)&=&-i\Theta(t)\Sigma_{\hat\rho\hat\rho}({\bf r},t)
\nonumber \\
\Sigma_{\hat\theta\rho}({\bf r},t)&=&
i\Theta(t)\Big(-\hat K*\Sigma_{\hat\theta\hat\rho}({\bf r},t)
+\partial_t \Sigma_{\hat\theta\hat\theta}({\bf r},t)\Big) \nonumber \\
\Sigma_{\hat\theta \theta}({\bf r},t)
&=& -i\Theta(t) \Sigma_{\hat\theta \hat\rho}({\bf r},t)
\la{eqn:3.76}
\eea

We obtain the similar FDRs for the adjoint elements using 
$\Sigma_{\al \hat\beta}(t)=0$ for $t>0$:
\bea
\Sigma_{\rho \hat\rho}({\bf r},-t)&=& 
-i\Theta(t)
\Big(\hat K*\Sigma_{\hat\rho\hat\rho}({\bf r},t)
+\partial_t \Sigma_{\hat\theta \hat\rho}({\bf r},t)\Big) \nonumber \\
\Sigma_{\rho \hat\theta}({\bf r},-t) &=&
i\Theta(t)\Big(\hat K*\Sigma_{\hat\rho \hat\theta}({\bf r},t) 
+\partial_t \Sigma_{\hat\theta \hat\theta}({\bf r},t) \Big) \nonumber \\
\Sigma_{\theta \hat\rho}({\bf r},-t) &=&
-i\Theta(t)\Sigma_{\hat\rho \hat\rho}({\bf r},t)=
\Sigma_{\hat\rho \theta}({\bf r},t) \nonumber \\
\Sigma_{\theta \hat\theta}({\bf r},-t)
&=& i\Theta(t) \Sigma_{\hat\rho \hat\theta}({\bf r},t)
\la{eqn:3.77}
\eea

Note from (\ref{eqn:3.25}) that
while $\Sigma_{\hat\rho \hat\rho}({\bf r},t)$ and 
$\Sigma_{\hat\theta \hat\theta}({\bf r},t)$ are symmetric under time-reversal, 
 $\Sigma_{\hat\rho \hat\theta}({\bf r},t)$ and 
$\Sigma_{\hat\theta \hat\rho}({\bf r},t)$ are anti-symmetric under time-reversal. 
The unhatted diagonal elements in (\ref{eqn:3.75}) vanish.

\subsection{Dynamical Equation}
The dynamic equations for the correlation and response functions are formally 
given by the matrix Schwinger-Dyson (SD) equation (see (\ref{eqn:3.59})):
\bea
G_0^{-1}({\bf 13})\cdot G({\bf 32})  =\delta ({\bf 12})+
 \Sigma ({\bf 13}) \cdot G({\bf 32})  
\la{eqn:3.78}
\eea
where the unperturbed propagator $G_0^{-1}({\bf 12})$ is 
given in (\ref{eqn:3.60}).
Setting ${\bf 1}\equiv ({\bf r},t)$, ${\bf 2} \equiv ({\bf 0},0)$, and
${\bf 3}\equiv ({\bf r}_s,s)$, and introducing the space Fourier-transform 
\be
\Sigma({\bf r},t)=\int_{{\bf k}} \Sigma({\bf k},t) e^{i{\bf k}\cdot {\bf r}}
\la{eqn:3.79}
\ee
where $ \int_{{\bf k}} \equiv \int d{\bf k}/(2\pi)^3$, we can express the above matrix SD equation as
\be
\int ds \, G_0^{-1}({\bf k},t-s) \cdot G ({\bf k},s) 
=\delta (t){\bf I}+\int ds \, \Sigma({\bf k},t-s) \cdot G ({\bf k},s) 
\la{eqn:3.80} 
\ee
where ${\bf I}$ is the $4\times4$ unit matrix, and the time-integration
ranges within $(-\infty, \infty)$.
The unperturbed inverse propagator $G_0^{-1}({\bf k},t)$ is calculated from
(\ref{eqn:3.60}) as 
\be
G_0^{-1}({\bf k},t)= \left(\begin{array}{rrrr}
0\qquad\qquad&\ i\Big(\partial_t-\rho_0T k^2 K({\bf k})\Big)\delta(t)&0\qquad\qquad&0\\
i\Big(-\partial_t-\rho_0Tk^2 K({\bf k})\Big)\delta(t)&2\rho_0Tk^2\delta(t)
\quad&\-i\rho_0Tk^2\delta(t)\quad&0\\
0\qquad\qquad&-i\rho_0Tk^2\delta(t)\quad&0\qquad\qquad&-i\delta(t)\\
0\qquad\qquad&0\qquad\qquad&-i\delta(t)\qquad\qquad&0
\end{array}\right)
\la{eqn:3.81}
\ee
Using (\ref{eqn:3.81}), we write down matrix element of lhs of (\ref{eqn:3.80}).
\bea
&{}&\int ds \, G_0^{-1}({\bf k},t-s)\cdot G({\bf k},s)= \nonumber \\
&&\left(\begin{array}{rrrr}
X_{+}G_{\hat\rho\rho}&0&X_{+}G_{\hat\rho\theta}&0\\
X_{-}G_{\rho\rho}+R(2G_{\hat\rho\rho}-iG_{\theta\rho}) \quad&
X_{-}G_{\rho\hat\rho}-iR G_{\theta\hat\rho}\quad&X_{-}G_{\rho\theta}
+R(2G_{\hat\rho\theta}-iG_{\theta\theta})\quad& 
X_{-}G_{\rho\hat\theta}-iR G_{\theta\hat\theta}\\
-iR G_{\hat\rho\rho}-iG_{\hat\theta\rho}&0&
-iR G_{\hat\rho\theta}-iG_{\hat\theta\theta}&0\\
-iG_{\theta\rho}&-iG_{\theta\hat\rho}&-iG_{\theta\theta}&-iG_{\theta\hat\theta}
\end{array}\right) \nonumber \\
\la{eqn:3.82}
\eea
with
\be
X_+ \equiv i\Big(\partial_t-\rho_0T k^2 K({\bf k})\Big), \quad
X_- \equiv i\Big(-\partial_t-\rho_0T k^2 K({\bf k})\Big),
\quad R \equiv \rho_0 T k^2
\la{eqn:3.83}
\ee

We now find the matrix elements of the first term in the rhs of (\ref{eqn:3.80}). 
First of all, 
suppressing the wave number and time integration for the moment and 
denoting $\al,\beta, \gamma =\rho,\theta$, 
we note the following 
\begin{equation*}
[\Sigma\cdot G]_{\hat\al\beta}=\Sigma_{\hat\al\gamma}G_{\gamma\beta}
+\Sigma_{\hat\al\hat\gamma}G_{\hat\gamma\beta}\neq 0, 
\end{equation*}
whereas 
\begin{equation*} 
[\Sigma\cdot G]_{\al\hat\beta}=
\Sigma_{\al\gamma}G_{\gamma\hat\beta}
+\Sigma_{\al\hat\gamma}G_{\hat\gamma\hat\beta}= 0 
\end{equation*}
since $\Sigma_{\al\gamma}\equiv 0$ and $ G_{\hat\gamma\hat\beta}\equiv0$ by causality. 
Therefore we get  
\be
\Sigma\cdot G=
\left(\begin{array}{rrrr}
\Sigma_{\rho\hat\rho}\cdot G_{\hat\rho\rho}+\Sigma_{\rho\hat\theta}
\cdot G_{\hat\theta\rho}\quad &0 \qquad \qquad &\Sigma_{\rho\hat\rho}\cdot 
G_{\hat\rho\theta}+\Sigma_{\rho\hat\theta}\cdot G_{\hat\theta\theta}&0\\
\big[\Sigma\cdot G \big]_{\hat\rho\rho}\qquad&\Sigma_{\hat\rho\rho}\cdot G_{\rho\hat\rho}+
\Sigma_{\hat\rho\theta}\cdot G_{\theta\hat\rho} &[\Sigma\cdot G]_{\hat\rho\theta}
& \quad \Sigma_{\hat\rho\rho}\cdot G_{\rho\hat\theta}+\Sigma_{\hat\rho\theta}
  \cdot G_{\theta\hat\theta}  \\ 
\Sigma_{\theta\hat\rho}\cdot G_{\hat\rho\rho}+\Sigma_{\theta\hat\theta}
\cdot G_{\hat\theta\rho}&0& \quad \Sigma_{\theta\hat\rho}\cdot G_{\hat\rho\theta}+
\Sigma_{\theta\hat\theta}\cdot G_{\hat\theta\theta}&0  \\
\big[\Sigma\cdot G \big]_{\hat\theta\rho} & \quad
\Sigma_{\hat\theta\rho} \cdot G_{\rho\hat\rho}+\Sigma_{\hat\theta\theta}\cdot 
G_{\theta\hat\rho} &\big[\Sigma\cdot G \big]_{\hat\theta\theta}&
 \quad \Sigma_{\hat\theta\rho}\cdot 
G_{\rho\hat\theta}+\Sigma_{\hat\theta\theta}\cdot G_{\theta\hat\theta}
\end{array}\right)
\la{eqn:3.84}
\ee
where
\bea
\big[\Sigma\cdot G \big]_{\hat\rho\rho}&=&\Sigma_{\hat\rho\rho}
\cdot G_{\rho\rho}+\Sigma_{\hat\rho\hat\rho}\cdot 
G_{\hat\rho\rho}+\Sigma_{\hat\rho\theta}\cdot G_{\theta\rho}+
\Sigma_{\hat\rho\hat\theta}\cdot G_{\hat\theta\rho}\nonumber\\
\big[\Sigma\cdot G\big]_{\hat\rho\theta}&=&\Sigma_{\hat\rho\rho}
\cdot G_{\rho\theta}+\Sigma_{\hat\rho\hat\rho}\cdot G_{\hat\rho\theta}+
\Sigma_{\hat\rho\theta}\cdot G_{\theta\theta}+\Sigma_{\hat\rho\hat\theta}\cdot
 G_{\hat\theta\theta}\nonumber\\
\big[\Sigma\cdot G \big]_{\hat\theta\rho}&=&\Sigma_{\hat\theta\rho}
\cdot G_{\rho\rho}+\Sigma_{\hat\theta\hat\rho}\cdot G_{\hat\rho\rho}+
\Sigma_{\hat\theta\theta}\cdot G_{\theta\rho}+\Sigma_{\hat\theta\hat\theta}
\cdot G_{\hat\theta \rho}\nonumber\\
\big[\Sigma\cdot G \big]_{\hat\theta\theta}&=&
\Sigma_{\hat\theta \rho}\cdot G_{\rho\theta}+
\Sigma_{\hat\theta \hat\rho}\cdot G_{\hat\rho\theta}+
\Sigma_{\hat\theta \theta}\cdot G_{\theta\theta}+
\Sigma_{\hat\theta \hat\theta}\cdot G_{\hat\theta\theta}
\la{eqn:3.85}
\eea
Using the equations of various matrices explicitly displayed above
we list below the relevant equations of motion where the matrix element 
is shown as $[\al\beta]$
\bea
%&[11]& \quad i\Big(\partial_t-\rho_0T k^2 K({\bf k})\Big) G_{\hat\rho\rho}({\bf k},t)
%-\int ds \, [\Sigma_{\rho\hat\rho}({\bf k},t-s) 
%G_{\hat\rho\rho}({\bf k},s)
%+\Sigma_{\rho\hat\theta}({\bf k},t-s) G_{\hat\theta\rho}({\bf k},s)]
%=\delta(t)\nonumber\\
%
%&[13]& \quad i\Big(\partial_t-\rho_0T k^2 K({\bf k})\Big) G_{\hat\rho\theta}({\bf k},t)
% -\int ds \, [\Sigma_{\rho\hat\rho}({\bf k},t-s) G_{\hat\rho\theta}({\bf k},s)
%+\Sigma_{\rho\hat\theta}({\bf k},t-s)  G_{\hat\theta\theta}({\bf k},s)]= 0 \nonumber\\
&[21]& \quad i\Big(-\partial_t-\rho_0T k^2 K({\bf k})\Big)G_{\rho\rho}({\bf k},t)
+2\rho_0Tk^2G_{\hat\rho\rho}({\bf k},t)-i\rho_0k^2G_{\theta \rho}({\bf k},t) 
\nonumber \\
&-& \int ds\, \Big[\Sigma_{\hat\rho\rho}({\bf k},t-s)
 G_{\rho\rho}({\bf k},s)+\Sigma_{\hat\rho\hat\rho}({\bf k},t-s)
 G_{\hat\rho\rho}({\bf k},s) \nonumber \\
&+& \Sigma_{\hat\rho\theta}({\bf k},t-s) G_{\theta\rho}({\bf k},s)
+\Sigma_{\hat\rho\hat\theta}({\bf k},t-s) G_{\hat\theta\rho}({\bf k},s)
\Big]={\bf 0} \nonumber\\
&[22]&\quad i\Big(-\partial_t-\rho_0T k^2 K({\bf k})\Big)G_{\rho\hat\rho}({\bf k},t)
-i\rho_0Tk^2 G_{\theta\hat\rho}({\bf k},t)\nonumber \\
&-&\int ds \, \Big[ \Sigma_{\hat\rho\rho}({\bf k},t-s) G_{\rho\hat\rho}({\bf k},s)
+\Sigma_{\hat\rho\theta}({\bf k},t-s) G_{\theta\hat\rho}({\bf k},s)\Big]
=\delta(t) \nonumber\\
&[23]&\quad i\Big(-\partial_t-\rho_0T k^2 K({\bf k})\Big)G_{\rho\theta}({\bf k},t)
+2\rho_0Tk^2 G_{\hat\rho\theta}({\bf k},t)-i\rho_0Tk^2 G_{\theta\theta}({\bf k},t)
\nonumber \\
&-&\int ds \, \Big[ \Sigma_{\hat\rho\rho}({\bf k},t-s)
 G_{\rho\theta}({\bf k},s)+\Sigma_{\hat\rho\hat\rho}({\bf k},t-s)
 G_{\hat\rho\theta}({\bf k},s) \nonumber \\
&+&\Sigma_{\hat\rho\theta}({\bf k},t-s) G_{\theta\theta}({\bf k},s)
+\Sigma_{\hat\rho\hat\theta}({\bf k},t-s)
 G_{\hat\theta\theta}({\bf k},s)\Big] =0 \nonumber\\
&[24]&\quad  i\Big(-\partial_t-\rho_0T k^2 K({\bf k})\Big)G_{\rho\hat\theta}({\bf k},t)
-i\rho_0Tk^2 G_{\theta\hat\theta}({\bf k},t) \nonumber \\
&-&\int ds \,\Big[\Sigma_{\hat\rho\rho}({\bf k},t-s) G_{\rho\hat\theta}({\bf k},s)
+\Sigma_{\hat\rho\theta}({\bf k},t-s) G_{\theta\hat\theta}({\bf k},s)\Big]=0\nonumber\\
%
%&[31]&\quad -i\rho_0Tk^2G_{\hat\rho\rho}({\bf k},t)-iG_{\hat\theta\rho}({\bf k},t)
% -\int ds \,\Big[\Sigma_{\theta\hat\rho}({\bf k},t-s) G_{\hat\rho\rho}({\bf k},s)
%+\Sigma_{\theta\hat\theta}({\bf k},t-s) G_{\hat\theta\rho}({\bf k},s)\Big]=0\nonumber\\
%
%&[33]&\quad -i\rho_0Tk^2 G_{\hat\rho\theta}({\bf k},t)
%-iG_{\hat\theta\theta}({\bf k},t)
%-\int ds \, \Big[ \Sigma_{\theta\hat\rho}({\bf k},t-s)G_{\hat\rho\theta}({\bf k},s)
%+\Sigma_{\theta\hat\theta}({\bf k},t-s) G_{\hat\theta\theta}({\bf k},s)\Big]
%=\delta(t)\nonumber\\
%
&[41]&\quad -iG_{\theta\rho}({\bf k},t)-\int ds\, 
\Big[ \Sigma_{\hat\theta\rho}({\bf k},t-s) G_{\rho\rho}({\bf k},s)
+\Sigma_{\hat\theta\hat\rho}({\bf k},t-s) G_{\hat\rho\rho}({\bf k},s) \nonumber \\
&+& \Sigma_{\hat\theta\theta}({\bf k},t-s) G_{\theta\rho}({\bf k},s)
+\Sigma_{\hat\theta\hat\theta}({\bf k},t-s) G_{\hat\theta\rho}({\bf k},s)\Big]=0\nonumber\\
&[42]&\quad -iG_{\theta\hat\rho}({\bf k},t)
-\int ds\, \Big[ \Sigma_{\hat\theta\rho}({\bf k},t-s) G_{\rho\hat\rho}({\bf k},s)
+\Sigma_{\hat\theta\theta}({\bf k},t-s) G_{\theta\hat\rho}({\bf k},s)\Big]= 0 \nonumber\\
&[43]&\quad -iG_{\theta\theta}({\bf k},t)
-\int ds\, \Big[ \Sigma_{\hat\theta\rho}({\bf k},t-s) G_{\rho\theta}({\bf k},s)+
\Sigma_{\hat\theta\hat\rho}({\bf k},t-s) G_{\hat\rho\theta}({\bf k},s) \nonumber \\
&+&\Sigma_{\hat\theta \theta}({\bf k},t-s) G_{\theta\theta}({\bf k},s)+
\Sigma_{\hat\theta \hat\theta}({\bf k},t-s) G_{\hat\theta\theta}({\bf k},s)\Big]
= 0 \nonumber\\
&[44]&\quad -iG_{\theta \hat\theta}({\bf k},t)
-\int ds \, \Big[\Sigma_{\hat\theta\rho}({\bf k},t-s) G_{\rho\hat\theta}({\bf k},s)
+\Sigma_{\hat\theta\theta}({\bf k},t-s) G_{\theta\hat\theta}({\bf k},s)\Big]
=\delta(t)
\la{eqn:3.86}
\eea

We consider [21] element in (\ref{eqn:3.86}) for $t>0$.
Since $G_{\hat\rho \rho}({\bf k},t)=0$ for $t>0$ by causality, 
one can rewrite [21] element as
\bea
 \partial_t G_{\rho\rho}({\bf k},t)
&=& -\rho_0Tk^2 \Big( K({\bf k}) G_{\rho \rho}({\bf k},t)+G_{\theta \rho}({\bf k},t) \Big)
  \nonumber \\
&+&i \int_{-\infty}^t  ds\, \Big[\Sigma_{\hat\rho\rho}({\bf k},t-s) G_{\rho\rho}({\bf k},s)
+ \Sigma_{\hat\rho\theta}({\bf k},t-s) G_{\theta\rho}({\bf k},s)\Big]\nonumber \\
&+& i\int_{-\infty}^0 ds \, \Big[
\Sigma_{\hat\rho\hat\rho}({\bf k},t-s) G_{\hat\rho\rho}({\bf k},s)
+\Sigma_{\hat\rho\hat\theta}({\bf k},t-s) G_{\hat\theta\rho}({\bf k},s)\Big]
\la{eqn:3.87}
\eea
where we set the upper limits of the integrals using  
the causality properties of the self-energies (response functions) in the first (last)
two integrations: $\Sigma_{\hat\al \beta}({\bf k},t-s)=0$ for $s>t$ and 
$G_{\hat\al \beta}({\bf k},s)=0$ for $s>0$
where $\al, \beta=\rho, \theta$. 
One can simplify the integrals in (\ref{eqn:3.87})
 using the FDRs for the self-energies.
For the first  integral, using (\ref{eqn:3.76}) we obtain
\be
i \int_{-\infty}^t  ds\, \Sigma_{\hat\rho\rho}({\bf k},t-s) 
G_{\rho\rho}({\bf k},s)
=\int_{-\infty}^t  ds\, 
 \Big( K({\bf k})\Sigma_{\hat\rho \hat\rho}({\bf k},t-s) 
- \partial_t \Sigma_{\hat\rho \hat\theta}({\bf k},t-s)  \Big) G_{\rho\rho}({\bf k},s)
\la{eqn:3.88}
\ee
The second term in the rhs of (\ref{eqn:3.88}) can be rewritten by integration by parts
as 
\bea
&&-\int_{-\infty}^t  ds\,\partial_t \Sigma_{\hat\rho \hat\theta}({\bf k},t-s)
 G_{\rho\rho}({\bf k},s)=\int_{-\infty}^t  ds\, \big[\partial_s 
\Sigma_{\hat\rho \hat\theta}({\bf k},t-s)\big] G_{\rho\rho}({\bf k},s) \nonumber \\
 &=& \Sigma_{\hat\rho\hat\theta}({\bf k},0)G_{\rho\rho}({\bf k},t)
- \Sigma_{\hat\rho\hat\theta}({\bf k},\infty)G_{\rho\rho}({\bf k},-\infty)
-\int_{-\infty}^t  ds\, 
\Sigma_{\hat\rho \hat\theta}({\bf k},t-s) \partial_s G_{\rho\rho}({\bf k},s)\nonumber \\
&=& -\int_{-\infty}^t  ds\, 
\Sigma_{\hat\rho \hat\theta}({\bf k},t-s) \partial_s G_{\rho\rho}({\bf k},s)
\la{eqn:3.89}
\eea
where the boundary terms vanish since 
$\Sigma_{\hat\rho\hat\theta}({\bf k},0)=0$ due to the oddness of 
$\Sigma_{\hat\rho\hat\theta}({\bf k},t)$ in time 
(see the $\hat\rho \hat\theta$-element in (\ref{eqn:3.75})), 
and $G_{\rho\rho}({\bf k},-\infty)=0$. 
The second integral in (\ref{eqn:3.87}) can be simplified by use of 
(\ref{eqn:3.76}) as 
\be
i \int_{-\infty}^t  ds\, \Sigma_{\hat\rho\theta}({\bf k},t-s) 
G_{\theta\rho}({\bf k},s) = \int_{-\infty}^t  ds\, 
\Sigma_{\hat\rho\hat\rho}({\bf k},t-s) G_{\theta\rho}({\bf k},s)
\la{eqn:3.90}
\ee
Therefore we obtain for the first two integrals in (\ref{eqn:3.87})
\bea
&i& \int_{-\infty}^t  ds\, \Big[\Sigma_{\hat\rho\rho}({\bf k},t-s) 
G_{\rho\rho}({\bf k},s)
+\Sigma_{\hat\rho\theta}({\bf k},t-s) G_{\theta\rho}({\bf k},s)\Big] \nonumber \\
&=&\int_{-\infty}^t  ds\, 
\Big[\Sigma_{\hat\rho \hat\rho}({\bf k},t-s)
\Big(K({\bf k}) G_{\rho\rho}({\bf k},s)+G_{\theta \rho}({\bf k},s)\Big) 
-\Sigma_{\hat\rho \hat\theta}({\bf k},t-s)\partial_s G_{\rho \rho}({\bf k},s)\Big]
\la{eqn:3.91}
\eea
We now show that  $(-\infty,0)$-contributions of (\ref{eqn:3.91}) cancel
the last two integrals in (\ref{eqn:3.87}).
The third integral in  (\ref{eqn:3.87}) involves the response function
$G_{\hat\rho \rho}(s)$ with $s<0$. 
Using the $\hat\rho \rho$-element of (\ref{eqn:3.71}), we have
\be
G_{\hat\rho \rho}({\bf k},-t)=i \Big(K({\bf k})G_{\rho \rho}({\bf k},t)
+G_{\theta \rho}({\bf k},t)\Big), \quad \mbox{for} \quad t>0
\la{eqn:3.92}
\ee
Setting $t=-s (s<0)$ in (\ref{eqn:3.92}), we obtain
\bea
G_{\hat\rho \rho}({\bf k},s)&=& i \Big(K({\bf k})G_{\rho \rho}({\bf k},-s)
+G_{\theta \rho}({\bf k},-s)\Big), \quad \mbox{for} \quad s<0 \nonumber \\
&=& i \Big(K({\bf k})G_{\rho \rho}({\bf k},s)
+G_{\theta \rho}({\bf k},s)\Big), \quad \mbox{for} \quad s<0 
\la{eqn:3.93}
\eea
where we use the fact that the $\rho$ and $\theta$ field do not change under 
time reversal.
Substituting (\ref{eqn:3.93}) into the third integral in (\ref{eqn:3.87}), we 
obtain
\be
i\int_{-\infty}^0 ds \, \Sigma_{\hat\rho\hat\rho}({\bf k},t-s) 
G_{\hat\rho\rho}({\bf k},s)
= -\int_{-\infty}^0 ds \, \Sigma_{\hat\rho\hat\rho}({\bf k},t-s) 
\Big( K({\bf k})G_{\rho \rho}({\bf k},s)+G_{\theta \rho}({\bf k},s) \Big) 
\la{eqn:3.94}
\ee 
This cancels the $(-\infty,0)$ parts of the first two integrals in (\ref{eqn:3.91}).
Similarly, for the last integral in (\ref{eqn:3.87}), we use 
the FDR, the Fourier transform of the [$\hat\theta\rho$]  element of (\ref{eqn:3.71})
\be
G_{\hat\theta \rho}({\bf k},-t)=i \partial_t G_{\rho \rho}({\bf k},t)
\quad \mbox{for} \quad t>0
\la{eqn:3.95}
\ee
Again setting $t=-s$ in (\ref{eqn:3.95}), we have
\be
G_{\hat\theta \rho}({\bf k},s)=-i \partial_s G_{\rho \rho}({\bf k},-s)
=-i \partial_s G_{\rho \rho}({\bf k},s)
\la{eqn:3.96}
\ee
Substituting (\ref{eqn:3.96}) into the last integral in (\ref{eqn:3.87}), 
we obtain 
\be
i \int_{-\infty}^0 ds \, 
\Sigma_{\hat\rho\hat\theta}({\bf k},t-s) G_{\hat\theta\rho}({\bf k},s)
=\int_{-\infty}^0 ds \, \Sigma_{\hat\rho\hat\theta}({\bf k},t-s)
\partial_s G_{\rho \rho}({\bf k},s), 
\la{eqn:3.97}
\ee
which cancels the $(-\infty, 0)$ part of the  last integral in (\ref{eqn:3.91}).
Using these cancellations, we can write down the final form for the 
dynamic equation for $G_{\rho \rho}({\bf k},t)$ for $t>0$ as 
\bea
 &&\partial_t G_{\rho\rho}({\bf k},t)
= -\rho_0Tk^2 \Big( K({\bf k}) G_{\rho \rho}({\bf k},t)
 + G_{\theta \rho}({\bf k},t) \Big) \nonumber \\
&+&\int_0^t  ds\, \Big[\Sigma_{\hat\rho \hat\rho}({\bf k},t-s)
\Big(K({\bf k}) G_{\rho\rho}({\bf k},s)+G_{\theta \rho}({\bf k},s)\Big) 
-\Sigma_{\hat\rho \hat\theta}({\bf k},t-s)\partial_s G_{\rho \rho}({\bf k},s)\Big]
\la{eqn:3.98}
\eea 

In similar ways, one can obtain  dynamic equations
for the remaining elements of $G$ in (\ref{eqn:3.86}).
The eq. (\ref{eqn:3.98}) is coupled to the eq. for 
$G_{\theta \rho}({\bf k},t)$ which is given by
\bea
G_{\theta \rho}({\bf k},t)&=& \Sigma_{\hat\theta \hat\theta}({\bf k},0)
G_{\rho \rho}({\bf k},t)+ \int_0^t ds \,
   \Sigma_{\hat\theta \hat\rho}({\bf k}, t-s)
   \Big( K({\bf k}) G_{\rho \rho}({\bf k},s)
+  G_{\theta \rho}({\bf k},s) \Big) \nonumber \\
&-& \int_0^t ds \, \Sigma_{\hat\theta \hat\theta}({\bf k}, t-s)
\partial_s G_{\rho \rho}({\bf k},s)
\la{eqn:3.99}
\eea
Likewise, we can obtain the two coupled eqs. for 
$G_{\rho\theta}({\bf k},t)$ and $G_{\theta \theta}({\bf k},t)$:
\bea
\partial_t G_{\rho\theta}({\bf k},t)&=&-\rho_0T k^2
\Big( K({\bf k})G_{\rho\theta}({\bf k},t)+G_{\theta\theta}({\bf k},t) \Big) 
+\int_0^t ds \, \Sigma_{\hat\rho \hat\rho}({\bf k},t-s)
\Big(K({\bf k}) G_{\rho\theta}({\bf k},s) +G_{\theta\theta}({\bf k},s)\Big) 
\nonumber \\
&-&\int_0^t ds \, \Sigma_{\hat\rho\hat\theta}({\bf k},t-s)
\partial_s G_{\rho\theta}({\bf k},s), \nonumber \\
G_{\theta \theta}({\bf k},t)&=& \Sigma_{\hat\theta \hat\theta}({\bf k},0)
G_{\rho \theta}({\bf k},t) 
+ \int_0^t ds \,  \Sigma_{\hat\theta \hat\rho}({\bf k}, t-s)
        \Big( K({\bf k}) G_{\rho \theta}({\bf k},s)+G_{\theta \theta}({\bf k},s)\Big)
        \nonumber \\
 &-& \int_0^t ds \, \Sigma_{\hat\theta \hat\theta}({\bf k}, t-s)
\partial_s G_{\rho \theta}({\bf k},s)
\la{eqn:3.100}
\eea 
In the same manner, we obtain two sets of coupled dynamic eqs. for 
the response functions:
\bea
\partial_t G_{\rho \hat\rho}({\bf k},t)&=&-\rho_0Tk^2K({\bf k})
G_{\rho \hat\rho}({\bf k},t)
-\rho_0Tk^2 G_{\theta\hat\rho}({\bf k},t) \nonumber \\
&+&i \int_0^t ds \, \Big[ \Sigma_{\hat\rho \hat\rho}({\bf k},t-s)
\Big( K^2({\bf k})G_{\rho \rho}({\bf k},s)
+2K({\bf k}) G_{\rho\theta}({\bf k},s)+G_{\theta \theta}({\bf k},s)\Big) \nonumber \\
&-& \partial_t \Sigma_{\hat\rho \hat\theta}({\bf k},t-s)
\Big( K({\bf k})G_{\rho \rho}({\bf k},s)+G_{\rho \theta}({\bf k},s)\Big) \Big]
 \nonumber \\
G_{\theta \hat\rho}({\bf k},t) &=& i \int_0^t ds \, 
\Big[ \Sigma_{\hat\theta \hat\rho}({\bf k},t-s)
\Big( K^2({\bf k})G_{\rho \rho}({\bf k},s)
+2K({\bf k}) G_{\rho\theta}({\bf k},s)+G_{\theta \theta}({\bf k},s)\Big) \nonumber \\
&-& \partial_t \Sigma_{\hat\theta \hat\theta}({\bf k},t-s)
\Big( K({\bf k})G_{\rho \rho}({\bf k},s)+G_{\rho \theta}({\bf k},s)\Big) \Big]
\nonumber \\
\partial_t G_{\rho\hat\theta}({\bf k},t)&=&-\rho_0T k^2 K({\bf k})
G_{\rho\hat\theta}({\bf k},t)
-\rho_0Tk^2 G_{\theta\hat\theta}({\bf k},t) \nonumber \\
&+&i\int_0^t ds \,\Big[ \Sigma_{\hat\rho \hat\rho}({\bf k},t-s)
\Big( K({\bf k})\partial_s G_{\rho \rho}({\bf k},s)
+\partial_s G_{\theta\rho}({\bf k},s)  \Big)
-\partial_t \Sigma_{\hat\rho \hat\theta}({\bf k},t-s)
\partial_s G_{\rho\rho}({\bf k},s)\Big] \nonumber \\
G_{\theta \hat\theta}({\bf k},t)&=& i\delta(t)
+i\int_0^t ds \,\Big[ \Sigma_{\hat\theta \hat\rho}({\bf k},t-s)
\Big(K({\bf k})\partial_s G_{\rho \rho}({\bf k},s)
+\partial_s G_{\theta\rho}({\bf k},s)  \Big) \nonumber \\
&-&\partial_t \Sigma_{\hat\rho \hat\theta}({\bf k},t-s)
\partial_s G_{\rho\rho}({\bf k},s)\Big] \nonumber \\
\la{eqn:3.101}
\eea

\subsection{Static input}
In the present work we are  considering the situation in which the system (from some arbitrary initial state) 
has already evolved into the equilibrium state excluding, however, the crystalline state here and after.
That is, the system is in equilibrium state for $ t \geq 0$, and the present theory aims to describe 
the dynamics of the equilibrium fluctuations given the static information as {\em input}. 
Therefore the static informations of the system enter through the initial values of the dynamic 
correlation functions:
\be
G_{\rho\rho}({\bf k},0)=S_{\rho \rho}({\bf k}), \qquad
G_{\theta\rho}({\bf k},0)=S_{ \theta\rho}({\bf k}), \qquad
G_{\theta\theta}({\bf k},0)=S_{\theta \theta}({\bf k})
\la{eqn:3.102}
\ee
where $S_{\rho \rho}({\bf k})$ etc. are the equilibrium correlation functions.

In Appendix A, we derived the following relation between the equilibrium correlation functions:
\be
K({\bf k})S_{\rho \rho}({\bf k})+S_{\theta \rho}({\bf k})=1.
\la{eqn:3.103}
\ee
We also obtain the following relations by setting $t=0$ in the dynamic equations (\ref{eqn:3.99}) and (\ref{eqn:3.100})
\bea
G_{\theta\rho}({\bf k},0)&=&\Sigma_{\hat\theta \hat\theta}({\bf k},0)
G_{\rho\rho}({\bf k},0) \nonumber \\
G_{\theta \theta}({\bf k},0)&=& \Sigma_{\hat\theta \hat\theta}({\bf k},0)
G_{\rho\theta}({\bf k},0)
\la{eqn:3.104}
\eea
Using the static input (\ref{eqn:3.102}) and the equilibrium relation (\ref{eqn:3.103}), one can 
express $\Sigma_{\hat\theta \hat\theta}({\bf k},0)$ and 
$G_{\theta \theta}({\bf k},0)$ from (\ref{eqn:3.104}) in terms of the static structure factor 
$S_{\rho \rho}({\bf k}) \equiv S({\bf k}) $ as 
\bea
\Sigma_{\hat\theta \hat\theta}({\bf k},0)&=& \frac{S_{\theta \rho}({\bf k})}{S_{\rho \rho}({\bf k})}=\frac{1}{S({\bf k})}-K({\bf k}),  \nonumber \\
G_{\theta \theta}({\bf k},0) &=& S_{\theta \theta}({\bf k})=\big(\frac{1}{S({\bf k})}-K({\bf k}) \big)^2 S({\bf k})
\la{eqn:3.105}
\eea

In Appendix A, we showed that $S_{\rho \theta}({\bf k})=S_{\theta \rho}({\bf k})=0$ in the absence of interaction, $U=0$. 
This means that the initial value of the self-energy $\Sigma_{\hat\theta \hat\theta}({\bf k},t)$ should vanish
for the noninteracting case:
\be
\Sigma_{\hat\theta \hat\theta}({\bf k},0)=0 \quad \mbox{for} \quad U=0
\la{eqn:3.106}
\ee 

If one uses the RY free energy (\ref{eqn:2.3}), then as shown in Appendix A, 
$K({\bf k})$ is equal to the inverse of the static structure factor: $K({\bf k})=S^{-1}({\bf k})$. 
For this case, it implies from (\ref{eqn:3.104}) and (\ref{eqn:3.105}) that 
$\Sigma_{\hat\theta \hat\theta}({\bf k},0)=G_{\theta \rho}({\bf k}, 0)=G_{\theta \theta}({\bf k}, 0)=0 $ 
even for the interacting case.

\subsection{Further nonperturbative results}
We first note that defining
\bea
X({\bf k},t)&\equiv& K({\bf k})G_{\rho \rho}({\bf k},t)+G_{\theta \rho}({\bf k},t)
\nonumber \\
Y({\bf k},t)&\equiv& K({\bf k})G_{\rho \theta}({\bf k},t)+G_{\theta \theta}({\bf k},t), 
\la{eqn:3.107}
\eea
one can rewrite the sets of eqs. (\ref{eqn:3.98}), (\ref{eqn:3.99}) and 
(\ref{eqn:3.100}) in terms of 
$G_{\rho \rho}({\bf k},t)$ and $X({\bf k},t)$,  and 
$G_{\rho \theta}({\bf k},t)$ and $Y({\bf k},t)$ respectively as 
\bea
&&\partial_t G_{\rho\rho}({\bf k},t)
= -\rho_0 T k^2 X({\bf k},t)
+\int_0^t  ds\, \Big[\Sigma_{\hat\rho \hat\rho}({\bf k},t-s)X({\bf k},s)
-\Sigma_{\hat\rho \hat\theta}({\bf k},t-s)\partial_s G_{\rho \rho}({\bf k},s)\Big],
\nonumber \\
&& X({\bf k},t)= \frac{1}{S({\bf k})}G_{\rho \rho}({\bf k},t)+ 
\int_0^t ds \, \Big[  \Sigma_{\hat\theta \hat\rho}({\bf k}, t-s) X({\bf k},s)
- \Sigma_{\hat\theta \hat\theta}({\bf k}, t-s)
\partial_s G_{\rho \rho}({\bf k},s)\Big]
\la{eqn:3.108}
\eea
and 
\bea
&&\partial_t G_{\rho\theta}({\bf k},t)
= -\rho_0 T k^2 Y({\bf k},t)
+\int_0^t  ds\, \Big[\Sigma_{\hat\rho \hat\rho}({\bf k},t-s)Y({\bf k},s)
-\Sigma_{\hat\rho \hat\theta}({\bf k},t-s)\partial_s G_{\rho \theta}({\bf k},s)\Big],
\nonumber \\
&& Y({\bf k},t)= \frac{1}{S({\bf k})}G_{\rho \theta}({\bf k},t)+ 
\int_0^t ds \, \Big[  \Sigma_{\hat\theta \hat\rho}({\bf k}, t-s) Y({\bf k},s)
- \Sigma_{\hat\theta \hat\theta}({\bf k}, t-s)
\partial_s G_{\rho \theta}({\bf k},s)\Big]
\la{eqn:3.109}
\eea
In (\ref{eqn:3.108}) and (\ref{eqn:3.109}), the first member of (\ref{eqn:3.105}) was 
used. One can notice that these two sets of equations share the same structure, 
which leads to interesting results,  as shown below. Defining the Laplace transform 
\be
G^L_{\rho \rho}({\bf k},z)\equiv \int_0^{\infty}dt \,  e^{-zt} \, 
G_{\rho \rho}({\bf k},t), \quad \mbox{etc.}, 
\la{eqn:3.110}
\ee
we  readily obtain the Laplace transform $G^L_{\rho \rho}({\bf k},z)$ from 
(\ref{eqn:3.108}) as
\bea
G^L_{\rho \rho}({\bf k},z)&=& S({\bf k})\cdot 
\Big[z+ \frac{\rho_0Tk^2}{S({\bf k})}\cdot \frac{1}{{\cal D}(\Sigma^L(z))}\Big]^{-1}, 
   \nonumber \\
&{}& \nonumber \\
{\cal D}(\Sigma^L(z))&\equiv& 
\frac{\big(1+\Sigma^L_{\hat\rho \hat\theta}({\bf k},z)\big)
\big(1-\Sigma^L_{\hat\theta \hat\rho}({\bf k},z) \big)}
{1-\Sigma^L_{\hat\rho \hat\rho}({\bf k},z)/\rho_0Tk^2}
-\rho_0Tk^2 
\Sigma^L_{\hat\theta \hat\theta}({\bf k},z)
\la{eqn:3.111}
\eea
Likewise from another set (\ref{eqn:3.109}) we obtain
\bea
G^L_{\rho \theta}({\bf k},z)&=& S_{\rho \theta}({\bf k}) \cdot 
\Big[z+ \frac{\rho_0Tk^2}{S({\bf k})}\cdot \frac{1}{{\cal D}(\Sigma^L(z))}\Big]^{-1}
\la{eqn:3.112}
\eea
The two equations (\ref{eqn:3.111}) and (\ref{eqn:3.112}) imply that the two correlation functions 
$G_{\rho \rho}({\bf k},t)$  and $G_{\rho \theta}({\bf k},t) $ turn out to be the same when each is normalized 
with its own initial value (i.e., equilibrium correlation):
\be
G_{\rho \theta}({\bf k},t)=\frac{S_{\rho \theta}({\bf k})}{S({\bf k})} \cdot G_{\rho \rho}({\bf k},t)
=\Big(\frac{1}{ S({\bf k})}-K({\bf k}) \Big) G_{\rho \rho}({\bf k},t)
\la{eqn:3.113}
\ee
where (\ref{eqn:3.103}) is used.
This relation in turn leads to 
\be
X({\bf k},t)\equiv  K({\bf k})G_{\rho \rho}({\bf k},t)+ G_{\theta \rho}({\bf k},t)
=K({\bf k})G_{\rho \rho}({\bf k},t)+ G_{\rho \theta }({\bf k},t) = \frac{1}{S({\bf k})}G_{\rho \rho}({\bf k},t)
 \la{eqn:3.114}
\ee
where the time-reversal symmetric relation $G_{\theta \rho}({\bf k},t)= G_{\rho \theta}({\bf k},t)$ is used. 
In similar manner,  one obtains  
\bea
Y({\bf k},t) &\equiv&  K({\bf k}) G_{\rho \theta}({\bf k},t)
+G_{\theta \theta}({\bf k},t)=\frac{1}{S({\bf k})}G_{\rho \theta}({\bf k},t) 
=\frac{1}{S({\bf k})} \Big(\frac{1}{ S({\bf k})}-K({\bf k}) \Big) G_{\rho \rho}({\bf k},t),  \nonumber \\
G_{\theta \theta}({\bf k},t)&=&\Big(\frac{1}{S({\bf k})}-K({\bf k})\Big)
G_{\rho \theta}({\bf k},t)=\Big(\frac{1}{S({\bf k})}-K({\bf k})\Big)^2 G_{\rho \rho}({\bf k},t)
\la{eqn:3.115}
\eea
The relations (\ref{eqn:3.114}) and (\ref{eqn:3.115}) with $t=0$ are of course fully consistent with 
(\ref{eqn:3.105}). Note also that (\ref{eqn:3.113}) implies that 
\be
G_{\rho\theta}({\bf k},t)= 0 \quad \mbox{for} \quad U=0 
\la{eqn:3.116}
\ee
since the relation $K({\bf k})=1/S({\bf k})$ holds for the noninteracting case.
Due to the time reversal symmetric relation, we also have 
\bea
G_{\theta \rho}({\bf k},t)= G_{\rho\theta }({\bf k},t)=G_{\theta\theta}({\bf k},t)
= 0 \quad \mbox{for} \quad U=0 
\la{eqn:3.117}
\eea

We further point out that if the RY free energy is employed from the outset, then  we have from (\ref{eqn:3.113})
 and (\ref{eqn:3.115}) that $G_{\rho \theta}({\bf k},t)=G_{\theta \theta}({\bf k},t)=0$ even for the interacting
 case since $K({\bf k})=1/S({\bf k})$. We also note in passing that the second line of  (\ref{eqn:3.108}) 
and (\ref{eqn:3.114}) leads to a nonperturbative identity 
\be
\int_0^t ds \, \Big[\Sigma_{\hat\theta \hat\rho}({\bf k}, t-s) 
\frac{G_{\rho \rho}({\bf k},s)}{S({\bf k})}
- \Sigma_{\hat\theta \hat\theta}({\bf k}, t-s)
\partial_s G_{\rho \rho}({\bf k},s)\Big] =0
\la{eqn:3.118}
\ee

Using (\ref{eqn:3.114}),  one can rewrite the dynamic eq. for $G_{\rho \rho}({\bf k},t)$, (\ref{eqn:3.108}) as
\be 
\partial_t G_{\rho\rho}({\bf k},t)
= -\frac{\rho_0 T k^2}{S({\bf k})} G_{\rho\rho}({\bf k},t)
+\int_0^t  ds\, \Big[\Sigma_{\hat\rho \hat\rho}({\bf k},t-s)\frac
{G_{\rho\rho}({\bf k},s)}{S({\bf k})}
-\Sigma_{\hat\rho \hat\theta}({\bf k},t-s)\partial_s G_{\rho \rho}({\bf k},s)\Big]
\la{eqn:3.119}
\ee
Note that the equation now becomes a closed equation for the density correlation function $G_{\rho \rho}({\bf k},t)$ 
alone since the self-energies $\Sigma_{\hat\rho \hat\rho}({\bf k},t) $ and 
$\Sigma_{\hat\rho \hat\theta}({\bf k},t)$ can be expressed solely in terms of $G_{\rho\rho}({\bf k},t)$ 
via the FDRs (\ref{eqn:3.73}) and the relations (\ref{eqn:3.113}) and the second relation of (\ref{eqn:3.115}).

\section{One-loop reults and MCT equation} 
\subsection{One-loop calculations of the self-energies}
\setcounter{equation}{0}
We now need to express the self energies $\Sigma_{\hat\al \hat\beta}$ 
appearing in (\ref{eqn:3.119})  in terms of $G_{\rho \rho}({\bf k})$. 
Here we obtain this expression up to the one-loop order in the loop expansion
of the dynamic action. As shown in Fig.~4, we have two types of one-loop diagrams. 
But the first diagram which involves $4$-point vertex does not contribute since 
 self-energies appearing in (\ref{eqn:3.119})  are of the form $\Sigma_{\hat\al \hat\beta}({\bf 12})$, and 
there is no 4-point vertex involving two hatted variables. Therefore we only need to consider 
the second one-loop diagrams in Fig.~4, which can be generically written as
\be
\Sigma_{\hat\al \hat\beta}({\bf 12})=
\frac{1}{2} V_{\hat\al \gamma \delta}({\bf 134})V_{\hat\beta \gamma' \delta'}({\bf 256})
G_{\gamma \gamma'}({\bf 35})G_{\delta \delta'}({\bf 46}) 
\la{eqn:4.1}
\ee
where indices $\gamma,\gamma',\delta$ and $\delta'$ include hatted variable indices as well. 

We write down individual expressions of $\Sigma_{\hat\al \hat\beta}({\bf 12})$ diagrams, 
starting with $\Sigma_{\hat\theta\hat\theta}({\bf 12})$
\bea
\Sigma_{\hat\theta\hat\theta}({\bf 12})&=&
\frac{1}{2}V_{\hat\theta\rho\rho}({\bf 134})V_{\hat\theta\rho\rho}({\bf 256})
G_{\rho\rho}({\bf 35})G_{\rho\rho}({\bf 46})
\la{eqn:4.2}
\eea
The diagrammatic expression for (\ref{eqn:4.2}) is shown in Fig.~5.
Note that the symmetry factor generated from the exchange ${\bf 3 \leftrightarrow 4}$ and  
${\bf 5 \leftrightarrow 6}$ is already contained in the definition of the vertex $V_{\hat\theta\rho\rho}$. 
We keep this convention for other symmetry-possessing vertices 
$V^{int, id}_{\hat\rho \rho\rho}({\bf 123})$ and $V_{\hat\rho \rho \hat\rho}({\bf 123})$. 
We also note that since the vertex $ V_{\hat\theta\rho\rho}$ does not involve $U$, 
  one-loop expression (\ref{eqn:4.2}) at $t=0$ 
does not satisfy the nonperturbative requirement (\ref{eqn:3.106}) 
which should hold for the noninteracting case.
Next we write down the expressions for the self-energy 
$\Sigma_{\hat\rho\hat\theta}({\bf 12})$:
\bea
\Sigma_{\hat\rho\hat\theta}({\bf 12})&=&
\sum_{j=1}^4\Sigma^{(j)}_{\hat\rho\hat\theta}({\bf 12})\nonumber \\
\Sigma^{(1)}_{\hat\rho\hat\theta}({\bf 12})&\equiv&
\frac{1}{2}V^{int}_{\hat\rho \rho\rho}({\bf 134})
 V_{\hat\theta\rho\rho}({\bf 256})G_{\rho\rho}({\bf 35})G_{\rho\rho}({\bf 46})
 \nonumber \\
 \Sigma^{(2)}_{\hat\rho\hat\theta}({\bf 12})&\equiv& 
\frac{1}{2}V^{id}_{\hat\rho \rho\rho}({\bf 134})
 V_{\hat\theta\rho\rho}({\bf 256})G_{\rho\rho}({\bf 35})G_{\rho\rho}({\bf 46})
 \nonumber \\
 \Sigma^{(3)}_{\hat\rho\hat\theta}({\bf 12})&\equiv& 
\frac{1}{2}V_{\hat\rho\rho\theta}({\bf 134}) V_{\hat\theta\rho\rho}({\bf 256})
G_{\rho\rho}({\bf 35})G_{\theta\rho}({\bf 46})\times 2 \nonumber\\
 \Sigma^{(4)}_{\hat\rho\hat\theta}({\bf 12})&\equiv&
\frac{1}{2}V_{\hat\rho\hat\rho\rho}({\bf 134}) V_{\hat\theta\rho\rho}({\bf 256})
G_{\hat\rho\rho}({\bf 35})G_{\rho\rho}({\bf 46})\times 2
\la{eqn:4.3}
\eea
We first point out that the diagram $\Sigma^{(4)}_{\hat\rho\hat\theta}({\bf 12}) $ vanishes 
since $G_{\hat\rho \rho}({\bf 35})=0$ by causality for $t_1=t_3=t_4=t$ and $t_2=t_5=t_6=0$ with $t>0$. 
Hence we will omit this type of diagrams from now on. The multiplication factor $2$ in the last two diagrams 
comes from, e.g., 
\bea
\Sigma^{(3)}_{\hat\rho\hat\theta}({\bf 12})&=& 
\frac{1}{2}V_{\hat\rho\rho\theta}({\bf 134}) V_{\hat\theta\rho\rho}({\bf 256})
G_{\rho\rho}({\bf 35})G_{\theta\rho}({\bf 46})
+ \frac{1}{2}V_{\hat\rho\theta\rho}({\bf 134}) V_{\hat\theta\rho\rho}({\bf 256})
G_{\theta\rho}({\bf 35})G_{\rho\rho}({\bf 46}) \nonumber \\
&=& \frac{1}{2}V_{\hat\rho\rho\theta}({\bf 134}) V_{\hat\theta\rho\rho}({\bf 256})
G_{\rho\rho}({\bf 35})G_{\theta\rho}({\bf 46})\times 2
\la{eqn:4.4}
\eea 
The last line follows from the fact that under the exchanges of dummy variables 
${\bf 3} \leftrightarrow {\bf 4}$ and ${\bf 5} \leftrightarrow {\bf 6}$ 
in the second term, the two terms in rhs of the first lines become the same
 since $V_{\hat\rho\theta\rho}({\bf 143})=V_{\hat\rho\rho\theta}({\bf 134})$ 
and $V_{\hat\theta\rho\rho}({\bf 265})=V_{\hat\theta\rho\rho}({\bf 256})$. 
The diagrams for  non-vanishing  $\Sigma_{\hat\rho\hat\theta}({\bf 12})$ is shown in Fig.~6.

We now move on to the self-energies $\Sigma_{\hat\theta \hat\rho}({\bf 12})$ 
and $\Sigma_{\hat\rho \hat\rho}({\bf 12})$.
\bea
\Sigma_{\hat\theta \hat\rho}({\bf 12})&=&
\sum_{j=1}^{4}\Sigma^{(j)}_{\hat\theta \hat\rho}({\bf 12})\nonumber \\
\Sigma^{(1)}_{\hat\theta \hat\rho}({\bf 12})&\equiv&
\frac{1}{2}V_{\hat\theta\rho\rho}({\bf 134})V^{int}_{\hat\rho \rho\rho}({\bf 256})
 G_{\rho\rho}({\bf 35})G_{\rho\rho}({\bf 46}) \nonumber \\
 \Sigma^{(2)}_{\hat\theta \hat\rho}({\bf 12})&\equiv& 
\frac{1}{2}V_{\hat\theta\rho\rho}({\bf 134})V^{id}_{\hat\rho \rho\rho}({\bf 256})
 G_{\rho\rho}({\bf 35})G_{\rho\rho}({\bf 46}) \nonumber \\
 \Sigma^{(3)}_{\hat\theta \hat\rho}({\bf 12})&\equiv& 
\frac{1}{2}V_{\hat\theta\rho\rho}({\bf 134})V_{\hat\rho \rho\theta}({\bf 256})
 G_{\rho\rho}({\bf 35})G_{\rho\theta}({\bf 46}) \times 2 \nonumber \\
\Sigma^{(4)}_{\hat\theta \hat\rho}({\bf 12})&\equiv&  
\frac{1}{2}V_{\hat\theta\rho\rho}({\bf 134}) V_{\hat\rho \rho\hat\rho}({\bf 256}) 
G_{\rho\rho}({\bf 35})G_{\rho\hat\rho}({\bf 46})\times 2 
\la{eqn:4.5}
\eea
The self-energy $\Sigma_{\hat\rho \hat\rho}({\bf 12})$ has $14$ diagrams: 
\bea
\Sigma_{\hat\rho\hat\rho}({\bf 12})&=& \sum_{j=1}^{14}
\Sigma^{(j)}_{\hat\rho\hat\rho}({\bf 12})\nonumber \\
\Sigma^{(1)}_{\hat\rho\hat\rho}({\bf 12})&\equiv&
\frac{1}{2}V^{int}_{\hat\rho\rho\rho}({\bf 134})
V^{int}_{\hat\rho\rho\rho}({\bf 256})G_{\rho\rho}({\bf 35})
G_{\rho\rho}({\bf 46}) \nonumber \\
\Sigma^{(2)}_{\hat\rho\hat\rho}({\bf 12})&\equiv&
\frac{1}{2}V^{int}_{\hat\rho\rho\rho}({\bf 134})
V^{id}_{\hat\rho\rho\rho}({\bf 256})G_{\rho\rho}({\bf 35})
G_{\rho\rho}({\bf 46}) \nonumber \\
\Sigma^{(3)}_{\hat\rho\hat\rho}({\bf 12})&\equiv&
\frac{1}{2}V^{int}_{\hat\rho\rho\rho}({\bf 134})
V_{\hat\rho\rho \hat\rho}({\bf 256})G_{\rho\rho}({\bf 35})
G_{\rho\hat\rho}({\bf 46})\times 2 \nonumber \\
\Sigma^{(4)}_{\hat\rho\hat\rho}({\bf 12})
&\equiv& \frac{1}{2}V^{id}_{\hat\rho\rho\rho}({\bf 134})
V^{int}_{\hat\rho\rho\rho}({\bf 256})G_{\rho\rho}({\bf 35})
G_{\rho\rho}({\bf 46}) \nonumber \\
\Sigma^{(5)}_{\hat\rho\hat\rho}({\bf 12})
&\equiv& \frac{1}{2}V^{id}_{\hat\rho\rho\rho}({\bf 134})
V^{id}_{\hat\rho\rho\rho}({\bf 256})G_{\rho\rho}({\bf 35})
G_{\rho\rho}({\bf 46}) \nonumber \\
\Sigma^{(6)}_{\hat\rho\hat\rho}({\bf 12})&\equiv&
\frac{1}{2}V^{id}_{\hat\rho\rho\rho}({\bf 134})
V_{\hat\rho\rho \hat\rho}({\bf 256})G_{\rho\rho}({\bf 35})
G_{\rho\hat\rho}({\bf 46})\times 2 \nonumber \\
\Sigma^{(7)}_{\hat\rho\hat\rho}({\bf 12})&\equiv&
\frac{1}{2}V^{int}_{\hat\rho\rho\rho}({\bf 134})
V_{\hat\rho\rho\theta}({\bf 256})G_{\rho\rho}({\bf 35})
G_{\rho\theta}({\bf 46}) \times 2 \nonumber \\
\Sigma^{(8)}_{\hat\rho\hat\rho}({\bf 12})&\equiv&
\frac{1}{2}V^{id}_{\hat\rho\rho\rho}({\bf 134})
V_{\hat\rho\rho\theta}({\bf 256})G_{\rho\rho}({\bf 35})
G_{\rho\theta}({\bf 46}) \times 2 \nonumber \\
\Sigma^{(9)}_{\hat\rho\hat\rho}({\bf 12})&\equiv&
\frac{1}{2}V_{\hat\rho\rho\theta}({\bf 134})
V^{int}_{\hat\rho\rho\rho}({\bf 256})
G_{\rho\rho}({\bf 35})
G_{\theta\rho}({\bf 46}) \times 2 \nonumber \\
\Sigma^{(10)}_{\hat\rho\hat\rho}({\bf 12})&\equiv&
\frac{1}{2}V_{\hat\rho\rho\theta}({\bf 134})
V^{id}_{\hat\rho\rho\rho}({\bf 256})
G_{\rho\rho}({\bf 35})
G_{\theta\rho}({\bf 46}) \times 2 \nonumber \\
\Sigma^{(11)}_{\hat\rho\hat\rho}({\bf 12})&\equiv&
\frac{1}{2}V_{\hat\rho\rho\theta}({\bf 134})
V_{\hat\rho\rho\theta}({\bf 256})
G_{\rho\rho}({\bf 35})
G_{\theta\theta}({\bf 46}) \times 2 \nonumber \\
\Sigma^{(12)}_{\hat\rho\hat\rho}({\bf 12})&\equiv&
\frac{1}{2}V_{\hat\rho\rho\theta}({\bf 134})
V_{\hat\rho\theta\rho}({\bf 256})
G_{\rho\theta}({\bf 35})
G_{\theta\rho}({\bf 46}) \times 2 \nonumber \\
\Sigma^{(13)}_{\hat\rho\hat\rho}({\bf 12})&\equiv&
\frac{1}{2}V_{\hat\rho\rho\theta}({\bf 134})
V_{\hat\rho\rho \hat\rho}({\bf 256})
G_{\rho\rho}({\bf 35})
G_{\theta \hat\rho}({\bf 46}) \times 2 \nonumber \\
\Sigma^{(14)}_{\hat\rho\hat\rho}({\bf 12})&\equiv&
\frac{1}{2}V_{\hat\rho\rho\theta}({\bf 134})
V_{\hat\rho \hat\rho \rho}({\bf 256})
G_{\rho \hat\rho}({\bf 35})
G_{\theta \rho}({\bf 46}) \times 2 
\la{eqn:4.6}
\eea
The diagrams for $\Sigma_{\hat\theta \hat\rho}({\bf 12}) $ and 
$\Sigma_{\hat\rho\hat\rho}({\bf 12})$ are shown respectively in Fig.~7 and Fig.~8.

It is convenient to classify  the entire one-loop diagrams into three classes: the diagrams whose 
vertex biproducts possessing no $U$ (Class 0), the ones possessing single $U$ (Class 1), 
and the ones quadratic in $U$ (Class 2) where we note that $U$ occurs only in $V^{int}_{\hat\rho \rho\rho}$ 
(the vertex with filled circle in Fig.~6 through Fig.~8):
\bea
\mbox{Class 0}&=&\Sigma_{\hat\theta \hat\theta}({\bf 12}), 
\quad \Sigma^{(2)}_{\hat\rho \hat\theta}({\bf 12}), 
\quad \Sigma^{(3)}_{\hat\rho \hat\theta}({\bf 12}),
\quad \Sigma^{(2)}_{\hat\theta \hat\rho}({\bf 12}),
\quad \Sigma^{(3)}_{\hat\theta \hat\rho}({\bf 12}),  
\quad \Sigma^{(4)}_{\hat\theta \hat\rho}({\bf 12}),  \nonumber \\ 
&&  \Sigma^{(5)}_{\hat\rho \hat\rho}({\bf 12}),
\quad \Sigma^{(6)}_{\hat\rho \hat\rho}({\bf 12}), 
\quad \Sigma^{(8)}_{\hat\rho \hat\rho}({\bf 12}),
\quad \Sigma^{(10)}_{\hat\rho \hat\rho}({\bf 12}),
\quad \Sigma^{(11)}_{\hat\rho \hat\rho}({\bf 12}), \nonumber \\
&&  \Sigma^{(12)}_{\hat\rho \hat\rho}({\bf 12}),
\quad \Sigma^{(13)}_{\hat\rho \hat\rho}({\bf 12}),
\quad \Sigma^{(14)}_{\hat\rho \hat\rho}({\bf 12}) \nonumber \\
\mbox{Class 1}&=&
\Sigma^{(1)}_{\hat\rho \hat\theta}({\bf 12}), 
\quad \Sigma^{(1)}_{\hat\theta \hat\rho}({\bf 12}), 
\quad \Sigma^{(2)}_{\hat\rho \hat\rho}({\bf 12}), 
\quad \Sigma^{(3)}_{\hat\rho \hat\rho}({\bf 12}),
\quad \Sigma^{(4)}_{\hat\rho \hat\rho}({\bf 12}),
\quad \Sigma^{(7)}_{\hat\rho \hat\rho}({\bf 12}) 
\quad \Sigma^{(9)}_{\hat\rho \hat\rho}({\bf 12}) \nonumber \\
\mbox{Class 2}&=&
\Sigma^{(1)}_{\hat\rho \hat\rho}({\bf 12})
\la{eqn:4.7}
\eea

First we find it convenient to rewrite the convolution integral in 
(\ref{eqn:3.119}) using the identity (\ref{eqn:3.118}) as 
\bea
&&\int_0^t  ds\, \Big[\Sigma_{\hat\rho \hat\rho}({\bf k},t-s)\frac
{G_{\rho\rho}({\bf k},s)}{S({\bf k})}
-\Sigma_{\hat\rho \hat\theta}({\bf k},t-s)\partial_s G_{\rho \rho}({\bf k},s)\Big]
=\nonumber \\
&& \int_0^t  ds\, \Big[\Big(\Sigma_{\hat\rho \hat\rho}
-\rho_0Tk^2 \Sigma_{\hat\theta \hat\rho}\Big)({\bf k},t-s) \frac
{G_{\rho\rho}({\bf k},s)}{S({\bf k})} 
+\Big(\rho_0Tk^2 \Sigma_{\hat\theta \hat\theta}-\Sigma_{\hat\rho \hat\theta}
\Big)({\bf k},t-s)
\dot G_{\rho \rho}({\bf k},s)\Big]
\la{eqn:4.8}
\eea
Next, we note the following relation between the two vertices $V_{\hat\theta \rho\rho}({\bf 123})$
 and $V^{id}_{\hat\rho \rho\rho} ({\bf 123})$ (see (\ref{eqn:3.64})):
\be
\rho_0 T\nabla^2_1 V_{\hat\theta \rho\rho}({\bf 123})+
V^{id}_{\hat\rho \rho \rho}({\bf 123})=0
\la{eqn:4.9}
\ee
This relation reflects the fact that the auxililary field is 
designed to take care of the nonlinearity of the ideal-gas contribution. 
As shown below, this identity leads to the mutual cancellation of the 5 pairs of 
diagrams in the convolution integral in (\ref{eqn:3.80}).
We note the following relation between 
$\Sigma_{\hat\theta \hat\theta}({\bf 12})$ and 
$\Sigma^{(2)}_{\hat\rho \hat\theta}({\bf 12})$ in the convolution integral: 
\be
\rho_0 T \nabla^2_1 \Sigma_{\hat\theta \hat\theta}({\bf 12})
+\Sigma^{(2)}_{\hat\rho \hat\theta}({\bf 12})
=\frac{1}{4}\Big( \rho_0 T\nabla^2_1 V_{\hat\theta \rho\rho}({\bf 134})+
V^{id}_{\hat\rho \rho \rho}({\bf 134})\Big) V_{\hat\theta \rho\rho}
({\bf 256})G_{\rho\rho}({\bf 35}) G_{\rho\rho}({\bf 46})=0 
\la{eqn:4.10}
\ee
The Fourier transform of (\ref{eqn:4.10}) can be written as
\be
\Sigma^{(2)}_{\hat\rho \hat\theta}({\bf k},t)=
\rho_0Tk^2 \Sigma_{\hat\theta \hat\theta}({\bf k},t)
\la{eqn:4.11}
\ee
In the same manner we obtain the same type of relations for 
the other 4 pairs of  one-loop self energies as consequences of 
(\ref{eqn:4.9}):
\bea
\Sigma^{(4)}_{\hat\rho \hat\rho}({\bf k},t)&=&
\rho_0Tk^2 \Sigma^{(1)}_{\hat\theta \hat\rho}({\bf k},t), \qquad
\Sigma^{(5)}_{\hat\rho \hat\rho}({\bf k},t)=
\rho_0Tk^2 \Sigma^{(2)}_{\hat\theta \hat\rho}({\bf k},t),  \nonumber \\
\Sigma^{(6)}_{\hat\rho \hat\rho}({\bf k},t)&=&
\rho_0Tk^2 \Sigma^{(4)}_{\hat\theta \hat\rho}({\bf k},t), \qquad
\Sigma^{(8)}_{\hat\rho \hat\rho}({\bf k},t)=
\rho_0T k^2 \Sigma^{(3)}_{\hat\theta \hat\rho}({\bf k},t), 
\la{eqn:4.12}
\eea
Therefore the kernels $\Sigma_{\hat\theta \hat\rho}({\bf k},t)$ and $\Sigma_{\hat\theta \hat\theta}({\bf k},t)$
 are eliminated from the convolution integrals in (\ref{eqn:4.8}). 
There will be the corresponding cancellations in the higher loop calculations as well. 
Hence the convolution integral (\ref{eqn:4.8}) can now be rewritten as
\be
\int_0^t  ds\, \Big[\Sigma^R_{\hat\rho \hat\rho}({\bf k},t-s) \frac
{G_{\rho\rho}({\bf k},s)}{S({\bf k})} 
-\Big(\Sigma^{(1)}_{\hat\rho \hat\theta}+\Sigma^{(3)}_{\hat\rho \hat\theta}
\Big)({\bf k},t-s)\dot G_{\rho \rho}({\bf k},s)\Big]
\la{eqn:4.13}
\ee
where the superscript $R$ denotes the remaining diagrams in the kernel $\Sigma_{\hat\rho \hat\rho}$.

We now compute the remaining diagrams and list the results below. Details of calculation are given in Appendix B.
We first write down the results for Class 1 and Class 2 diagrams.
\bea
\Sigma^{(1)}_{\hat\rho \hat\theta}({\bf k},t)
&=& -\frac{1}{\rho_0^2} \int_{\bf q} {\bf k}\cdot {\bf q}\,
U({\bf q})G_{\rho \rho}({\bf q},t) G_{\rho \rho}({\bf k}-{\bf q},t) \nonumber \\
\Sigma^{(1)}_{\hat\rho \hat\rho}({\bf k},t)
&=& - \int_{\bf q} \Big[ ({\bf k}\cdot {\bf q})^2 \,
U^2({\bf q})+ ({\bf k}\cdot {\bf q}) ({\bf k}\cdot ({\bf k}- {\bf q}))U({\bf q})
U({\bf k}- {\bf q})) \Big] G_{\rho \rho}({\bf q},t) G_{\rho \rho}({\bf k}-{\bf q},t)
\nonumber \\ 
\Sigma^{(2)}_{\hat\rho \hat\rho}({\bf k},t)
&=& -\frac{T}{\rho_0}k^2 \int_{\bf q} {\bf k}\cdot {\bf q}\,
U({\bf q})G_{\rho \rho}({\bf q},t) G_{\rho \rho}({\bf k}-{\bf q},t) \nonumber \\
\Sigma^{(3)}_{\hat\rho \hat\rho}({\bf k},t)
&=& -2iT \int_{\bf q} \Big[ ({\bf k}\cdot {\bf q})^2 \,
U({\bf q})+ ({\bf k}\cdot {\bf q}) ({\bf k}\cdot ({\bf k}- {\bf q}))
U({\bf k}- {\bf q})) \Big] 
G_{\rho \hat\rho}({\bf q},t) G_{\rho \rho}({\bf k}-{\bf q},t) \nonumber \\
\Sigma^{(7)}_{\hat\rho \hat\rho}({\bf k},t)
&=& -T \int_{\bf q} \Big[ ({\bf k}\cdot {\bf q})^2 \,
U({\bf q})+ ({\bf k}\cdot {\bf q}) ({\bf k}\cdot ({\bf k}- {\bf q}))
U({\bf k}- {\bf q})) \Big] 
G_{\rho \theta}({\bf q},t) G_{\rho \rho}({\bf k}-{\bf q},t) \nonumber \\
\Sigma^{(9)}_{\hat\rho \hat\rho}({\bf k},t)
&=& -T \int_{\bf q} \Big[ ({\bf k}\cdot {\bf q})^2 \,
U({\bf q})+ ({\bf k}\cdot {\bf q}) ({\bf k}\cdot ({\bf k}- {\bf q}))
U({\bf k}- {\bf q})) \Big] 
G_{ \theta \rho }({\bf q},t) G_{\rho \rho}({\bf k}-{\bf q},t) 
\la{eqn:4.14}
\eea
where $\int_{\bf q} \equiv \int d^3 {\bf q}/(2\pi)^3$.
Noting that $\Sigma^{(7)}_{\hat\rho \hat\rho}({\bf k},t)
=\Sigma^{(9)}_{\hat\rho \hat\rho}({\bf k},t) $ (since 
$G_{\rho \theta}({\bf q},t)=G_{ \theta \rho}({\bf q},t)$), and 
$G_{\rho \hat\rho}({\bf q},t)=i \Big(K({\bf q})G_{\rho \rho}({\bf q},t)
+G_{\rho \theta}({\bf q},t) \Big)$ with $K({\bf q})\equiv 1/\rho_0 +U({\bf q})/T$, 
(FDR (\ref{eqn:3.73})), we obtain 
\bea
&& \Big(\Sigma^{(3)}_{\hat\rho \hat\rho}+\Sigma^{(7)}_{\hat\rho \hat\rho}+
\Sigma^{(9)}_{\hat\rho \hat\rho}\Big)({\bf k},t) \nonumber \\
&=& 2T \int_{\bf q} \Big[ ({\bf k}\cdot {\bf q})^2 \,
U({\bf q})+ ({\bf k}\cdot {\bf q}) ({\bf k}\cdot ({\bf k}- {\bf q}))
U({\bf k}- {\bf q})) \Big] 
K({\bf q}) G_{\rho \rho}({\bf q},t) G_{\rho \rho}({\bf k}-{\bf q},t) \nonumber \\
&=& 2\frac{T}{\rho_0} \int_{\bf q} \Big[ ({\bf k}\cdot {\bf q})^2 \,
U({\bf q})+ ({\bf k}\cdot {\bf q}) ({\bf k}\cdot ({\bf k}- {\bf q}))
U({\bf k}- {\bf q})) \Big]G_{\rho \rho}({\bf q},t) G_{\rho \rho}({\bf k}-{\bf q},t) 
\nonumber \\
&+& 2\int_{\bf q} \Big[ ({\bf k}\cdot {\bf q})^2 \,
U^2({\bf q})+ ({\bf k}\cdot {\bf q}) ({\bf k}\cdot ({\bf k}- {\bf q}))U({\bf q})
U({\bf k}- {\bf q})) \Big] G_{\rho \rho}({\bf q},t) G_{\rho \rho}({\bf k}-{\bf q},t) 
 \la{eqn:4.15}
\eea
where the last integral has the same structure as that of 
$\Sigma^{(1)}_{\hat\rho \hat\rho}({\bf k},t) $, (\ref{eqn:4.14}).
The first integral of (\ref{eqn:4.15}) can be simplified by use of
$$\int_{\bf q} ({\bf k}\cdot {\bf q}) ({\bf k}\cdot ({\bf k}- {\bf q}))
U({\bf k}- {\bf q}) G_{\rho \rho}({\bf q},t) G_{\rho \rho}({\bf k}-{\bf q},t)
=\int_{\bf q}({\bf k}\cdot {\bf q}) ({\bf k}\cdot ({\bf k}- {\bf q}))
U({\bf q})G_{\rho \rho}({\bf q},t) G_{\rho \rho}({\bf k}-{\bf q},t)$$ 
which is obtained by shifting the integration variable 
${\bf q} \rightarrow {\bf k}-{\bf q}$.
Using this feature, the first integral in (\ref{eqn:4.15}) is simplified as
\be
 2\frac{T}{\rho_0} \int_{\bf q}\Big[({\bf k}\cdot {\bf q})^2
+ ({\bf k}\cdot {\bf q}) ({\bf k}\cdot ({\bf k}- {\bf q}))\Big]
U({\bf q})G_{\rho \rho}({\bf q},t) G_{\rho \rho}({\bf k}-{\bf q},t) 
= 2\frac{T}{\rho_0} k^2 \int_{\bf q} {\bf k}\cdot {\bf q}\, 
U({\bf q})G_{\rho \rho}({\bf q},t) G_{\rho \rho}({\bf k}-{\bf q},t)
\la{eqn:4.16}
\ee
which is the same integral as that of $\Sigma^{(2)}_{\hat\rho \hat\rho}({\bf k},t) $, (\ref{eqn:4.14}).
Therefore, we obtain 
\bea
&& \Big(\Sigma^{(1)}_{\hat\rho \hat\rho}
+\Sigma^{(2)}_{\hat\rho \hat\rho}+\Sigma^{(3)}_{\hat\rho \hat\rho}+
\Sigma^{(7)}_{\hat\rho \hat\rho}+
\Sigma^{(9)}_{\hat\rho \hat\rho}\Big)({\bf k},t) 
= \frac{T}{\rho_0}k^2 \int_{\bf q} {\bf k}\cdot {\bf q}\,
U({\bf q})G_{\rho \rho}({\bf q},t) G_{\rho \rho}({\bf k}-{\bf q},t) \nonumber \\
&+& \int_{\bf q} \Big[ ({\bf k}\cdot {\bf q})^2 \,
U^2({\bf q})+ ({\bf k}\cdot {\bf q}) ({\bf k}\cdot ({\bf k}- {\bf q}))U({\bf q})
U({\bf k}- {\bf q})) \Big] G_{\rho \rho}({\bf q},t) G_{\rho \rho}({\bf k}-{\bf q},t) 
\la{eqn:4.17}
\eea

Now we calculate and examine Class 0 diagrams.
\bea
\Sigma^{(3)}_{\hat\rho \hat\theta}({\bf k},t)
&=& -\frac{T}{\rho_0^2} \int_{\bf q} {\bf k}\cdot {\bf q}\,
G_{\theta \rho}({\bf q},t) G_{\rho \rho}({\bf k}-{\bf q},t) \nonumber \\
\Sigma^{(10)}_{\hat\rho \hat\rho}({\bf k},t)
&=& - \frac{T^2}{\rho_0} k^2 \int_{\bf q} {\bf k}\cdot {\bf q}\,
G_{\theta \rho}({\bf q},t) G_{\rho \rho}({\bf k}-{\bf q},t) \nonumber \\
\Sigma^{(11)}_{\hat\rho \hat\rho}({\bf k},t)
&=& - T^2 \int_{\bf q} ({\bf k}\cdot {\bf q})^2\,
G_{\theta \theta}({\bf q},t) G_{\rho \rho}({\bf k}-{\bf q},t) \nonumber \\
\Sigma^{(12)}_{\hat\rho \hat\rho}({\bf k},t)
&=& - T^2 \int_{\bf q} ({\bf k}\cdot {\bf q}) ({\bf k}\cdot ({\bf k}- {\bf q}))
G_{\theta \rho}({\bf q},t) G_{\rho \theta}({\bf k}-{\bf q},t) \nonumber \\
\Sigma^{(13)}_{\hat\rho \hat\rho}({\bf k},t)
&=& - 2iT^2 \int_{\bf q} ({\bf k}\cdot {\bf q})^2\,
G_{\theta \hat\rho}({\bf q},t) G_{\rho \rho}({\bf k}-{\bf q},t) \nonumber \\
\Sigma^{(14)}_{\hat\rho \hat\rho}({\bf k},t)
&=& - 2i T^2 \int_{\bf q} ({\bf k}\cdot {\bf q}) ({\bf k}\cdot ({\bf k}- {\bf q}))
G_{\theta \rho}({\bf q},t) G_{\rho \hat\rho}({\bf k}-{\bf q},t)  
\la{eqn:4.18}
\eea

Using the FDR relation, $G_{\theta \hat\rho}({\bf q},t)
=i \Big( K({\bf q})G_{\theta \rho}({\bf q},t)+G_{\theta \theta}({\bf q},t)\Big)$, given in (\ref{eqn:3.73}), we have
\bea
\Big( \Sigma^{(11)}_{\hat\rho \hat\rho}+\Sigma^{(13)}_{\hat\rho \hat\rho}\Big)({\bf k},t)
&=& T^2 \int_{\bf q} ({\bf k}\cdot {\bf q})^2\, G_{\theta \theta}({\bf q},t) G_{\rho \rho}({\bf k}-{\bf q},t) 
+ 2 T^2 \int_{\bf q} ({\bf k}\cdot {\bf q})^2\, K({\bf q})
G_{\theta \rho}({\bf q},t) G_{\rho \rho}({\bf k}-{\bf q},t) \nonumber \\
&=& T^2 \int_{\bf q} ({\bf k}\cdot {\bf q})^2\, \Big( \frac{1}{S^2({\bf q})}-K^2({\bf q})\Big)
G_{\rho \rho}({\bf q},t) G_{\rho \rho}({\bf k}-{\bf q},t) 
\la{eqn:4.19}
\eea
where the last line follows from using (\ref{eqn:3.113}) and (\ref{eqn:3.115}):
\bea
2 K({\bf q})G_{\theta \rho}({\bf q},t)+G_{\theta \theta}({\bf q},t)
&=&\Big[ 2 K({\bf q})\Big(\frac{1}{S({\bf q})}-K({\bf q})\Big)
+ \Big(\frac{1}{S({\bf q})}-K({\bf q})\Big)^2 \Big]  G_{\rho \rho}({\bf q},t) \nonumber \\
&=&\Big( \frac{1}{S^2({\bf q})}-K^2({\bf q})\Big)G_{\rho \rho}({\bf q},t)
\la{eqn:4.20}
\eea
Likewise using the FDR $G_{\rho \hat\rho}({\bf q},t)
=i \Big( K({\bf q})G_{\rho \rho}({\bf q},t)+G_{\rho \theta}({\bf q},t)\Big)$, (\ref{eqn:3.71}), we have 
\bea
&&\Big( \Sigma^{(12)}_{\hat\rho \hat\rho}
+\Sigma^{(14)}_{\hat\rho \hat\rho}\Big)({\bf k},t)
= T^2 \int_{\bf q} ({\bf k}\cdot {\bf q}) ({\bf k}\cdot ({\bf k}-{\bf q}))
G_{\theta \rho}({\bf q},t) G_{\rho \theta}({\bf k}-{\bf q},t) \nonumber \\
&+& 2 T^2 \int_{\bf q} ({\bf k}\cdot {\bf q}) ({\bf k}\cdot ({\bf k}-{\bf q}))
 K({\bf k}-{\bf q}) G_{\theta \rho}({\bf q},t) G_{\rho \rho}({\bf k}-{\bf q},t) 
 \nonumber \\
 &=& T^2 \int_{\bf q} ({\bf k}\cdot {\bf q}) ({\bf k}\cdot ({\bf k}-{\bf q}))
\Big(\frac{1}{S({\bf q})}-K({\bf q}) \Big)
\cdot \Big( \frac{1}{S({\bf k}-{\bf q})}+K({\bf k}-{\bf q})\Big)  
G_{\rho \rho}({\bf q},t) G_{\rho \rho}({\bf k}-{\bf q},t) \nonumber \\
&=& T^2 \int_{\bf q} ({\bf k}\cdot {\bf q}) ({\bf k}\cdot ({\bf k}-{\bf q}))
\frac{1}{2}\Big[ \Big(\frac{1}{S({\bf q})}-K({\bf q}) \Big)
\cdot \Big( \frac{1}{S({\bf k}-{\bf q})}+K({\bf k}-{\bf q})\Big) \nonumber \\
&+& \Big(\frac{1}{S({{\bf k}-\bf q})}-K({{\bf k}-\bf q}) \Big)
\cdot \Big( \frac{1}{S({\bf q})}+K({\bf q})\Big)  \Big] 
 G_{\rho \rho}({\bf q},t) G_{\rho \rho}({\bf k}-{\bf q},t) \nonumber \\ 
&=& T^2 \int_{\bf q} ({\bf k}\cdot {\bf q}) ({\bf k}\cdot ({\bf k}-{\bf q}))
\Big(\frac{1}{S({\bf q})S({\bf k}-{\bf q})}-K({\bf q}) K({\bf k}-{\bf q})\Big) 
G_{\rho \rho}({\bf q},t) G_{\rho \rho}({\bf k}-{\bf q},t) 
 \la{eqn:4.21}
\eea
where the  relation (\ref{eqn:3.113}) was used.

Now we are going to add up all the diagrams of $\Sigma_{\hat\rho \hat\rho}
({\bf k},t)$ where $\Sigma_{\hat\rho \hat\rho}^{(j)}\,\,j=4,5,6,8$ are already taken care of through (\ref{eqn:4.12}).
One can recognize the terms sharing with the same structure in the integrand: \\
(a) $\Sigma^{(10)}_{\hat\rho \hat\rho}({\bf k},t)$ with the first integral of (\ref{eqn:4.17}),\\ 
(b) $\big(\Sigma^{(11)}_{\hat\rho \hat\rho} +\Sigma^{(13)}_{\hat\rho \hat\rho}\big)({\bf k},t) $ with
the first term in the second integral of (\ref{eqn:4.17}),\\ 
(c) $\big(\Sigma^{(12)}_{\hat\rho \hat\rho}
+\Sigma^{(14)}_{\hat\rho \hat\rho}\big)({\bf k},t) $ with
the second term in the second integral of (\ref{eqn:4.17}).\\
We first compute the following terms:
\bea
(a)  &\quad& U({\bf q})-T\Big(\frac{1}{S({\bf q})}-K({\bf q}) \Big)
=U({\bf q})-T\Big(\frac{1}{\rho_0}-c({\bf q})-\Big(\frac{1}{\rho_0}
+\frac{U({\bf q})}{T}\Big) \Big)=2U({\bf q})+Tc({\bf q}),  \nonumber \\
(b)  &\quad& U^2({\bf q})+T^2 \Big( \frac{1}{S^2({\bf q})}-K^2({\bf q})\Big)
=U^2({\bf q})+T^2 \Big[ \Big(\frac{1}{\rho_0}-c({\bf q}))^2
-\Big( \frac{1}{\rho_0}+\frac{U({\bf q})}{T}\Big)^2\Big] \nonumber \\
&=& T^2 c({\bf q})^2-2 \frac{T^2}{\rho_0} \Big(c({\bf q})+\frac{U({\bf q})}{T} \Big), 
\nonumber \\
(c) &\quad&  U({\bf q})U({\bf k}-{\bf q})
+T^2 \Big(\frac{1}{S({\bf q})S({\bf k}-{\bf q})}-K({\bf q}) K({\bf k}-{\bf q})\Big)
\nonumber \\ 
&=&T^2 c({\bf q}) c({\bf k}-{\bf q})-\frac{T^2}{\rho_0}
\Big[ \Big(c({\bf q})+\frac{U({\bf q})}{T} \Big)
+\Big(c({\bf k}-{\bf q})+\frac{U({\bf k}-{\bf q})}{T} \Big) \Big] 
\la{eqn:4.22}
\eea 
where the direct correlation function $c({\bf q})$ is related to the static structure 
factor $S({\bf q})$ as $S({\bf q})=\rho_0/(1-\rho_0 c({\bf q}))$. 
Using (\ref{eqn:4.22}), we obtain
\bea
&&\Sigma_{\hat\rho \hat\rho}({\bf k},t)=
T^2 \int_{\bf q} \Big[({\bf k}\cdot {\bf q})^2 c^2({\bf q}) 
 +({\bf k}\cdot {\bf q}) ({\bf k}\cdot ({\bf k}-{\bf q})) 
c({\bf q}) c({\bf k}-{\bf q})\Big]
G_{\rho \rho}({\bf q},t) G_{\rho \rho}({\bf k}-{\bf q},t) \nonumber \\
&-&2\frac{T^2}{\rho_0} \int_{\bf q} ({\bf k}\cdot {\bf q})^2  
 \Big(c({\bf q})+\frac{U({\bf q})}{T} \Big)
G_{\rho \rho}({\bf q},t) G_{\rho \rho}({\bf k}-{\bf q},t) \nonumber \\
&-& \frac{T^2}{\rho_0} \int_{\bf q} 
({\bf k}\cdot {\bf q}) ({\bf k}\cdot ({\bf k}-{\bf q}))   
 \Big[ \Big(c({\bf q})+\frac{U({\bf q})}{T} \Big)
 +\Big(c({\bf k}-{\bf q})+\frac{U({\bf k}-{\bf q})}{T} \Big) \Big]
G_{\rho \rho}({\bf q},t) G_{\rho \rho}({\bf k}-{\bf q},t) \nonumber \\
&+& \frac{T}{\rho_0} k^2 \int_{\bf q} {\bf k}\cdot {\bf q}\,
\Big[ 2 U({\bf q})+ T c({\bf q})\Big] G_{\theta \rho}({\bf q},t) 
G_{\rho \rho}({\bf k}-{\bf q},t)
\la{eqn:4.23}
\eea
One can further simplify the last three integrals by noting 
the third integral becomes by symmetry 
\bea
&&\int_{\bf q} 
({\bf k}\cdot {\bf q}) ({\bf k}\cdot ({\bf k}-{\bf q}))   
 \Big[ \Big(c({\bf q})+\frac{U({\bf q})}{T} \Big)
 +\Big(c({\bf k}-{\bf q})+\frac{U({\bf k}-{\bf q})}{T} \Big) \Big]
G_{\rho \rho}({\bf q},t) G_{\rho \rho}({\bf k}-{\bf q},t) \nonumber \\
&=& 2 \int_{\bf q} 
({\bf k}\cdot {\bf q}) ({\bf k}\cdot ({\bf k}-{\bf q}))   
\Big(c({\bf q})+\frac{U({\bf q})}{T} \Big)
G_{\rho \rho}({\bf q},t) G_{\rho \rho}({\bf k}-{\bf q},t)
\la{eqn:4.24}
\eea
Then the sum of the last three integrals is given by
\bea
&&\Big(\mbox{the sum of the last three integrals of (\ref{eqn:4.23})}\Big) \nonumber \\
&=& -2\frac{T^2}{\rho_0}k^2 \int_{\bf q} ({\bf k}\cdot {\bf q})
\Big(c({\bf q})+\frac{U({\bf q})}{T} \Big)
G_{\rho \rho}({\bf q},t) G_{\rho \rho}({\bf k}-{\bf q},t) \nonumber \\
&+&  \frac{T}{\rho_0}k^2 \int_{\bf q} ({\bf k}\cdot {\bf q}) 
\Big( 2 U({\bf q})+T c({\bf q}) \Big) 
G_{\rho \rho}({\bf q},t) G_{\rho \rho}({\bf k}-{\bf q},t) \nonumber \\
&=&- \frac{T^2}{\rho_0}k^2 \int_{\bf q} ({\bf k}\cdot {\bf q})
c({\bf q}) G_{\rho \rho}({\bf q},t) G_{\rho \rho}({\bf k}-{\bf q},t)
\la{eqn:4.25}
\eea 
One can also rewrite the first integral in (\ref{eqn:4.23}) as
\bea
&& T^2 \int_{\bf q} \Big[({\bf k}\cdot {\bf q})^2 c^2({\bf q}) 
 +({\bf k}\cdot {\bf q}) ({\bf k}\cdot ({\bf k}-{\bf q})) 
c({\bf q}) c({\bf k}-{\bf q})\Big]
G_{\rho \rho}({\bf q},t) G_{\rho \rho}({\bf k}-{\bf q},t) \nonumber \\
&=& T^2 \int_{\bf q} \Big[\frac{1}{2} ({\bf k}\cdot {\bf q})^2 c^2({\bf q}) 
+\frac{1}{2} ({\bf k}\cdot ({\bf k}-{\bf q}))^2 c^2({\bf k}-{\bf q}) \nonumber \\
 &+& ({\bf k}\cdot {\bf q}) ({\bf k}\cdot ({\bf k}-{\bf q})) 
c({\bf q}) c({\bf k}-{\bf q})\Big]
G_{\rho \rho}({\bf q},t) G_{\rho \rho}({\bf k}-{\bf q},t) \nonumber \\
&=& \frac{T^2}{2} \int_{\bf q} \Big[({\bf k}\cdot {\bf q}) c({\bf q}) 
+({\bf k}\cdot ({\bf k}-{\bf q})) c({\bf k}-{\bf q})\Big]^2
G_{\rho \rho}({\bf q},t) G_{\rho \rho}({\bf k}-{\bf q},t)
\la{eqn:4.26}
\eea
Using (\ref{eqn:4.25}) and (\ref{eqn:4.26}), we obtain final expression for $\Sigma_{\hat\rho \hat\rho}({\bf k},t)$ as
\bea
\Sigma_{\hat\rho \hat\rho}({\bf k},t)&=&
\frac{T^2}{2} \int_{\bf q} \Big[({\bf k}\cdot {\bf q}) c({\bf q}) 
+({\bf k}\cdot ({\bf k}-{\bf q})) c({\bf k}-{\bf q})\Big]^2
G_{\rho \rho}({\bf q},t) G_{\rho \rho}({\bf k}-{\bf q},t) \nonumber \\
&-& \frac{T^2}{\rho_0}k^2 \int_{\bf q} ({\bf k}\cdot {\bf q})
c({\bf q}) G_{\rho \rho}({\bf q},t) G_{\rho \rho}({\bf k}-{\bf q},t)
\la{eqn:4.27}
\eea

There are two remaining kernels,  $\Sigma^{(1)}_{\hat\rho \hat\theta}$, (\ref{eqn:4.14}) and 
$\Sigma^{(3)}_{\hat\rho \hat\theta}$, (\ref{eqn:4.18})   in the convolution integral (\ref{eqn:4.13}). 
The sum of these kernels is given by
\bea
\Big(\Sigma^{(1)}_{\hat\rho \hat\theta}+\Sigma^{(3)}_{\hat\rho \hat\theta}\Big)
({\bf k},t)
&=& -\frac{1}{\rho_0^2} \int_{\bf q} {\bf k}\cdot {\bf q}\,
U({\bf q})G_{\rho \rho}({\bf q},t) G_{\rho \rho}({\bf k}-{\bf q},t)
-\frac{T}{\rho_0^2} \int_{\bf q} {\bf k}\cdot {\bf q}\,
G_{\theta \rho}({\bf q},t) G_{\rho \rho}({\bf k}-{\bf q},t) \nonumber \\
&=& \frac{T}{\rho_0^2} \int_{\bf q} {\bf k}\cdot {\bf q}\,
c({\bf q})G_{\rho \rho}({\bf q},t) G_{\rho \rho}({\bf k}-{\bf q},t)
\la{eqn:4.28}
\eea
where (\ref{eqn:3.113}) and $U({\bf q})+T\big(\frac{1}{S({\bf q})}-K({\bf q}) \big)=-Tc({\bf q})$ is used.

Summing up, we have obtained the following one-loop expressions of the self-energies 
in the nonperturbative dynamic equation for $G_{\rho \rho}({\bf k},t)$ 
\bea
\partial_t G_{\rho\rho}({\bf k},t)
&=& -\frac{\rho_0 T k^2}{S({\bf k})} G_{\rho\rho}({\bf k},t)
+\int_0^t  ds\, \Big[\Sigma_{\hat\rho \hat\rho}({\bf k},t-s)\frac
{G_{\rho\rho}({\bf k},s)}{S({\bf k})}
-\Sigma_{\hat\rho \hat\theta}({\bf k},t-s)\dot G_{\rho \rho}({\bf k},s)\Big]
\nonumber \\
&{}& \nonumber \\
\Sigma_{\hat\rho \hat\rho}({\bf k},t)&=&
\frac{T^2}{2} \int_{\bf q} \Big[V^2({\bf k},{\bf q})-\frac{k^2}{\rho_0}V({\bf k},{\bf q})
\Big] \, G_{\rho \rho}({\bf q},t) G_{\rho \rho}({\bf k}-{\bf q},t), \nonumber \\
\Sigma_{\hat\rho \hat\theta}({\bf k},t)
&=& \frac{T}{2\rho_0^2} \int_{\bf q} V({\bf k},{\bf q}) \, 
G_{\rho \rho}({\bf q},t) G_{\rho \rho}({\bf k}-{\bf q},t), \nonumber \\
V({\bf k},{\bf q})&\equiv& \Big[({\bf k}\cdot {\bf q}) c({\bf q}) 
+({\bf k}\cdot ({\bf k}-{\bf q})) c({\bf k}-{\bf q})\Big]
\la{eqn:4.29}
\eea

\subsection{The MCT equation} 
The equation (\ref{eqn:4.29}) takes a different form from the standard MCT equation.
While the convolution integral in MCT contains only the time derivative of the density
correlation function, the first part of the convolution integral of (\ref{eqn:4.29}) 
involves the density correlation function itself, instead of its time derivative. 
Note that this structure is a {\em nonperturbative} feature, independent of one-loop calculations.
It is also very interesting that  in the convolution integral of (\ref{eqn:4.29}) there is an additional term which involves
the time derivative of the density correlation function. One can appreciate below
the importance of this term in recovering the standard MCT equation. 

The difference between (\ref{eqn:4.29}) and the standard MCT  is quite analogous to that between the 
reducible and irreducible memory functions \cite{cichohess} appearing  in the projection operator approach to 
the dissipative stochastic systems. It has been shown for a general class of dissipative stochastic systems 
obeying the detailed balance condition \cite{kk9597} that the conventional memory function in the exact equation 
for the correlation function, which is obtained from the projection operator method, can be further reduced to  
the so-called irreducible memory function. In particular, the derived exact dynamic equation 
for the correlation function of the slow variable $A(t)$ is given by
\bea
\partial_t C_A(t)&=&-|E_A| C_A(t)+\int_0^t ds \,  M_A(t-s) C_A(s), \nonumber \\
C^L_A(z)&=& C_A(0) \Big[z+|E_A|-M^L_A(z) \Big]^{-1}
\la{eqn:4.30}
\eea
where $|E_A|^{-1}$ is a characteristic  short time scale in the system, and $M_A(t)$ is the 
conventional memory function. The memory function $M_A(t)$ turns out to be further reducible to the 
irreducible memory function $M^{irr}_A(t)$: 
\bea
M_A(t)&=& M^{irr}_A(t)-|E_A|^{-1} \int_0^t ds\, M_A(t-s) M^{irr}_A(s), \nonumber \\
M^L_A(z)&=& \frac{M^{L,irr}_A(z)}{1+|E_A|^{-1}M^{L, irr}_A(z)}
\la{eqn:4.31}
\eea
 Note that $M_A^{L,irr}(z=0)$ can grow indefinitely 
when the global relaxation time grows indefinitely in contrast to $M_A^L(z=0)$.
The above two eqs. lead to the dynamic eq. for $C_A(t)$ 
\bea
\partial_t C_A(t)&=&-|E_A| C_A(t)-|E_A|^{-1} \int_0^t ds\,  M^{irr}_A(t-s) \dot C_A(s)
\nonumber \\
C^L_A(z)&=& C_A(0) \Big[z+\frac{|E_A|}{1+|E_A|^{-1} M^{L,irr}_A(z)} \Big]^{-1}
\la{eqn:4.32}
\eea

For dissipative systems with detailed balance like the one under consideration, 
the mode coupling approximation directly applied to the ususal memory kernel $M_A(t)$ in 
 (\ref{eqn:4.30}) can lead to absurd results, which is not the case for the irreducible memory kernel 
$M_A^{irr}(t)$ in (\ref{eqn:4.32}). See \cite{kk9597}. 
This is expected to be the same for loop expansions \cite{irred}: 
 the dynamic equation for $G_{\rho \rho}({\bf k},t)$ with  one-loop self-energies, (\ref{eqn:4.29}), is 
indeed likely to become unstable in the long-time region.

Invoking the irreducible memory function formulation,  we  rewrite (\ref{eqn:4.29}) into the corresponding 
 form of (\ref{eqn:4.32}) where the convolution integral involves only the time derivative of the density 
correlation function: 
\be
\partial_t G_{\rho\rho}({\bf k},t)= -\frac{\rho_0 T k^2}{S({\bf k})} G_{\rho\rho}({\bf k},t)
-\int_0^t  ds\, {\cal M}({\bf k},t-s)\dot G_{\rho\rho}({\bf k},s)
\la{eqn:4.33}
\ee
The kernel ${\cal M}({\bf k},t)$,  corresponding to the irreducible memory function 
(${\cal M}({\bf k},t)=|E_A|^{-1} M^{irr}_A(t)$ with $|E_A|=\rho_0Tk^2/S({\bf k})$), 
satisfies the following equation: 
\be
{\cal M}({\bf k},t) = \tilde \Sigma_{\hat\rho \hat\rho}({\bf k},t) +\Sigma_{\hat\rho \hat\theta}({\bf k},t)
+\int_0^{t} ds \,  {\cal M}({\bf k},t-s)\tilde \Sigma_{\hat\rho \hat\rho}({\bf k},s)
\la{eqn:4.34}
\ee
where $ \tilde \Sigma_{\hat\rho \hat\rho}({\bf k},t)  \equiv \Sigma_{\hat\rho \hat\rho}({\bf k},t)/(\rho_0Tk^2)$.
It is interesting to recognize the sum of the first two terms in (\ref{eqn:4.34}) is nothing but 
the standard mode coupling kernel: 
\bea
\Sigma_{MC}({\bf k},t) &\equiv& \tilde \Sigma_{\hat\rho \hat\rho}({\bf k},t)
+\Sigma_{\hat\rho \hat\theta}({\bf k},t) \nonumber \\
&=&  \frac{T}{2\rho_0} \int_{\bf q} \Big[(\hat{\bf k}\cdot {\bf q}) c({\bf q}) 
+(\hat{\bf k}\cdot ({\bf k}-{\bf q})) c({\bf k}-{\bf q})\Big]^2 G_{\rho \rho}({\bf q},t) G_{\rho \rho}({\bf k}-{\bf q},t)
\la{eqn:4.35}
\eea
where $\hat {\bf k} \equiv {\bf k}/k$. 

Now in (\ref{eqn:4.34}) when ${\cal M}({\bf k},t)$ is iterated, the convolution integral generates 
the terms $\int_0^{t} ds \,  \Sigma_{MC}({\bf k},t-s)\tilde\Sigma_{\hat\rho \hat\rho}({\bf k},s)+\cdots$.
All these terms are higher-loop terms. We thus see that the structure of the theory becomes so simple that 
only the first contribution  ${\cal M}({\bf k},t)=\Sigma_{MC}({\bf k},t)$
retains the one-loop two-particle irreducible structure. 
Therefore up to the one-loop order, it is perfectly legitimate to take this one-loop contribution only, 
ignoring the terms generated by the convolution integral in (\ref{eqn:3.34}). 
With ${\cal M}({\bf k},t)=\Sigma_{MC}({\bf k},t)$, the equation (\ref{eqn:4.33}) 
reduces to the standard MCT equation: 
\be
\partial_t G_{\rho\rho}({\bf k},t)
= -\frac{\rho_0 T k^2}{S({\bf k})} G_{\rho\rho}({\bf k},t)
-\int_0^t  ds\, \Sigma_{MC} ({\bf k},t-s)\dot G_{\rho\rho}({\bf k},s) 
\la{eqn:4.36}
\ee

\section{Summary and discussion}
\setcounter{equation}{0}
\subsection{Summary}
A field theoretical model for interacting Brownian particle systems is analyzed 
with a view to develop  a renormalized perturbation theory 
consistent with the FDR. 
This is made particularly transparent by introduction of a conjugate pair of 
auxiliary field variables thereby linearizing the TR transformation. 
There is a price to pay for this linearization: a logarithmic nonlinearity reappears in the new 
dynamical action , the second memeber of (\ref{eqn:2.32}), whereas the original dynamical action, 
the first memeber of (\ref{eqn:2.32}), contains only polynomial nonlinearities. 
For non-interacting particle systems we recover a simple diffusion law as one expects, 
which is not the case in some recent works \cite{miya,abl}. For interacting particle cases, we recover the standard MCT in the one-loop  order 
of the renormalized perturbation theory.	

\subsection{Discussion} 
Having worked out the one loop calculation, the next natural step is to undertake higher loop calculations. 
This is anticipated to be an audacious task: one needs to pay meticulous attention to
 all possible terms (or diagrams) up to desired higher order. 
Still this is worthwhile since no systematic calculation of corrections to 
the standard MCT is yet available which is crucial to theoretically  assess successes and inadequacies of the 
standard MCT, and to push the theory beyond the current MCT. 

In order to systematize higher order calculations, one needs to identify a proper smallness parameter 
(or parameters) of the expansion. Such an expansion parameter is denoted as $\lambda$ in (\ref{eqn:3.17}), 
but is equated to unity afterwards  \cite{note2}.

For a Kac-type system the inverse force range is such a smallness parameter. 
However, this cannot be a whole story as indicated by failure of this approach 
to meet requirements of TR-invariance and FDR that ensues. 
Non-trivial effects proportional to the absolute temperature 
arising from nonlinearities of the non-interacting part of the action 
have to be properly included. Development of systematic expansion scheme for 
such model systems would be quite instructive.
In future we may undertake extensions of our approach to treat 
multibody correlations \cite{bb04,bbmr06,iwss2,bb07,bbbkmr}. 
And we may also attempt to extend our approach to genuinely non-equilibrium problems like aging \cite{latz} 
and colloids under external shear flow \cite{bradder}.  

Theory of dynamic aspects, not static aspects, is common to old MCT. Its spirit is to predict dynamics with
statics as input since dynamics is profoundly more difficult. Even this modest aim has not been achieved to 
our satisfaction. 

Now, any satisfactory nonequilibrium theory must describe correctly the dynamics of equilibrium fluctuations.
Our work is only a  first step towards genuine nonequilibrium theory. 

We can compare situation with well-understood critical dynamics. 
There the dynamical renormalization group theory 
 (DRG) successfully combines Wilson's RG theory of statics with the old MCT of critical dynamics \cite{hh77}. 
The resulting DRG is the final theory where indeed statics and dynamics can be treated 
on equal footing and is capable of predicting even minute details. 
Extension of the present formulation to genuine non-equilibrium situation 
should be possible. The problems like nonlinear rheology are now within our reach.

It is our hope to develop non-equilibrium theory of glassy dynamics 
where statics and dynamics can be treated on equal footing eventually. 
However, for that purpose  one first needs a theory which correctly describes dynamics of 
equilibrium fluctuations.  Sophisticated stage of statics of the current liquid state theory \cite{hm06}  
gives us a hope to bring dynamics to comparable stage in future. 

Before we end our discussion, it is worthwhile to make  comments on the recent related works.
Szamel reports in a recent paper \cite{szamel07} a diagrammatic formulation of interacting Brownian 
particles. It is a weak coupling expansion scheme (i.e., the expansion in terms of the bare propagator)
for the hierarchical equations for the multi-density correlation functions. 
We understand possible usefulness of Szamel's approach which does not seem to have problem with the FDR.
In some way his (Andersen's \cite{andersen2} as well) is closer to the original MCT where equilibrium properties are
 included as input. Nonetheless, we wonder how convenient is the calculation. 
Moreover, some uncontrolled aspects enter in obtaining the final hierarchical
 set of equations. Also the higher order structure is not revealed in the formulation.
 Although the formulation uses 'one-loop approximation', it is not a genuine loop expansion
 theory in contrast to our field theory formulation. 

\appendix 
\section{Derivation of the diffusion equation from the ABL action in the absence of 
particle interaction}
We carefully look at the ABL action in the absence of interaction: 
\be
{\cal S}_{ABL, id}[\psi]
= \int \, d {\bf r} \int dt \,
 \Big\{ i\hat\rho \Big[ \partial_t \rho 
-T\nabla \cdot \Big( \rho \nabla \theta \Big)\Big]
  -T \rho ( \nabla \hat \rho)^2 + i\hat\theta 
\Big(\theta-\frac{\delta \rho}{\rho_0}-f(\delta \rho)\Big) \Big\}
\la{eq1}
\ee
First note that  
\be
T \nabla \cdot (\rho \nabla \theta) \equiv \nabla \cdot 
\big( \rho \nabla \frac{\delta F_{id}}{\delta \rho} \big)= T\nabla^2 \rho 
\la{eq2} 
\ee
We rewrite the action (\ref{eq1}) as 
\be
{\cal S}_{ABL, id}[\psi]
= \int \, d {\bf r} \int dt \,
 \Big\{ i\hat\rho \Big[ \partial_t \rho -T\nabla^2 \rho + \u{T\nabla^2 \rho}
-\u{T\nabla \cdot \Big( \rho \nabla \theta \Big)}\Big]
  -T \rho ( \nabla \hat \rho)^2 + i\hat\theta 
\Big(\theta-\frac{\delta \rho}{\rho_0}-f(\delta \rho)\Big) \Big\}
\la{eq3}
\ee
where the two underlined terms cancel due to (\ref{eq2}).

We make use of the following identities
\be
\Big<\delta \rho({\bf 2}) \frac{\delta {\cal S}_{ABL, id}[\psi]}
{\delta {\hat \rho} ({\bf 1})} \Big> =0,
\qquad
\Big< \delta \rho({\bf 2}) \frac{\delta {\cal S}_{ABL, id}[\psi]}
{\delta  \theta ({\bf 1})} \Big> =0
\la{eq4}
\ee
where ${\bf 1} \equiv ({\bf r}, t)$ and ${\bf 2} \equiv ({\bf 0}, 0)$.
The first identity can be written explicitly as
\be
0=\Big<\delta \rho({\bf 2}) \frac{\delta {\cal S}_{id}[\psi]}
{\delta \hat \rho ({\bf 1})} \Big>
= i\Big(\frac{\partial}{\partial t} - T \nabla^2  \Big)G_{\rho \rho}
({\bf1}-{\bf 2}) +2T\rho_0 \nabla^2 \big< \hat \rho({\bf 1})
 \delta \rho({\bf 2}) \big> 
+ 2T \big< \delta \rho ({\bf 2}) \nabla \cdot \big(\delta \rho ({\bf 1}) \nabla
\hat \rho ({\bf 1}) \big)\big>
\la{eq5}
\ee
where we used the fact that the sum of the two underlined terms in 
(\ref{eq3}) vanishes. 
Similarly, using the second identity in (\ref{eq4}),
we obtain
\be
0=\Big<\delta \rho({\bf 2}) \frac{\delta
{\cal S}_{id}[\psi]}{\delta \theta ({\bf 1})} \Big>
=-i\rho_0 T \nabla^2 \big< \hat \rho ({\bf 1}) \delta \rho({\bf 2}) \big>
+i\big<\hat \theta({\bf 1}) \delta \rho({\bf 2}) \big> 
-i T \big< \delta \rho ({\bf 2}) \nabla \cdot \big(\delta \rho ({\bf 1}) \nabla
\hat \rho ({\bf 1}) \big)\big>
\la{eq6}
\ee
where cancellation of the underlined terms was not used.
Since in (\ref{eq6}), 
$\big< \hat \rho ({\bf 1}) \delta \rho({\bf 2}) \big>
=\big<\hat \theta({\bf 1}) \delta \rho({\bf 2}) \big>=0$ 
for $t > 0$ by causality, we obtain
\be
\big< \delta \rho ({\bf 2}) \nabla \cdot \big(\delta \rho ({\bf 1}) \nabla
\hat \rho ({\bf 1}) \big)\big>=0   \quad \mbox{for} \quad t >0
\la{eq7}
\ee
The eqs (\ref{eq6}), (\ref{eq7}),  and causality leads to 
\be
\partial_t G_{\rho \rho}({\bf r},t)=T\nabla^2 G_{\rho \rho}({\bf r},t), \quad
\mbox{for} \quad t>0.
\la{eq8}
\ee
In this way, we recover the diffusion eq. from the ABL action.

In Sec. II.G, we have obtained ((II.46))
\be
\partial_t G_{\rho \rho}({\bf r},t)=T\nabla^2 G_{\rho \rho}({\bf r},t)
+\rho_0 T \nabla^2 \Big<f(\del \rho({\bf r},t)\del \rho({\bf 0},0) \Big>
+T\nabla \cdot 
\Big<\del \rho({\bf r},t)\nabla \theta ({\bf r},t) \del \rho({\bf 0},0)\Big> 
\la{eq9}
\ee
Here if the constraint $\theta \equiv \delta \rho/\rho_0+ f(\delta \rho)$ is used, the last two terms cancel:
\bea
\rho_0 T \nabla^2 f(\delta \rho)+T\nabla \cdot \big( \delta \rho \nabla \theta \big)
&=& \rho_0 T \nabla \cdot \Big( \nabla f + \frac{\delta \rho}{\rho_0} \nabla \theta \Big) 
= \rho_0 T \nabla \cdot \Big( \nabla \theta -\frac{\nabla \rho}{\rho_0}
+\frac{\delta \rho}{\rho_0} \nabla \theta \Big) \nonumber \\ 
&=& \rho_0 T \nabla \cdot \Big( \frac{\rho}{\rho_0}\nabla \theta -\frac{\nabla \rho}{\rho_0}\Big)=0
\la{eq10}
\eea
where the last equality holds due to (\ref{eq2}). Therefore (\ref{eq9}) reduces to the diffusion eq. (\ref{eq8}).
It is extremely puzzling then why ABL's own analysis does not yield (\ref{eq8}) in the 
absence of particle interaction.

\section{Equal time correlations}
\setcounter{equation}{0}
\subsection{Non-interacting particles}
We begin with the microscopic density 
 $\hat\rho({\bf r})\equiv \sum_j\delta({\bf r-r}_j)$.
This should not be confused with the field conjugate to $\rho$. 
Then
\bea
&{}&<\hat\rho({\bf r})\hat\rho({\bf r}')>=
\sum_{jl}<\delta({\bf r-r}_j)\delta({\bf r'-r}_l)>
=\sum_{j=l}\cdots +\sum_{j\neq l}\cdots\nonumber\\
&=&N\delta({\bf r-r}')<\delta({\bf r-r}_1)>
+N(N-1)<\delta({\bf r-r}_1)><\delta({\bf r-r}_2)>\nonumber \\
&=&\rho_0\delta({\bf r-r}')+\rho_0^2
\la{eqn:a1}
\eea
Since $<\hat \rho({\bf r})>=\rho_0$ we get
\be
S_{\rho\rho}({\bf r-r'})=<\delta\hat\rho({\bf r})\delta\hat\rho({\bf r'})>
=\rho_0\delta({\bf r-r'})
\la{eqn:a2}
\ee

\subsection{Interacting particles}
We use (\ref{eqn:2.24})
\be
\theta({\bf r})=\beta\frac{\delta F_{id}}{\delta\rho({\bf r})}
-\frac{\delta\rho({\bf r})}{\rho_0}
\la{eqn:a3}
\ee
where $\beta\equiv 1/T$. We first show that the variable $\theta$ has zero average:
\bea
<\theta({\bf r})>&=&\beta<\frac{\delta F_{id}}{\delta \rho({\bf r})}>
=\beta<\frac{\delta F}{\delta \rho({\bf r})}>
-\beta<\frac{\delta F_{int}}{\delta \rho({\bf r})}>
=\beta<\frac{\delta F}{\delta \rho({\bf r})}>\nonumber \\
&=&-\int d\{\rho\}\frac{\delta}{\delta \rho({\bf r})}e^{-\beta F}=0
\la{eqn:a4}
\eea
 where we have used the fact that 
$\delta F_{int}/\delta \rho({\bf r})$ is linear in $\delta\rho({\bf r})$ 
whose average vanishes. 
Next we turn to another correlation:
\bea
S_{\rho\theta}({\bf r-r'})&\equiv& <\delta\rho({\bf r})\theta ({\bf r}')>=
\Big<\delta\rho({\bf r})
\Big(\beta\frac{\delta F_{id}}{\delta \rho({\bf r}')}
-\frac{\delta \rho({\bf r}')}{\rho_0} \Big)\Big>  \nonumber \\
&=&\Big<\delta\rho({\bf r})\beta\frac{\delta F}{\delta \rho({\bf r}')}\Big>
-\Big<\delta\rho({\bf r}) \hat K*\delta\rho({\bf r}')\Big>\nonumber\\
&=&\delta({\bf r-r}')-\hat K*S_{\rho\rho}({\bf r-r}')
\la{eqn:a5}
\eea
where $K({\bf r})\equiv \Big(\delta({\bf r})/\rho_0+\beta U({\bf r})\Big)$.
The Fourier transform of (\ref{eqn:a6}) is given by 
\bea
&& K({\bf k}) S_{\rho \rho}({\bf k})+S_{\rho \theta}({\bf k})=1 \nonumber \\
&& K({\bf k})=\frac{1}{\rho_0}+\frac{1}{T}U({\bf k})
\la{eqn:a6}
\eea
Note that since $S_{\rho \rho}({\bf k})=\rho_0$ ((\ref{eqn:a2}))
in the noninteracting system ($U\equiv0$), $S_{\rho\theta}({\bf k})$ vanishes for 
the noninteracting case:
\be
S_{\rho\theta}({\bf k})=1-\frac{1}{\rho_0}\rho_0 =0 \quad \mbox{for} \quad U=0
\la{eqn:a7}
\ee

If at the outset  the RY free energy functional (\ref{eqn:2.3}) was used, then the function $K({\bf k})$ 
in (\ref{eqn:a6}) is the inverse of the static structure factor $S_{\rho\rho}({\bf k})$ since 
\be
K({\bf k})=\frac{1}{\rho_0}-c({\bf k})=\frac{1}{S_{\rho\rho}({\bf k})}
\la{eqn:a8}
\ee
This relation implies from (\ref{eqn:a6}) that the equal-time correlation $S_{\rho \theta}({\bf k})$ vanishes 
even in the presence of the particle interaction:
\be
S_{\rho\theta}({\bf k})=0
\la{eqn:a9}
\ee

Finally we consider 
\bea
S_{\theta\theta}({\bf r-r}')&=& \Big< \theta({\bf r})\theta({\bf r}') \Big>
= \Big< \Big(\beta\frac{\delta F}{\delta \rho({\bf r})}
- \hat K*\delta\rho({\bf r})\Big)
\Big(\beta\frac{\delta F}{\delta \rho({\bf r}')}
-\hat K*\delta\rho({\bf r}')\Big)  \Big> \nonumber\\
&=&\beta^2\Big<\frac{\delta F}{\delta \rho({\bf r})}
\frac{\delta F}{\delta \rho({\bf r}')}\Big>
-2 \hat K*\delta({\bf r-r}')
+ \hat K*\hat K*S_{\rho\rho}({\bf r-r}')\nonumber\\
&=&\beta\Big<\frac{\delta^2 F}{\delta \rho({\bf r}) \delta \rho({\bf r}')}\Big>
-2 K({\bf r-r}')
+ \hat K*\hat K*S_{\rho\rho}({\bf r-r}')
\la{eqn:a11}
\eea
This is different from the direct correlation functions as defined by
\be
c^{(2)}({\bf r-r}')\equiv -\beta\frac{\delta^2 {\cal F}^{ex}}{\delta \rho({\bf r})\delta({\bf r}')}
\la{eqn:a12}
\ee
where ${\cal F}^{ex}={\cal F}-{\cal F}^{id}$ is the interaction part (excess part) of the full 
(i.e. renormalized with respect to fluctuations) density functional.

\section{Calculations of $\Sigma_{\hat\rho \hat\rho}$ 
and $\Sigma_{\hat\rho \hat\theta}$}
\setcounter{equation}{0}
Here we derive the equations (\ref{eqn:4.14}) and (\ref{eqn:4.18}).
Referring to (\ref{eqn:4.3}) for $\Sigma_{\hat\rho\hat\theta} $ 
and to (\ref{eqn:3.63}) for the $V$'s we find
\subsection {Derivation of (\ref{eqn:4.14})}
\bea
&&\Sigma^{(1)}_{\hat\rho \hat\theta}({\bf 12})=
\frac{1}{2} \int_{\bf 3456} V^{int}_{\hat\rho \rho\rho}({\bf 134})
V_{\hat\theta \rho\rho}({\bf 256})G_{\rho\rho}({\bf 35})G_{\rho\rho}({\bf 46})
\nonumber \\
&=&\frac{1}{2} \int_{\bf 3456} (-i)\nabla_1 \cdot \big[\del({\bf 13})
\nabla_1 U({\bf 14})+\del({\bf 14})\nabla_1 U({\bf 13})\big]\cdot
\frac{i}{\rho^2_0}\del({\bf 25})\del({\bf 26})
G_{\rho\rho}({\bf 35})G_{\rho\rho}({\bf 46})\nonumber \\
&=& \frac{1}{2\rho^2_0}\int_{\bf 34}\nabla_1 \cdot \big[\del({\bf 13})
\nabla_1 U({\bf 14})+\del({\bf 14})\nabla_1 U({\bf 13})\big]
G_{\rho\rho}({\bf 32})G_{\rho\rho}({\bf 42})\nonumber \\
&=& \frac{1}{\rho^2_0}\int_{\bf 34}\nabla_1 \cdot \big[\del({\bf 13})
\nabla_1 U({\bf 14})\big]G_{\rho\rho}({\bf 32})G_{\rho\rho}({\bf 42}) \nonumber \\
&=& \frac{1}{\rho^2_0} \nabla_1 \cdot \big[G_{\rho\rho}({\bf 12})
\nabla_{1}\hat U*G_{\rho\rho}({\bf 12})\big]
\la{eqn:b.1}
\eea
where $\int_{\bf 3456}\equiv \int d{\bf 3} \int d{\bf 4} \int d{\bf 5}
 \int d{\bf 6}$, etc.
With ${\bf 1}=({\bf r},t)$ and ${\bf 2}=({\bf 0},0)$, 
the spatial Fourier transform of (\ref{eqn:b.1}) is given by
\be
\Sigma^{(1)}_{\hat\rho \hat\theta}({\bf k},t)
=-\frac{1}{\rho^2_0} \int_{\bf q} {\bf k}\cdot {\bf q} \, 
U({\bf q}) G_{\rho \rho}({\bf q},t) G_{\rho \rho}({\bf k- q},t)
\la{eqn:b.2}
\ee
which is the first member of (\ref{eqn:4.14}). 
Next we find similarly from (\ref{eqn:4.6}) and (\ref{eqn:3.63})
\bea
&&\Sigma^{(1)}_{\hat\rho \hat\rho}({\bf 12})
=\frac{1}{2} \int_{\bf 3456} V^{int}_{\hat\rho \rho\rho}({\bf 134})
V^{int}_{\hat\rho \rho\rho}({\bf 256})G_{\rho\rho}({\bf 35})G_{\rho\rho}({\bf 46})
\nonumber \\
&=& \frac{1}{2} \int_{\bf 3456} 
(-i)\nabla_1 \cdot \big[\del({\bf 13})
\nabla_1 U({\bf 14})+\del({\bf 14})\nabla_1 U({\bf 13})\big] \nonumber \\
&\cdot& (-i)\nabla_2 \cdot \big[\del({\bf 25})
\nabla_2 U({\bf 26})+\del({\bf 26})\nabla_2 U({\bf 25})\big]
 G_{\rho\rho}({\bf 35})G_{\rho\rho}({\bf 46}) \nonumber \\
&=& -\int_{\bf 3456} \nabla^j_1\big[\del({\bf 13})\nabla^j_1 U({\bf 14})\big]
\nabla^l_2\big[\del({\bf 25})\nabla^l_2 U({\bf 26})\big] \nonumber \\
&\cdot& \big[G_{\rho\rho}({\bf 35})G_{\rho\rho}({\bf 46})
+G_{\rho\rho}({\bf 36})G_{\rho\rho}({\bf 45}) \big]
\la{eqn:b.3}
\eea  
The first term of (\ref{eqn:b.3}) is computed as
\bea
&&\Sigma^{(1.a)}_{\hat\rho \hat\rho}({\bf 12})\equiv
-\int_{\bf 3456} \nabla^j_1\big[\del({\bf 13})\nabla^j_1 U({\bf 14})\big]\,\,
\nabla^l_2\big[\del({\bf 25})\nabla^l_2 U({\bf 26})\big] 
  \, G_{\rho\rho}({\bf 35})G_{\rho\rho}({\bf 46}) \nonumber \\
&=& -\int_{\bf 56} \nabla^j_1\big[G_{\rho\rho}({\bf 15})\nabla^j_1 
\hat U*G_{\rho\rho}({\bf 16})\big] 
\nabla^l_2\big[\del({\bf 25})\nabla^l_2 U({\bf 26})\big] \nonumber \\
&=& -\nabla^j_1 \nabla^l_2 \big[ G_{\rho\rho}({\bf 12})
\nabla^l_2 \nabla^j_1 \hat U* \hat G_{\rho\rho}*U ({\bf 12}) \big]
\nonumber \\
&=& -\nabla^j_1 \nabla^l_1 \big[ G_{\rho\rho}({\bf 12})
\nabla^l_1 \nabla^j_1 \hat U* \hat G_{\rho\rho}*U ({\bf 12}) \big]
\la{eqn:b.4}
\eea
where $\hat G*$ implies convolution with the function $G({\bf r})$ 
as in the case of $\hat U*$.
The second term of (\ref{eqn:b.3}) is similarly computed as
\bea
&&\Sigma^{(1.b)}_{\hat\rho \hat\rho}({\bf 12})\equiv
-\int_{\bf 3456} \nabla^j_1\big[\del({\bf 13})\nabla^j_1 U({\bf 14})\big]\,\,
\nabla^l_2\big[\del({\bf 25})\nabla^l_2 U({\bf 26})\big] 
  \, G_{\rho\rho}({\bf 36})G_{\rho\rho}({\bf 45}) \nonumber \\
&=&-\int_{\bf 56} \nabla^j_1 \big[G_{\rho\rho}({\bf 16})\nabla^j_1 
\hat U*G_{\rho\rho}({\bf 15})\big] 
\nabla^l_2\big[\del({\bf 25})\nabla^l_2 U({\bf 26})\big] \nonumber \\
&=& -\nabla^j_1 \nabla^l_2 
\big[ \nabla^j_1 \hat U*G_{\rho\rho}({\bf 12})
\nabla^l_2 \hat G_{\rho\rho}*U ({\bf 12}) \big]
\nonumber \\
&=& -\nabla^j_1 \nabla^l_1 
\big[ \nabla^j_1 \hat U*G_{\rho\rho}({\bf 12})
\nabla^l_1 \hat G_{\rho\rho}*U ({\bf 12}) \big]
\la{eqn:b.5}
\eea
Adding up the above two terms we obtain
\bea
&&\Sigma^{(1)}_{\hat\rho \hat\rho}({\bf 12})
= \Sigma^{(1.a)}_{\hat\rho \hat\rho}({\bf 12})+
\Sigma^{(1.b)}_{\hat\rho \hat\rho}({\bf 12}) \nonumber \\
&=& 
-\nabla^j_1 \nabla^l_1 \big[ G_{\rho\rho}({\bf 12})
\nabla^l_1 \nabla^j_1 \hat U* \hat G_{\rho\rho}*U ({\bf 12}) \big] 
-\nabla^j_1 \nabla^l_1 \big[ \nabla^j_1 \hat U*G_{\rho\rho}({\bf 12})
\nabla^l_1 \hat G_{\rho\rho}*U ({\bf 12}) \big]
\la{eqn:b.6}
\eea
whose Fourier transform is given by
\be
\Sigma^{(1)}_{\hat\rho \hat\rho}({\bf k},t)
= - \int_{\bf q} \Big[ ({\bf k}\cdot {\bf q})^2 \,
U^2({\bf q})+ ({\bf k}\cdot {\bf q}) ({\bf k}\cdot ({\bf k}- {\bf q}))U({\bf q})
U({\bf k}- {\bf q})) \Big] G_{\rho \rho}({\bf q},t) G_{\rho \rho}({\bf k}-{\bf q},t)
\la{eqn:b.7}
\ee
which is the second member of (\ref{eqn:4.14}).
Next we have
\bea
\Sigma^{(2)}_{\hat\rho \hat\rho}({\bf 12})&=&
\frac{1}{2} \int_{\bf 3456} V^{int}_{\hat\rho \rho\rho}({\bf 134})
V^{id}_{\hat\rho \rho\rho}({\bf 256})G_{\rho\rho}({\bf 35})G_{\rho\rho}({\bf 46})
\nonumber \\
&=&\frac{1}{2} \int_{\bf 3456} 
(-i)\nabla_1 \cdot \big[\del({\bf 13})
\nabla_1 U({\bf 14})+\del({\bf 14})\nabla_1 U({\bf 13})\big]\nonumber \\
& \times & \Big(\frac{-iT}{\rho_0}\Big)\nabla^2_2\big[ \del({\bf 25})\del({\bf 26})\big]
G_{\rho\rho}({\bf 35})G_{\rho\rho}({\bf 46}) \nonumber \\
&=&\Big(-\frac{T}{\rho_0} \Big)\int_{\bf 34}
\nabla_1 \cdot \big[\del({\bf 13})\nabla_1 U({\bf 14})\big]
\nabla^2_2\big[G_{\rho\rho}({\bf 32})G_{\rho\rho}({\bf 42})\big]\nonumber \\
&=& \Big(-\frac{T}{\rho_0} \Big) \nabla_1 \cdot 
\nabla^2_2\big[ G_{\rho\rho}({\bf 12})\nabla_1 \hat U*G_{\rho\rho}({\bf 12})\big]
\nonumber \\
&=& \Big(-\frac{T}{\rho_0} \Big)\nabla^2_1 \nabla_1 \cdot
\big[ G_{\rho\rho}({\bf 12})\nabla_1 \hat U*G_{\rho\rho}({\bf 12})\big]
\la{eqn:b.8}
\eea
The Fourier transform of (\ref{eqn:b.8}) is given by
\be
\Sigma^{(2)}_{\hat\rho \hat\rho}({\bf k},t)=
-\frac{T}{\rho_0}k^2 \int_{\bf q} {\bf k}\cdot {\bf q}\,
U({\bf q})G_{\rho \rho}({\bf q},t) G_{\rho \rho}({\bf k}-{\bf q},t) 
\la{eqn:b.9}
\ee
which is the third member of (\ref{eqn:4.14}).

\bea
&&\Sigma^{(3)}_{\hat\rho \hat\rho}({\bf 12})
= \int_{\bf 3456} V^{int}_{\hat\rho \rho\rho}({\bf 134})
V_{\hat\rho \rho \hat\rho}({\bf 256})G_{\rho\rho}({\bf 35})G_{\rho\hat\rho}({\bf 46})
\nonumber \\
&=& \int_{\bf 3456} (-i)\nabla_1 \cdot \big[\del({\bf 13})
\nabla_1 U({\bf 14})+\del({\bf 14})\nabla_1 U({\bf 13})\big]\nonumber \\
& \times & (-2T)\nabla_2\del({\bf 25})\cdot [\nabla_6\del({\bf 56})]
G_{\rho\rho}({\bf 35})G_{\rho\hat\rho}({\bf 46})\nonumber \\
&=& 2iT \int_{\bf 3456}
\nabla_1 \cdot \big[\del({\bf 13})\nabla_1 U({\bf 14})\big]
\nabla_2\del({\bf 25})\cdot [\nabla_6\del({\bf 56})]\nonumber \\
& \times &\big[G_{\rho\rho}({\bf 35})G_{\rho \hat\rho}({\bf 46})
+G_{\rho\rho}({\bf 45})G_{\rho\hat\rho}({\bf 36})\big]
\la{eqn:b.10}
\eea
The first integral of (\ref{eqn:b.10}) is calculated as
\bea
&&\Sigma^{(3.a)}_{\hat\rho \hat\rho}({\bf 12})
=2iT\int_{\bf 3456}
\nabla^j_1 \big[\del({\bf 13})\nabla^j_1 U({\bf 14})\big]
\nabla^l_2\del({\bf 25}) [\nabla^l_6\del({\bf 56})] G_{\rho\rho}({\bf 35})
G_{\rho \hat\rho}({\bf 46})\nonumber \\
&=& -2iT \int_{\bf 345}
\nabla^j_1 \big[\del({\bf 13})\nabla^j_1 U({\bf 14})\big]
\nabla^l_2\del({\bf 25}) G_{\rho\rho}({\bf 35})
\nabla^l_5 G_{\rho \hat\rho}({\bf 45}) \nonumber \\
&=&-2iT\int_{\bf 34}
\nabla^j_1 \big[\del({\bf 13})\nabla^j_1 U({\bf 14})\big]
\nabla^l_2\big[ G_{\rho\rho}({\bf 32})
\nabla^l_2 G_{\rho \hat\rho}({\bf 42})\big] \nonumber \\
&=&-2iT \nabla^j_1 \nabla^l_2 \big[G_{\rho\rho}({\bf 12})
\nabla^l_2 \nabla^j_1 \hat U*G_{\rho \hat\rho}({\bf 12}) \big] \nonumber \\
&=&-2iT \nabla^j_1 \nabla^l_1 \big[G_{\rho\rho}({\bf 12})
\nabla^l_1 \nabla^j_1 \hat U*G_{\rho \hat\rho}({\bf 12}) \big]
\la{eqn:b.11}
\eea
The second integral of (\ref{eqn:b.10}) is similarly calculated as
\bea
&&\Sigma^{(3.b)}_{\hat\rho \hat\rho}({\bf 12})
=2iT\int_{\bf 3456}
\nabla^j_1 \big[\del({\bf 13})\nabla^j_1 U({\bf 14})\big]
\nabla^l_2\del({\bf 25})[\nabla^l_6\del({\bf 56})] G_{\rho \rho}({\bf 45})
G_{\rho \hat\rho}({\bf 36})\nonumber \\
&=& -2iT\int_{\bf 345}
\nabla^j_1 \big[\del({\bf 13})\nabla^j_1 U({\bf 14})\big]
\nabla^l_2\del({\bf 25}) G_{\rho\rho}({\bf 45})
\nabla^l_5 G_{\rho \hat\rho}({\bf 35}) \nonumber \\
&=& -2iT \int_{\bf 34}
\nabla^j_1 \big[\del({\bf 13})\nabla^j_1 U({\bf 14})\big]
\nabla^l_2\big[ G_{\rho\rho}({\bf 42})
\nabla^l_2 G_{\rho \hat\rho}({\bf 32})\big] \nonumber \\
&=&-2iT
\nabla^j_1 \nabla^l_2 \big[\nabla^j_1 \hat U*G_{\rho\rho}({\bf 12})
\nabla^l_2 G_{\rho \hat\rho}({\bf 12}) \big] \nonumber \\
&=&-2iT \nabla^j_1 \nabla^l_1 \big[\nabla^j_1 \hat U*G_{\rho\rho}({\bf 12})
\nabla^l_1 G_{\rho \hat\rho}({\bf 12}) \big]
\la{eqn:b.12}
\eea
Adding up (\ref{eqn:b.11})          and (\ref{eqn:b.12}), we obtain
\bea
&&\Sigma^{(3)}_{\hat\rho \hat\rho}({\bf 12})
=\Sigma^{(3.a)}_{\hat\rho \hat\rho}({\bf 12})
+\Sigma^{(3.b)}_{\hat\rho \hat\rho}({\bf 12}) \nonumber \\
&=&  -2iT \nabla^j_1 \nabla^l_1 \big[G_{\rho\rho}({\bf 12})
\nabla^l_1 \nabla^j_1 \hat U*G_{\rho \hat\rho}({\bf 12}) \big]
-2iT \nabla^j_1 \nabla^l_1 \big[\nabla^j_1 \hat U*G_{\rho\rho}({\bf 12})
\nabla^l_1 G_{\rho \hat\rho}({\bf 12}) \big]
\la{eqn:b.13}
\eea
The Fourier transform of (\ref{eqn:b.13}) is given by
\be
\Sigma^{(3)}_{\hat\rho \hat\rho}({\bf k},t)
= -2iT \int_{\bf q} \Big[ ({\bf k}\cdot {\bf q})^2 \,
U({\bf q})+ ({\bf k}\cdot {\bf q}) ({\bf k}\cdot ({\bf k}- {\bf q}))
U({\bf k}- {\bf q})) \Big] 
G_{\rho \hat\rho}({\bf q},t) G_{\rho \rho}({\bf k}-{\bf q},t)
\la{eqn:b.14}
\ee
This is the 4th member of (\ref{eqn:4.14}).

\bea
&&\Sigma^{(7)}_{\hat\rho \hat\rho}({\bf 12})=
\int_{\bf 3456} V^{int}_{\hat\rho \rho\rho}({\bf 134})
V_{\hat\rho \rho\theta}({\bf 256})G_{\rho\rho}({\bf 35})G_{\rho\theta}({\bf 46})
\nonumber \\
&=& \int_{\bf 3456} 
(-i)\nabla_1 \cdot \big[\del({\bf 13})
\nabla_1 U({\bf 14})+\del({\bf 14})\nabla_1 U({\bf 13})\big]
\cdot (-iT) \nabla^l_2\Big[ \del({\bf 25})\nabla^l_2 \del({\bf 26}) \Big]
G_{\rho\rho}({\bf 35})G_{\rho\theta}({\bf 46}) \nonumber \\
&=&(-T)\int_{\bf 3456} \nabla^j_1 \Big[\del({\bf 13})\nabla^j_1 U({\bf 14}) \Big]
\nabla^l_2\Big[ \del({\bf 25})\nabla^l_2 \del({\bf 26})
\Big[G_{\rho\rho}({\bf 35})G_{\rho\theta}({\bf 46})
+G_{\rho\rho}({\bf 45})G_{\rho\theta}({\bf 36}) \Big]
\la{eqn:b.15}
\eea
The first integral of (\ref{eqn:b.15}) is computed as
\bea
&&\Sigma^{(7.a)}_{\hat\rho \hat\rho}({\bf 12})
=(-T)\int_{\bf 3456} \nabla^j_1 \Big[\del({\bf 13})\nabla^j_1 U({\bf 14}) \Big]
\nabla^l_2\Big[ \del({\bf 25})\nabla^l_2 \del({\bf 26})\Big]
G_{\rho\rho}({\bf 35})G_{\rho\theta}({\bf 46}) \nonumber \\
&=& (-T)\int_{\bf 56}\, \nabla^j_1 \Big[ G_{\rho\rho}({\bf 15})
 \nabla^j_1 \hat U*G_{\rho\theta}({\bf 16}) \Big]
\nabla^l_2\Big[ \del({\bf 25})\nabla^l_2 \del({\bf 26})\Big] \nonumber \\
 &=& (-T)\nabla^j_1 \nabla^l_2 \Big[G_{\rho\rho}({\bf 12})
\nabla^l_2 \nabla^j_1 \hat U*G_{\rho\theta}({\bf 12}) \Big] \nonumber \\
&=& (-T)\nabla^j_1 \nabla^l_1 \Big[G_{\rho\rho}({\bf 12})
\nabla^l_1 \nabla^j_1 \hat U*G_{\rho\theta}({\bf 12}) \Big]
\la{eqn:b.16}
\eea
Likewise the second integral of (\ref{eqn:b.15}) is computed as
\bea
&&\Sigma^{(7.b)}_{\hat\rho \hat\rho}({\bf 12})
=(-T)\int_{\bf 3456} \nabla^j_1 \Big[\del({\bf 13})\nabla^j_1 U({\bf 14}) \Big]
\nabla^l_2\Big[ \del({\bf 25})\nabla^l_2 \del({\bf 26})\Big]
G_{\rho\rho}({\bf 45})G_{\rho\theta}({\bf 36}) \nonumber \\
&=&(-T) \nabla^j_1 \nabla^l_1 \Big[ \nabla^l_1 G_{\rho\theta}({\bf 12})
 \nabla^j_1 \hat U*G_{\rho\rho}({\bf 12}) \Big]
 \la{eqn:b.17}
\eea
Adding up (\ref{eqn:b.16}) and (\ref{eqn:b.17}), we obtain
\bea
&&\Sigma^{(7)}_{\hat\rho \hat\rho}({\bf 12})=
\Sigma^{(7.a)}_{\hat\rho \hat\rho}({\bf 12})+
\Sigma^{(7.b)}_{\hat\rho \hat\rho}({\bf 12}) \nonumber \\
&=& (-T)\nabla^j_1 \nabla^l_1 \Big[G_{\rho\rho}({\bf 12})
\nabla^l_1 \nabla^j_1 \hat U*G_{\rho\theta}({\bf 12}) \Big]
+(-T)\nabla^j_1 \nabla^l_1 \Big[ \nabla^l_1 G_{\rho\theta}({\bf 12})
 \nabla^j_1 \hat U*G_{\rho\rho}({\bf 12}) \Big]
 \la{eqn:b.18}
\eea
The Fourier transform of (\ref{eqn:b.18}) is given by
\be
\Sigma^{(7)}_{\hat\rho \hat\rho}({\bf k},t)
= -T \int_{\bf q} \Big[ ({\bf k}\cdot {\bf q})^2 \,
U({\bf q})+ ({\bf k}\cdot {\bf q}) ({\bf k}\cdot ({\bf k}- {\bf q}))
U({\bf k}- {\bf q})) \Big] 
G_{\rho \theta}({\bf q},t) G_{\rho \rho}({\bf k}-{\bf q},t)
\la{eqn:b.19}
\ee
This is the 5th member of (\ref{eqn:4.14}).

\bea
&&\Sigma^{(9)}_{\hat\rho \hat\rho}({\bf 12})=
\int_{\bf 3456} V_{\hat\rho \rho\theta}({\bf 134})
V^{int}_{\hat\rho \rho\rho}({\bf 256})
G_{\rho\rho}({\bf 35})G_{\theta \rho}({\bf 46})
\nonumber \\
&=& \int_{\bf 3456} 
(-iT) \nabla^l_1\Big[ \del({\bf 13})\nabla^l_1 \del({\bf 14}) \Big]
(-i)\nabla_2 \cdot \big[\del({\bf 25})
\nabla_2 U({\bf 26})+\del({\bf 26})\nabla_1 U({\bf 25})\big]
\cdot 
G_{\rho\rho}({\bf 35})G_{\theta\rho}({\bf 46}) \nonumber \\
&=&(-T)\int_{\bf 3456} \nabla^j_1 \Big[\del({\bf 13})\nabla^j_1 \del({\bf 14}) \Big]
\nabla^l_2\Big[ \del({\bf 25})\nabla^l_2 U({\bf 26})\Big]
\Big[G_{\rho\rho}({\bf 35})G_{\theta\rho}({\bf 46})
+G_{\rho\rho}({\bf 36})G_{\theta\rho}({\bf 45}) \Big]
\la{eqn:b.20}
\eea
The first integral of (\ref{eqn:b.20}) is calculated as
\bea
&& \Sigma^{(9.a)}_{\hat\rho \hat\rho}({\bf 12})=
(-T)\int_{\bf 3456} \nabla^j_1 \Big[\del({\bf 13})\nabla^j_1 \del({\bf 14}) \Big]
\nabla^l_2\Big[ \del({\bf 25})\nabla^l_2 U({\bf 26})\Big]
G_{\rho\rho}({\bf 35})G_{\theta\rho}({\bf 46}) \nonumber \\
&=& (-T)\int_{\bf 56} 
\nabla^j_1 \Big[G_{\rho\rho}({\bf 15}) \nabla^j_1 G_{\theta\rho}({\bf 16}) \Big]
\nabla^l_2\Big[ \del({\bf 25})\nabla^l_2 U({\bf 26})\Big] \nonumber \\
&=& (-T) \nabla^j_1 \nabla^l_2 \Big[ G_{\rho\rho}({\bf 12})
 \nabla^l_2 \nabla^j_1 \hat G_{\theta\rho}*U({\bf 12}) \Big] \nonumber \\
&=& (-T) \nabla^j_1 \nabla^l_1 \Big[ G_{\rho\rho}({\bf 12})
\nabla^l_1 \nabla^j_1 \hat G_{\theta\rho}*U({\bf 12}) \Big] 
\la{eqn:b.21}
\eea
The second integral of (\ref{eqn:b.20}) is calculated as 
\bea
&& \Sigma^{(9.b)}_{\hat\rho \hat\rho}({\bf 12})=
(-T)\int_{\bf 3456} \nabla^j_1 \Big[\del({\bf 13})\nabla^j_1 \del({\bf 14}) \Big]
\nabla^l_2\Big[ \del({\bf 25})\nabla^l_2 U({\bf 26})\Big]
G_{\rho\rho}({\bf 36})G_{\theta\rho}({\bf 45}) \nonumber \\
&=& (-T)\int_{\bf 56} 
\nabla^j_1 \Big[G_{\rho\rho}({\bf 16}) \nabla^j_1 G_{\theta\rho}({\bf 15}) \Big]
\nabla^l_2\Big[ \del({\bf 25})\nabla^l_2 U({\bf 26})\Big] \nonumber \\
&=& (-T) \nabla^j_1 \nabla^l_2 
\Big[   \nabla^j_1 G_{\theta\rho}({\bf 12})
\nabla^l_2 \hat G_{\rho\rho}*U({\bf 12}) \Big] \nonumber \\
&=& (-T) \nabla^j_1 \nabla^l_1 
\Big[   \nabla^j_1 G_{\theta\rho}({\bf 12})
\nabla^l_1 \hat G_{\rho\rho}*U({\bf 12}) \Big]
\la{eqn:b.22}
\eea
Adding up (\ref{eqn:b.21}) and (\ref{eqn:b.22}) leads to
\bea
&& \Sigma^{(9)}_{\hat\rho \hat\rho}({\bf 12})=
 \Sigma^{(9.a)}_{\hat\rho \hat\rho}({\bf 12})+
\Sigma^{(9.b)}_{\hat\rho \hat\rho}({\bf 12}) \nonumber \\
&=& (-T) \nabla^j_1 \nabla^l_1 \Big[ G_{\rho\rho}({\bf 12})
\nabla^l_1 \nabla^j_1 \hat G_{\theta\rho}*U({\bf 12}) \Big]
+(-T) \nabla^j_1 \nabla^l_1 
\Big[   \nabla^j_1 G_{\theta\rho}({\bf 12})
\nabla^l_1 \hat G_{\rho\rho}*U({\bf 12}) \Big]
\la{eqn:b.23}
\eea
The Fourier transform of (\ref{eqn:b.23}) is given by
\be
\Sigma^{(9)}_{\hat\rho \hat\rho}({\bf k},t)
= -T \int_{\bf q} \Big[ ({\bf k}\cdot {\bf q})^2 \,
U({\bf q})+ ({\bf k}\cdot {\bf q}) ({\bf k}\cdot ({\bf k}- {\bf q}))
U({\bf k}- {\bf q})) \Big] 
G_{ \theta \rho }({\bf q},t) G_{\rho \rho}({\bf k}-{\bf q},t)
\la{eqn:b.24}
\ee
This completes the derivation of (\ref{eqn:4.14}).
We move on to the derivation of (\ref{eqn:4.18}).

\subsection{Derivation of (\ref{eqn:4.18})}
We start with $\Sigma^{(3)}_{\hat\rho \hat\theta}({\bf 12})$, (\ref{eqn:4.4}).
\bea
&&\Sigma^{(3)}_{\hat\rho \hat\theta}({\bf 12}) 
=\int_{{\bf 3456}} 
V_{\hat\rho \rho\theta}({\bf 134}) V_{\hat\theta \rho\rho}({\bf 256})
G_{\rho\rho}({\bf 35}) G_{\theta\rho}({\bf 46}) \nonumber \\
&=& \int_{{\bf 3456}} (-iT)\nabla^j_1 \Big[\del({\bf 13}) \nabla^j_1 \del({\bf 14})
  \Big] \cdot \frac{i}{\rho^2_0} \del({\bf 25})\del({\bf 26})
  G_{\rho\rho}({\bf 35})G_{\theta\rho}({\bf 46}) \nonumber \\
&=& \frac{T}{\rho^2_0} \int_{{\bf 34}} 
\nabla^j_1 \Big[\del({\bf 13}) \nabla^j_1 \del({\bf 14}) \Big]  
G_{\rho\rho}({\bf 32})G_{\theta\rho}({\bf 42}) \nonumber \\
  &=& \frac{T}{\rho^2_0} \nabla_1 \cdot 
\Big[ G_{\rho\rho}({\bf 12})\nabla_1 G_{\theta\rho}({\bf 12})  \Big]
\la{eqn:b.25}
\eea
The Fourier transform of (\ref{eqn:b.25}) is given by
\be
\Sigma^{(3)}_{\hat\rho \hat\theta}({\bf k},t)
= -\frac{T}{\rho_0^2} \int_{\bf q} {\bf k}\cdot {\bf q}\,
G_{\theta \rho}({\bf q},t) G_{\rho \rho}({\bf k}-{\bf q},t) 
\la{eqn:b.26}
\ee
which is the first member of (\ref{eqn:4.18}).

The qauntity $\Sigma^{(10)}_{\hat\rho \hat\rho}({\bf 12})$
can be obtained from $\Sigma^{(9)}_{\hat\rho \hat\rho}({\bf 12})$, (\ref{eqn:b.4}), 
since the vertex $V^{id}_{\hat\rho \rho\rho}({\bf 256})$ is obtained from 
$ V^{int}_{\hat\rho \rho\rho}({\bf 256})$ with $U({\bf 12})$ being 
replaced by $(T/\rho_0)\del({\bf 12})$. 
We thus have 
\bea
&&\Sigma^{(10)}_{\hat\rho \hat\rho}({\bf k},t)=
\Big[\Sigma^{(9)}_{\hat\rho \hat\rho}({\bf k},t)\Big]_{U({\bf q})
=U({\bf k}-{\bf q})=\frac{T}{\rho_0}} \nonumber \\
&=& -\frac{T^2}{\rho_0} \int_{\bf q} \Big[ ({\bf k}\cdot {\bf q})^2 \,
+ ({\bf k}\cdot {\bf q}) ({\bf k}\cdot ({\bf k}- {\bf q})) \Big] 
G_{ \theta \rho }({\bf q},t) G_{\rho \rho}({\bf k}-{\bf q},t) \nonumber \\
&=& -\frac{T^2}{\rho_0}k^2 \int_{\bf q} {\bf k}\cdot {\bf q} \,
G_{ \theta \rho }({\bf q},t) G_{\rho \rho}({\bf k}-{\bf q},t)
\la{eqn:b.27}
\eea 
which is the second member of (\ref{eqn:4.18}).

From (\ref{eqn:4.6})
\bea
&&\Sigma^{(11)}_{\hat\rho \hat\rho}({\bf 12})
=\int_{{\bf 3456}} V_{\hat\rho \rho\theta}({\bf 134})
V_{\hat\rho \rho\theta}({\bf 256})G_{\rho\rho}({\bf 35})
G_{ \theta \theta }({\bf 46}) \nonumber \\
&=& \int_{{\bf 3456}} (-iT) \nabla^j_1 \Big[\del({\bf 13})
\nabla^j_1 \del({\bf 14}) \Big] (-iT) \nabla^l_2 \Big[\del({\bf 25})
\nabla^l_2 \del({\bf 26}) \Big] G_{\rho\rho}({\bf 35})
G_{ \theta \theta }({\bf 46}) \nonumber \\
&=& (-T^2) \int_{{\bf 34}}\nabla^j_1 \Big[\del({\bf 13})
\nabla^j_1 \del({\bf 14}) \Big] \nabla^l_2 
\Big[ G_{\rho\rho}({\bf 32}) \nabla^l_2 G_{ \theta \theta }({\bf 42}) \Big]
\nonumber \\
&=& (-T^2) \nabla^j_1 \nabla^l_1 \Big[ G_{\rho\rho}({\bf 12}) 
\nabla^j_1 \nabla^l_1 G_{ \theta \theta }({\bf 12})\Big] 
\la{eqn:b.28}
\eea
whose Fourier transform is given by
\be
\Sigma^{(11)}_{\hat\rho \hat\rho}({\bf k},t)
= - T^2 \int_{\bf q} ({\bf k}\cdot {\bf q})^2\,
G_{\theta \theta}({\bf q},t) G_{\rho \rho}({\bf k}-{\bf q},t) 
\la{eqn:b.29}
\ee
This is the third member of (\ref{eqn:4.18}).

Again from (\ref{eqn:4.6})
\bea
&&\Sigma^{(12)}_{\hat\rho \hat\rho}({\bf 12})
=\int_{{\bf 3456}} V_{\hat\rho \rho\theta}({\bf 134})
V_{\hat\rho \theta\rho}({\bf 256})G_{\rho\theta}({\bf 35})
G_{ \theta \rho }({\bf 46}) \nonumber \\
&=& \int_{{\bf 3456}} V_{\hat\rho \rho\theta}({\bf 134})
V_{\hat\rho \rho\theta}({\bf 256})G_{ \rho \theta }({\bf 36})
G_{ \theta \rho }({\bf 45}) \nonumber \\
&=& (-T^2)  \int_{{\bf 3456}}  \nabla^j_1 \Big[\del({\bf 13})
\nabla^j_1 \del({\bf 14}) \Big]  \nabla^l_2 \Big[\del({\bf 25})
\nabla^l_2 \del({\bf 26}) \Big] G_{ \rho \theta }({\bf 36})
G_{ \theta \rho }({\bf 45}) \nonumber \\
&=& (-T^2)  \int_{{\bf 34}}  \nabla^j_1 \Big[\del({\bf 13})
\nabla^j_1 \del({\bf 14}) \Big]  
\nabla^l_2 \Big[ G_{\theta \rho }({\bf 42}) 
\nabla^l_2 G_{ \rho \theta }({\bf 32})\Big] \nonumber \\
&=& (-T^2) \nabla^j_1 \nabla^l_1 \Big[\nabla^j_1 G_{\theta\rho}({\bf 12}) 
 \nabla^l_1 G_{ \rho \theta }({\bf 12})\Big] 
 \la{eqn:b.30}
\eea
whose Fourier transform is given by
\be
\Sigma^{(12)}_{\hat\rho \hat\rho}({\bf k},t)
= - T^2 \int_{\bf q} ({\bf k}\cdot {\bf q}) ({\bf k}\cdot ({\bf k}- {\bf q}))
G_{\theta \rho}({\bf q},t) G_{\rho \theta}({\bf k}-{\bf q},t)
\la{eqn:b.31}
\ee
which is the 4th member of (\ref{eqn:4.18}).

\bea
\Sigma^{(13)}_{\hat\rho\hat\rho}({\bf 12})&=&
\int_{{\bf 3456}} V_{\hat\rho\rho\theta}({\bf 134})
V_{\hat\rho\rho \hat\rho}({\bf 256})
G_{\rho\rho}({\bf 35}) G_{\theta \hat\rho}({\bf 46}) \nonumber \\
&=& \int_{{\bf 3456}}(-iT) \nabla^j_1 \Big[\del({\bf 13})
\nabla^j_1 \del({\bf 14}) \Big] \cdot (-2T) \nabla^l_2 \del({\bf 25})
\nabla^l_6\del({\bf 56}) G_{\rho \rho }({\bf 35})G_{\theta \hat\rho }({\bf 46}) 
\nonumber \\
&=& (-2iT^2) \int_{{\bf 34}} \nabla^j_1 \Big[\del({\bf 13})
\nabla^j_1 \del({\bf 14}) \Big] \nabla^l_2 \Big[ 
  G_{\rho \rho }({\bf 32}) \nabla^l_2 G_{\theta \hat\rho }({\bf 42})\Big] 
  \nonumber \\
  &=& (-2iT^2) \nabla^j_1 \nabla^l_1 \Big[ G_{\rho \rho }({\bf 12})
 \nabla^j_1 \nabla^l_1 G_{\theta \hat\rho }({\bf 12})\Big]
 \la{eqn:b.32}
\eea
The Fourier transform of (\ref{eqn:b.32}) is given by
\be
\Sigma^{(13)}_{\hat\rho \hat\rho}({\bf k},t)
= - 2iT^2 \int_{\bf q} ({\bf k}\cdot {\bf q})^2\,
G_{\theta \hat\rho}({\bf q},t) G_{\rho \rho}({\bf k}-{\bf q},t)
\la{eqn:b.33}
\ee
which is the 5th member of (\ref{eqn:4.18}).

Finally, 
\bea
\Sigma^{(14)}_{\hat\rho\hat\rho}({\bf 12})&=&
\int_{{\bf 3456}} V_{\hat\rho \rho\theta}({\bf 134})
V_{\hat\rho \hat\rho \rho }({\bf 256})
G_{\rho \hat\rho}({\bf 35}) G_{\theta \rho}({\bf 46}) \nonumber \\
&=& \int_{{\bf 3456}} V_{\hat\rho \rho\theta}({\bf 134})
V_{\hat\rho \rho \hat\rho  }({\bf 256}) 
G_{\rho \hat\rho}({\bf 36}) G_{\theta \rho}({\bf 45}) \nonumber \\
&=& \int_{{\bf 3456}}(-iT) \nabla^j_1 \Big[\del({\bf 13})
\nabla^j_1 \del({\bf 14}) \Big] \cdot (-2T) \Big[\nabla^l_2 \del({\bf 25})
\nabla^l_6\del({\bf 56})\Big] G_{\rho \hat\rho}({\bf 36}) G_{\theta \rho}({\bf 45}) 
\nonumber \\
&=& (-2iT^2) \int_{{\bf 34}} \nabla^j_1 \Big[\del({\bf 13})
\nabla^j_1 \del({\bf 14}) \Big] \nabla^l_2 \Big[ 
  G_{\theta \rho }({\bf 42}) \nabla^l_2 G_{\rho \hat\rho }({\bf 32})\Big] 
  \nonumber \\
  &=& (-2iT^2) \nabla^j_1 \nabla^l_1 \Big[ \nabla^l_1 G_{\rho \hat\rho }({\bf 12})
 \nabla^j_1  G_{\theta \rho }({\bf 12})\Big]
 \la{eqn:b.34}
\eea
The Fourier transform of (\ref{eqn:b.34}) is given by
\be
\Sigma^{(14)}_{\hat\rho \hat\rho}({\bf k},t)
= - 2i T^2 \int_{\bf q} ({\bf k}\cdot {\bf q}) ({\bf k}\cdot 
({\bf k}- {\bf q}))G_{\theta \rho}({\bf q},t) 
G_{\rho \hat\rho}({\bf k}-{\bf q},t) 
\la{eqn:b.35}
\ee
which is the last member of (\ref{eqn:4.18}).
This completes the derivation of (\ref{eqn:4.18}).

\acknowledgements
KK was supported by Grant-in-Aid for Scientific Research (C) Grant 17540366 by Japan Society for the Promotion 
of Science (JSPS). BK thanks Prof. Fumio Hirata, Prof. Song-Ho Chong, and other group members for warm 
hospitality and many valuable discussions on the liquid dynamics during his sabbatical stay at the Institute 
for Molecular Science. 
BK acknowledges financial support by the Institute for Molecular Science.

%\end{document}

%\newpage
\centerline {\bf FIGURE CAPTIONS}
 
\renewcommand{\theenumi}{Figure~1}
\begin{enumerate}
\item
Diagrammatic expression for $W_2[J]$ ((\ref{eqn:3.30})). The line denotes 
the propagator ${\cal G}(\psi_c)$, and the crossing points with $3$-branching lines 
and with $4$-branching lines denote respectively the vertices $S^{(3)}_c$ 
and $S^{(4)}_c$. While the first two diagrams are 1PI diagrams, the last one is 1PR diagram.
\end{enumerate}

\renewcommand{\theenumi}{Figure~2}
\begin{enumerate}
\item
Two and three loop diagrams for $\Gamma_{1PI}[\phi]$. The line denotes 
the propagator ${\cal G}(\phi)$, and the crossing points with $3$-branching lines 
and with $4$-branching lines denote respectively the vertices $S^{(3)}(\phi)$ 
and $S^{(4)}(\phi)$. The two loop diagrams are both 1PI and 2PI diagrams.
Other 2PI diagrams are the 3rd, 5th, and 6th diagrams.
\end{enumerate}

\renewcommand{\theenumi}{Figure~3}
\begin{enumerate}
\item
Two and three loop diagrams for $\Gamma_{2PI}[\phi,G]$. 
Via the double Legendre transform the 2PR diagrams in Fig.~2 are eliminated in $\Gamma_{2PI}[\phi,G]$.
Here the line denotes the full propagator $G$.
In all figures, the vertices with more than $4$-legs are not shown. 
These higher vertices do not contribute to the two-loop results for both $\Gamma_{1PI}[\phi]$ and 
$\Gamma_{2PI}[\phi]$, and hence do not contribute to the one-loop result for the self-energy.
\end{enumerate}

\renewcommand{\theenumi}{Figure~4}
\begin{enumerate}
\item
One and two loop diagrams for the self-energy $\Sigma({\bf 12})$.
 These diagrams are obtained from those in Fig.~3 by cutting a single line.
\end{enumerate}

\renewcommand{\theenumi}{Figure~5}
\begin{enumerate}
\item
The one-loop diagram for $\Sigma_{\hat\theta \hat\theta}({\bf 12})$.
\end{enumerate}

\renewcommand{\theenumi}{Figure~6}
\begin{enumerate}
\item
The one-loop diagrams for $\Sigma_{\hat\rho \hat\theta}({\bf 12})$. 
The filled circle in the vertex $\hat\rho\rho\rho$ denotes the vertex
$V^{int}_{\hat\rho \rho\rho}$, which is to be distinguished from the vertex
$V^{id}_{\hat\rho \rho\rho}$ (without filled circle)
\end{enumerate}

\renewcommand{\theenumi}{Figure~7}
\begin{enumerate}
\item
The one-loop diagrams for $\Sigma_{ \hat\theta \hat\rho}({\bf 12})$.
\end{enumerate}

\renewcommand{\theenumi}{Figure~8}
\begin{enumerate}
\item
The one-loop diagrams for $\Sigma_{ \hat\rho \hat\rho}({\bf 12})$.
\end{enumerate}


\begin{thebibliography}{99}
\bibitem{ideal} Throughout the paper
the term ''ideal gas" is meant only for static behavior.
Dynamics for our system is of course different from the ideal gas
behavior even without inter-particle interactions.
\bibitem{ideal2} In \cite{br07} which we saw after submission of our paper, a statement is 
made that our demonstration of the diffusion law for non-interacting systems ignores vertex corrections. 
However our demonstration  here is a general result independent of loop expansion.
\bibitem{free}  We remind that separation into
Gaussian and non-Gaussian parts do not coincide
for the action $\cal S$ and for the free energy functional $F[\rho]$.
\bibitem{onepi} One-particle reducible (irreducible) diagram is the one which  can (cannot) be disconnected by 
cutting single line in the diagram.
\bibitem{note2} If $F[\rho]$ were independent of $T$, $\lambda$ can be identified with $T$ by change of variable as 
$\hat\rho\rightarrow \hat\rho/T$. However, this does not work since $F_{id}[\rho]$ depends on $T$ which brings in complications. 
If we can find some new non-perturbative results involving ideal gas such as the one
 we met around (\ref{eqn:3.26}), that would be most helpful to discover smallness parameter (parameters).
\end{thebibliography}
\end{document}